\documentclass[prc,reprint, superscriptaddress,twocolumn,showpacs]{revtex4-1}
\usepackage{amsmath,amssymb}
\usepackage[colorlinks=true,citecolor=blue,urlcolor=blue,linkcolor=blue]{hyperref}
\usepackage{graphicx}

\usepackage{color}
\newcommand{\colour}[1]{}

\begin{document}

\title{Shell Model States in the Continuum}
\author{A. M. Shirokov}
\affiliation{Skobeltsyn Institute of Nuclear Physics, Lomonosov Moscow State University, 
 Moscow 119991, Russia}
\affiliation{Department of Physics and Astronomy, Iowa State University, Ames, Iowa 50011, USA}
\affiliation{Department of Physics, Pacific National University, Khabarovsk 680035, Russia}
\author{A. I. Mazur}
\affiliation{Department of Physics, Pacific National University, Khabarovsk 680035, Russia}
\author{I. A. Mazur}
\affiliation{Department of Physics, Pacific National University, Khabarovsk 680035, Russia}
\author{J. P. Vary}
\affiliation{Department of Physics and Astronomy, Iowa State University, Ames, Iowa 50011, USA}

\begin{abstract}
We suggest a method for calculating scattering phase shifts and energies and widths of resonances which utilizes only eigenenergies obtained in variational calculations with oscillator basis and their dependence on oscillator basis spacing~$\hbar\Omega$. We make use of simple 
expressions for the $S$-matrix at
eigenstates of a finite (truncated) Hamiltonian matrix in the oscillator basis obtained in the HORSE ($J$-matrix) formalism of quantum
scattering theory. The validity of the suggested approach is verified in calculations with model Woods--Saxon potentials and applied to calculations of  $n\alpha$ resonances and non-resonant scattering using the no-core shell model.
\end{abstract}

\pacs{21.10.Tg, 21.60.Cs, 21.60.De, 24.10.-i, 24.30.Gd, 25.40.Dn, 27.10.+h}

\maketitle

\section{Introduction}

 To calculate energies of  nuclear ground states and other bound states within various shell model approaches, 
one conventionally starts 
by calculating the $\hbar \Omega$-dependence of the energy 
$E_{\nu}(\hbar\Omega)$ of the bound state $\nu$
in some model space. The minimum of~$E_{\nu}(\hbar\Omega)$ is correlated with the 
energy of the  state $\nu$. The convergence of calculations and accuracy of the energy prediction is  estimated
by comparing with the results obtained in neighboring model spaces. To improve the accuracy of theoretical predictions,
various extrapolation techniques have been suggested 
recently~{\colour{red}\cite{VMS,Coon, CoonNTSE12,Dick,More, CoonNTSE13,IT-extrap,Forssen2014, Furnstahl-S-matrix, Furnstahl14, Furnstahl15, Furnstahl-HH, Sid16}} 
which make it possible to
estimate the binding energies in the complete infinite shell-model basis space. The studies of extrapolations to the infinite model spaces 
reveal general trends  of convergence patterns of shell model calculations.

Is it possible to study nuclear states in the continuum, low-energy scattering and
resonant states in particular, in the shell model using bound state techniques? A 
conventional belief is that the energies of shell-model states in the continuum should be associated with the resonance
energies. It was shown however in Ref.~\cite{PRC79, ApplMathInfSci} that the energies of shell-model states may appear well above the
energies of resonant states, especially for broad resonances. Moreover, 
the analysis of Refs.~\cite{PRC79,ApplMathInfSci} clearly 
demonstrated that the shell model should also generate some states in a non-resonant nuclear continuum. The 
nuclear resonance properties 
can be studied in the Gamow  shell model, including the {\em ab initio} no-core Gamow  shell model (NCGSM)~\cite{JimmyNTSE13,Jimmy}. 
Another option is to combine the shell model with resonating group method (RGM). An impressive progress in the
description of various nuclear reactions was achieved by means of the combined no-core shell model/RGM (NCSM/RGM) 
approach~\cite{NavrJPG09,NavratilNTSE13,QuagNav1, Vary2013,QuagNav2, PhysScr}. 
Both NCGSM and NCSM/RGM complicate essentially the shell model calculations. Is it possible
to get some information about unbound nuclear states directly from the results of calculations in
NCSM or in other versions of the nuclear shell model without introducing additional Berggren basis states as in NCGSM or 
additional RGM calculations as in the NCSM/RGM approach?

The general
behaviour of shell model eigenstates at positive energies (or just at the energies above various thresholds) is not well-studied and there is no
well-established extrapolation technique to the infinite basis space for resonances. Generally, a 
 complete study of the nuclear continuum can be performed by extending the nuclear shell model with the
$J$-matrix formalism of scattering theory. The $J$-matrix formalism has been suggested in atomic physics~\cite{Hell,YaFi}.
Later it was independently rediscovered in nuclear physics~\cite{Fill,NeSm} and was successfully used in shell-model
applications~\cite {M1}. The $J$-matrix approach utilizes diagonalization of the Hamiltonian in one of two bases: the so-called
Laguerre basis that is of a particular interest for atomic physics applications and the oscillator basis that is appropriate for nuclear physics.
The version of the $J$-matrix formalism with the oscillator basis is also sometimes referred to as an Algebraic Version of RGM~\cite{Fill} or
as a HORSE (Harmonic Oscillator Representation of Scattering Equations) method~\cite{Bang}~--- we shall use the latter
nomenclature in what follows.

We note that a direct implementation of the HORSE formalism in modern large-scale
shell-model calculations is very complicated and unpractical: the HORSE\linebreak 
method requires calculation of a huge number of eigenstates while
modern shell-model codes usually utilize the Lanczos algorithm which provides only the few lowest Hamiltonian eigenstates.
Furthermore, the HORSE method needs also the weight of the highest component of the 
wave function of each eigenstate which is usually obtained with a low precision. On the other hand, the HORSE 
formalism can be used for a simple calculation of the scattering phase shift or $S$-matrix at a single energy $E_{\nu}(\hbar\Omega)$ 
which is an eigenstate of the shell-model Hamiltonian. In this case, the HORSE phase shift calculation requires only the value of the 
energy~$E_{\nu}(\hbar\Omega)$ and the basis parameters (the $\hbar\Omega$ value and the basis truncation). We shall refer to such a simplified approach as a Single State HORSE (SS-HORSE)
method. Varying the shell-model parameter $\hbar\Omega$ and using results from a set of basis spaces, 
we generate a variation of~$E_{\nu}(\hbar\Omega)$ in some energy range and hence we can calculate the phase shifts in that energy range. 

Calculations of scattering phase shifts at the eigenenergies of the Hamiltonian in the oscillator basis and obtaining the phase shift energy dependence by variation of basis parameters, was recently performed in Ref.~\cite{More} using another (not the HORSE) technique. A detailed study of
scattering phase shifts at eigenenergies of the Hamiltonian in arbitrary finite~$\mathcal{L}^{2}$ basis was performed in Ref.~\cite{Rub1}. This study was based on the theory of spectral shift functions introduced by I.~M.~Lifshitz 
nearly 70~years ago~\cite{Lif} and later forgotten by physicists though used up to now by mathematicians  (see Ref.~\cite{Rub1} and references therein).

Another method to obtain scattering phase shifts from bound state calculations in a harmonic oscillator basis features the use of an additional 
harmonic oscillator potential~\cite{Luu}.  The method was demonstrated with nucleon-nucleon scattering where it reveals a challenge of needing 
a large basis to access the low-energy scattering region.

It is worth noting here that approximate resonant widths can be extracted from 
bound state approaches to many-body nuclear systems using a relation between the partial width in a specified breakup channel and an integral 
over the ``interaction region'' where all of the nucleons are close to each other. This method was described in detail in Ref.~\cite{Nollett}
where it was used to evaluate widths of resonances in light nuclei based on the variational Monte Carlo calculations. It has been used before
in combination with other many-body approaches (see Ref.~\cite{Nollett} for the list of respective references), in particular, it can be utilized
within the nuclear shell model {\colour{red}\cite{Timofeyuk}}. 
However this approach is applicable to narrow enough resonances only and is unable to provide  information
about non-resonant scattering.

In this contribution, we suggest a simpler and more powerful approach. 
We \mbox{formulate} below a method for calculating low-energy phase shifts and for 
extracting resonant energies~$E_{r}$ and widths~$\Gamma$  from the shell model results, or, generally,
from results of any variational calculation with a finite oscillator basis. We
apply the SS-{HORSE } formalism to calculate the $S$-matrix in the energy interval of variation of one of the Hamiltonian
eigenenergies~$E_{\nu}(\hbar\Omega)$ due to variation of~$\hbar\Omega$ and truncation boundary of the Hamiltonian matrix. We use
either a low-energy expansion of the $S$-matrix or
express the $S$-matrix as a pole term plus slowly varying with energy background terms and fit the expansion parameters to describe
the $S$-matrix behaviour in the above energy interval. The low-energy phase shifts~$\delta_{\ell}$, the
resonant energy~$E_{r}$ and width~$\Gamma$  appear as a result of this fit.
We obtain relations describing the general behaviour of shell-model states associated with a resonance or with a non-resonant continuum as functions 
of~$\hbar\Omega$ and truncation boundary of the Hamiltonian matrix. This approach is tested in calculations of 
phase shifts and resonance parameters
of two-body scattering with model potential. Next we apply the SS-HORSE method to the calculation of resonances 
and of non-resonant continuum in the 
neutron-$\alpha$ scattering based on No-core Shell Model (NCSM) results obtained with the JISP16 $NN$ interaction~\cite {PLB644,JISP16-web}.

In our earlier study~\cite{RP2013}, we evaluated resonant energies~$E_{r}$ and widths~$\Gamma$ using the SS-{HORSE}
and Breit--Wigner formula for the description of resonances. The Breit--Wigner formula describes the phase shifts and $S$-matrix 
only in the case of narrow resonances and only in a narrow energy interval in the vicinity of the resonance. As a result, the approach
of Ref.~\cite{RP2013} can be used only in rare cases when the eigenenergies of the truncated Hamiltonian are obtained very close
to the resonant energy~$E_{r}$ and cannot provide an accurate description of resonant parameters even in these rare cases.
This drawback is eliminated in the current study.

{\colour{red}We present here an {ab initio} study of the neutron-$\alpha$ elastic scattering within the {NCSM-SS-HORSE} approach using
the JISP16 $NN$ interaction which was shown~\cite{NNandNNN} to provide a good description of $s$- and $p$-shell nuclei. 
{\em Ab initio} studies of the same reaction with various other modern inter-nucleon interactions  were performed within Quantum 
Monte Carlo approach in Ref.~\cite{QMC} and within the NCSM/RGM in Refs.~\cite{NavQua09,NavQua10,NavQua13,PhysScr}.
}

The paper is organized as follows. We present in Section~\ref{SS-HORSEtheory} the basic relations of the HORSE formalism, derive the SS-HORSE method and
present all equations needed to calculate phase shifts, $S$-matrix and resonant parameters~$E_{r}$ and~$\Gamma$. The SS-HORSE
approach to the calculation of resonant energy and width is verified in Section~\ref{modelproblem} using a two-body scattering with a model potential. Section~\ref{sec:5He}
is devoted to calculations of resonances in $n\alpha$ scattering based on NCSM calculations of $^{5}$He  with
JISP16 $NN$ interaction. Conclusions are presented in Section~\ref{sec:Conclusions}.

\section{SS-HORSE approach to calculation of low-energy scattering and resonant parameters\label{SS-HORSEtheory}}
\subsection{HORSE formalism}
The $J$-matrix approach and HORSE in particular are widely used in various applications. Some of the recent applications together with
pioneering papers where  the $J$-matrix has been suggested, can be found in the book~\cite{Arab}. We sketch here the basic relations
and ideas of the
HORSE formalism for the two-body single-channel scattering following our papers~\cite{true,trueZ,Bang, PRC04}.

The radial wave function $u_{\ell}(k,r)$ describing the relative motion in the partial wave with orbital momentum $\ell$
is expanded within the HORSE formalism in an infinite series of radial oscillator functions $R_{N\ell}(r)$,
  \begin{equation}
  u_{\ell}(k,r) = \sum_{N=N_0,N_{0}+2,...,\infty} a_{N\ell}(k)\, R_{N\ell}(r) , 
  \label{row}
  \end{equation}
  where
 \begin{multline}
 R_{N\ell}(r)=(-1)^{(N-\ell)/2} \,
\sqrt{ \frac{2 \Gamma (N/2-\ell/2+1)}{r_0
\Gamma(N/2 +\ell/2+3/2) } }  \\
           \times    \left(\frac{r}{r_0}\right)^{\!\ell+1} \! \exp\!{\left(-\frac {r^2}{2r_0^2}\right)} \,
        L_{(N-\ell)/2}^{\ell+\frac12}\!\left(\frac {r^2}{r_0^2}\right)\!.
\label{e24}
\end{multline}
Here $k$ is the relative motion momentum, $L_n^{\alpha}(z)$ are associated Laguerre polynomials,
the oscillator radius~$r_{0}=\sqrt{\frac{\hbar}{m\Omega}}$, $m$ is the reduced mass of colliding particles, 
$\hbar\Omega$ is the oscillator level spacing,
$N=2n+\ell$ is the oscillator quanta while $n$ is the oscillator principal quantum number, the minimal value of
oscillator quanta~$N_{0}=\ell$.
%
Using the expansion~\eqref{row}
we transform the radial Schr\"odinger equation
 \begin{equation}
  H^\ell \,u_{\ell}(k,r)=E\,u_{\ell}(k,r)
  \label{Sh_1ch}
  \end{equation}
into an infinite set of linear algebraic equations,
\begin{multline}
  \sum_{N'=N_0,N_{0}+2,...,\infty}(H_{NN'}^{\ell}-\delta_{NN'}E)\, a_{N'\ell}(k)=0 , \\ 
   N=N_0, N_0+2, ...\, ,  
  \label{Infsys}
  \end{multline}
where  
$H_{NN'}^{\ell} =T_{NN'}^{\ell} + V_{NN'}^{\ell}$ are matrix elements of the Hamiltonian $H^{\ell}$ in the oscillator basis, and
$T_{NN'}^{\ell}$ and $V_{NN'}^{\ell}$ are kinetic and potential energy matrix elements respectively. 
 
The kinetic energy matrix elements $T_{NN'}^{\ell}$ are known to form a tridiagonal matrix, i.\:e., the only non-zero
matrix elements are
  \begin{subequations}
  \label{Tnn}
  \begin{align}
  \label{Tnnd}
  T_{NN}^{\ell}& = \frac{1}{2}\hbar\Omega(N+3/2), \\ 
  T_{N,N+2}^{\ell}& =T_{N+2,N}^{\ell}= - \frac{1}{4}\hbar\Omega\sqrt{(N-\ell+2)(N+\ell+3)}.
  \label{Tnn2}
  \end{align}
  \end{subequations}
These matrix elements are seen to increase linearly with $N$ for large $N$. On the other hand, the potential energy 
matrix elements $V_{NN'}^{\ell}$ decrease as $N,N' \to \infty$. Hence the kinetic energy dominates in 
the Hamiltonian matrix at large enough $N$ 
and/or~$N'$. Therefore a reasonable approximation is to truncate the 
potential energy matrix at large $N$ 
\strut and/or~$N'$, i.\,e., to approximate the interaction $V$ by a nonlocal
separable potential $\tilde{V}$   of the rank  ${\mathcal N} = (\mathbb N - N_0)/2 +1$ with matrix elements
  \begin{equation}
    \tilde{V}_{NN'}^{\ell} = \left\{
    \begin{array}{lll}
          V_{NN'}^{\ell}   &\mbox{if} \ & N\leq \mathbb N\ \mbox{and}\ N'\leq \mathbb N;\\
          0 &\mbox{if}& N>\mathbb N\ \ \mbox{or}\ \ N'>\mathbb N .
     \end{array} \right.
  \label{trunc}
  \end{equation}
The approximation \eqref{trunc} is the only approximation within the HORSE method; for the separable interaction of
the type~\eqref{trunc}, the HORSE formalism suggests exact solutions. Note, the kinetic energy matrix is not
truncated within the HORSE theory contrary to conventional variational 
approaches like the shell model.  Hence the HORSE formalism suggests a natural generalization of the shell model.

The complete infinite harmonic oscillator basis space can be divided into two subspaces according to truncation~\eqref{trunc}:
an internal subspace spanned by oscillator functions with~$N \leq \mathbb N$ where the interaction $V$ is accounted for and
an asymptotic subspace spanned by oscillator functions with~$N > \mathbb N$ associated with the free motion.

Algebraic equations~\eqref{Infsys} in the asymptotic subspace take the form of a second order finite-difference equation:
  \begin{multline}
  T_{N,N-2}^{\ell}\,a_{N-2,\ell}^{ass}(E) + (T^{\ell}_{NN}-E)\,a_{N\ell}^{ass}(E)  \\
  + T_{N,N+2}^{\ell}\,a_{N+2,\ell}^{ass}(E) = 0 .
  \label{TRSMaz}
  \end{multline}
Any solution $a_{N\ell}^{ass}(E)$ of Eq.~\eqref{TRSMaz} can be expressed as a superposition of regular $S_{N\ell}(E)$ and
irregular $C_{N\ell}(E)$ solutions,
   \begin{equation}
  a_{N\ell}^{ass}(E) = \cos  \delta_{\ell}\, S_{N\ell}(E) + \sin \delta_{\ell}\, C_{N\ell}(E)  ,\quad N\geq \mathbb N,
  \label{stwave}
  \end{equation}
 where $\delta_{\ell}$ is the scattering phase shift. The solutions $S_{N\ell}(E)$ and $C_{N\ell}(E)$ have simple analytical
 expressions~\cite{YaFi,NeSm,Bang,true,trueZ}:  
\begin{multline}
S_{N\ell}(E)  = \sqrt{\frac{\pi  \Gamma(N/2-\ell/2+1)}{\Gamma(N/2+\ell/2+3/2)}}\: q^{\ell+1} \\
\times \exp{\!\left(\!-\frac{q^2}{2}\right)}\,L_{(N-\ell)/2}^{\ell+1/2}(q^2),  
\label{Snl}
\end{multline}
\begin{multline}
C_{N\ell}(E)  = (-1)^{\ell} \sqrt{\frac{\pi  \Gamma(N/2-\ell/2+1)}{\Gamma(N/2+\ell/2+3/2)}}\,
\frac{q^{-\ell}}{\Gamma(-\ell+1/2)}\\
\times \exp{\!\left(\!-\frac{q^2}{2}\right)}\,\Phi(-N/2-\ell/2-1/2,-\ell+1/2;q^2) ,  
\label{Cnl}
\vspace{-3ex}
\end{multline}
where $\Phi(a,b;z)$ is a confluent hypergeometric function and $q$ is a dimensionless momentum,\vspace{-.5ex}
\begin{equation}
q=\sqrt{\frac{2E}{\hbar\Omega}}.
\label{dimlessq}
\end{equation} 

The solutions $a_{N\ell}(E) $ of the algebraic set~\eqref{Infsys} in the internal subspace $N \leq \mathbb N$ are expressed through the
solutions $a_{N\ell}^{ass}(E)$ in the asymptotic subspace $N \geq \mathbb N$:\vspace{-.5ex}
\begin{multline}
  a_{N\ell}(E) = {\mathcal G}_{N\mathbb N}(E)\,T^{\ell}_{\mathbb N, \mathbb N+2}\,a_{\mathbb N+2,\, \ell}^{ass}(E), \\
   N=N_0, N_0+2, ...\, , \mathbb N. 
  \label{cnin}
  \end{multline}
Here the matrix elements\vspace{-.5ex} 
  \begin{equation}
  {\cal G}_{NN'}(E) = -\sum_{\nu =0}^{{\cal N}-1}
  \frac{ \langle N\ell| \nu \rangle \langle \nu |N'\ell\rangle }
  { E_{\nu }-E }
  \label{oscrm}
  \vspace{-1ex}
  \end{equation}
  are  related to the Green's function of the Hamiltonian $H^{\mathbb N}$ which is the Hamiltonian~$H^{\ell}$ truncated to the internal subspace,
and are expressed through eigen\-energies~$E_{\nu}$, ${\nu=0,1,2,...\, ,\,{\cal N}-1}$ ($\cal N$ is the dimensionality of the basis)
and respective eigenvectors $\langle N\ell | \nu \rangle$ of the Hamiltonian $H^{\mathbb N}$:\vspace{-.5ex}
\begin{multline}
\sum_{N'=N_0,N_{0}+2,...,\mathbb N} H_{NN'}^{\ell} \langle N' \ell |\nu\rangle
= E_\nu\langle N\ell |\nu\rangle, \\
\qquad N=N_0,N_{0}+2,...,\mathbb N. 
\label{Alge}
\end{multline}

A relation for calculation of the scattering phase shifts $\delta_\ell$ can be obtained through the matching condition
 \begin{equation}
  a_{\mathbb N\ell}(E) = a_{\mathbb N\ell}^{ass}(E) .
  \label{glu}
  \end{equation}
Using Eqs.~\eqref{stwave}, \eqref{cnin} and~\eqref{glu} it is easy to obtain~\cite{YaFi,NeSm,Bang,true,trueZ}
 \begin{gather}
 \tan{\delta_\ell (E)}  
 = -\frac{S_{\mathbb N\ell}(E)-{\mathcal G}_{\mathbb N\mathbb N}(E)\,T^{\ell}_{\mathbb N, \mathbb N+2}\,S_{\mathbb N+2,\ell}(E)}
 {C_{\mathbb N\ell}(E)-{\mathcal G}_{\mathbb N\mathbb N}(E)\,T^{\ell}_{\mathbb N, \mathbb N+2}\,C_{\mathbb N+2,\ell}(E)} .
\label{tandelt}
\end{gather}
The respective expression for the $S$-matrix reads
 \begin{equation}
 S (E)  
 = \frac{C^{(-)\!}_{\mathbb N\ell}(E)-{\mathcal G}_{\mathbb N\mathbb N}(E)\,T^{\ell}_{\mathbb N, \mathbb N+2}\,C^{(-)}_{\mathbb N+2,\ell}(E)}
 {C^{(+)\!}_{\mathbb N\ell}(E)-{\mathcal G}_{\mathbb N\mathbb N}(E)\,T^{\ell}_{\mathbb N, \mathbb N+2}\,C^{(+)}_{\mathbb N+2,\ell}(E)} ,
\label{S-matr}
\vspace{-1ex}
\end{equation}
where\vspace{-1ex}
\begin{gather}
C^{(\pm)\!}_{ N\ell}(E)=C_{ N\ell}(E)\pm S_{ N\ell}(E).
\label{Cpm}
\end{gather}

We are using here the single-channel version of the HORSE formalism described above. The multi-channel HORSE formalism is
discussed in detail in Refs.~\cite{YaFi,true,trueZ,Bang}.

\subsection{SS-HORSE method}

A direct HORSE extension of modern large-scale shell-model calculations is unpractical. Note, Eq.~\eqref{oscrm} involves a sum over all
shell-model eigenstates of a given spin-parity, i.\,e., over millions or even billions of states in  modern NCSM applications. These states should be 
accurately separated from those having center-of-mass excitations. Unfortunately one cannot restrict the sum in Eq.~\eqref{oscrm} to some
small enough set of eigenstates: even for the energies~$E$ close enough to one of the low-lying eigenstates~$E_{\nu}$, the contribution of some
high-lying eigenstates to the sum in Eq.~\eqref{oscrm} can be essential: in model two-body problems describing, e.\,g., $n\alpha$
scattering, the growth of the denominator in the r.h.s. of~ Eq.~\eqref{oscrm} is compensated by the growth of the numerator; in NCSM calculations of
$^{5}$He, the many-body eigenstates concentrate around the eigenstates of the model two-body Hamiltonian and though the contribution
of each particular NCSM eigenstate is small, the sum of their contributions is large and close to the contribution of the respective 
state of the model Hamiltonian. A calculation of a large number of many-body eigenstates is too computationally expensive. Note, in
many-body applications, one also needs to calculate the components~$\langle\mathbb N\ell| \nu \rangle$ of the wave function which should
be projected on the scattering channel of interest; this projection requires numerous applications  of Talmi--Moshinsky transformations which
increase the computational cost and makes it very difficult  to achieve a reasonable accuracy of the final sum in Eq.~\eqref{oscrm} due 
to computer noise. 

To avoid these difficulties, we propose the SS-HORSE approach which requires calculations of the $S$-matrix or phase shifts only
at~$E=E_{\nu}$, i.\,e., at the energy equal to one of the lowest eigenstates lying above the reaction threshold.  Equations~\eqref{tandelt} 
and~\eqref{S-matr} are essentially simplified in this case and reduce to
\begin{gather}
 \tan{\delta_\ell (E_{\nu})}  
 = -\frac{S_{\mathbb N+2,\ell}(E_{\nu})}  {C_{\mathbb N+2,\ell}(E_{\nu})} 
\label{tandelt-SS}
\\ \intertext{and}\vspace{-2ex}
 S (E_{\nu})  
 = \frac{C^{(-)}_{\mathbb N+2,\ell}(E_{\nu})} {C^{(+)}_{\mathbb N+2,\ell}(E_{\nu})} .
\label{S-matr-SS}
\end{gather}
Varying~$\mathbb N$ and~$\hbar\Omega$ we obtain eigenvalues~$E_{\nu}$ and hence phase shifts and $S$-matrix in some energy interval.
An accurate parametrization of~$\delta_\ell (E)$ and $S$-matrix in this energy interval makes it possible to extrapolate them to a 
larger energy interval and to calculate the resonance energy and width. 

The use of Eqs.~\eqref{tandelt-SS} 
and~\eqref{S-matr-SS} drastically reduces the
computational burden in many-body calculations. Within this SS-HORSE approach we need  only one or probably very few
low-lying eigenstates which energies should be  calculated relative to the respective threshold, e.\,g., in the case of $n\alpha$ scattering
we need to subtract from the $^{5}$He energies the $^{4}$He ground state energy. Another interesting and important feature of the 
SS-HORSE technique is that the Eqs.~\eqref{tandelt-SS} 
and~\eqref{S-matr-SS} do not involve any information regarding the eigenvectors~$\langle N\ell |\nu\rangle$. This essentially simplifies
calculations, 
the information about a particular channel under consideration is present only in the  threshold energy used to
calculate the eigenenergies~$E_{\nu}$ and in the channel orbital momentum~$\ell$. Equations~\eqref{tandelt-SS} 
and~\eqref{S-matr-SS} establish some correlations between scattering in different channels when the channel coupling can be 
neglected, a topic that deserves further investigation but is outside the scope of the present work.

We use here Eqs.~\eqref{tandelt-SS} 
and~\eqref{S-matr-SS} to obtain phase shifts and $S$-matrix from Hamiltonian diagonalization results. However these equations
can be used in inverse manner: if the phase shifts are known from analysis of experimental scattering data, one can solve Eq.~\eqref{tandelt-SS} 
to obtain eigenenergies~$E_{\nu}$ which the shell model Hamiltonian should have to be  consistent with scattering data. The direct use
of Eq.~\eqref{tandelt-SS} essentially simplifies the inverse approach to nucleon-nucleus scattering suggested in Refs.~\cite{PRC79, ApplMathInfSci}.

We see that the scattering phase shifts are determined by the universal function
\begin{equation}
 f_{\mathbb N\ell} (E)  = -\arctan\! \left[ \frac{S_{\mathbb N+2,\ell}(E)}{C_{\mathbb N+2,\ell}(E)} \right]\! .
\label{fnl}
\end{equation}
This is a smooth monotonically decreasing function which drops down by~$n\pi$ as \mbox{energy}~$E$ varies from~0 to~$\infty$. At low energies
when
\begin{equation}
\label{if1}
E\ll\frac18\hbar\Omega\,(\mathbb N+2-\ell)^{2},
\end{equation}
one can replace  the functions~$S_{\mathbb N+2,\ell}(E)$
and~$C_{\mathbb N+2,\ell}(E)$ in Eq.~\eqref{fnl} by their asymptotic expressions at large~$\mathbb N$ (see Refs.~\cite{true,trueZ})
to obtain\vspace{-.6ex}
\begin{equation}
f_{\mathbb N\ell} (E) \approx f^{l.e.}_{\ell} (E)  =  \arctan\! \left[ \frac{j_\ell\bigl(2\sqrt{E/s}\bigr)}{n_\ell\bigl(2\sqrt{E/s}\bigr)} \right] \! ,
\label{fnl_as}
\vspace{-.7ex}
\end{equation}
where\vspace{-.7ex}
\begin{gather}
s = \frac{\hbar\Omega}{\mathbb{N}+7/2},
\label{sScale}
\end{gather}
and~$j_{l}(x)$ and~$n_{l}(x)$ are spherical Bessel and spherical Neumann  functions.
If additionally\vspace{-.7ex}
\begin{gather}
E\gg\frac14 s
\label{Enus1}= \frac{\hbar\Omega}{4(\mathbb N+7/2)},
\end{gather}
one can use asymptotic expressions for spherical Bessel and Neumann  functions in Eq.~\eqref{fnl_as} to get a very simple expression
for the function~$f_{\mathbb N\ell} (E) $:
\begin{equation}
\label{fassss}
f_{\mathbb N\ell} (E) \approx  -2\sqrt{\frac{E}{s}}+\frac{\pi \ell}{2}.
\end{equation}

\begin{figure}[t!]
\centerline{\includegraphics[width=\columnwidth]{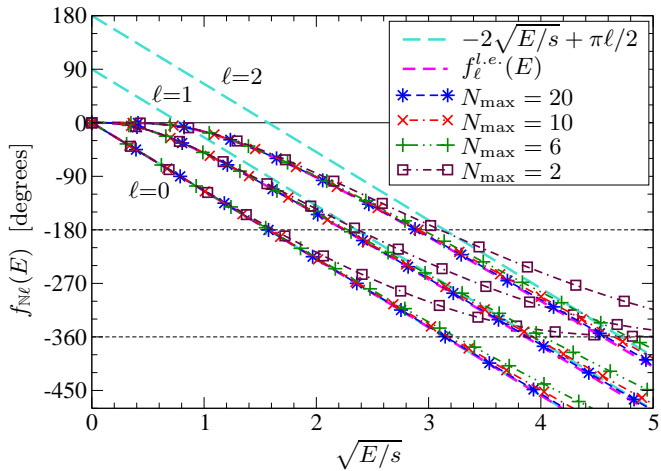}}
\caption{The function $f_{\mathbb N\ell} (E) $ (symbols) for different~$\mathbb N$ and~$\ell$ and its low-energy approximations~$f^{l.e.}_{\ell} (E)$ 
[see Eq.~\eqref{fnl_as}] and~$-2\sqrt{E/s}+\pi\ell/2$ [see Eq.~\eqref{fassss}].}
\label{fnleps}
\end{figure}

The universal function~$f_{\mathbb N\ell} (E)$ and its low-energy approximations~\eqref{fnl_as} and~\eqref{fassss} are shown in
Fig.~\ref{fnleps}. The basis space in shell model applications in conventionally labeled by the maximal oscillator excitation quanta~$N_{\max}$, and we
use~$N_{\max}$ in Fig.~\ref{fnleps} to distinguish functions~$f_{\mathbb N\ell} (E)$ corresponding to different basis sizes. Obviously,
\begin{gather}
\mathbb N= N_{\max}+\ell 
\label{bbN-Nmax}
\end{gather}
in the two-body scattering problem.
The approximation~\eqref{fnl_as} is seen to be very accurate at low energies even for small~$N_{\max}$. This 
low-energy approximation, as expected, deviates from the function~$f_{\mathbb N\ell} (E)$ as the energy~$E$
increases; the energy interval where the approximation~\eqref{fnl_as} accurately describes~$f_{\mathbb N\ell} (E)$ increases
with~$\mathbb N$ or~$N_{\max}$	in accordance with inequality~\eqref{if1}.
In the case~$\ell=0$, the simple expression~\eqref{fassss} is equivalent to the Eq.~\eqref{fnl_as} and therefore describes the
function~$f_{\mathbb N\ell} (E)$ with the same accuracy. For~$\ell>0$ the simplified approximation~\eqref{fassss} deviates from
the approximation~\eqref{fnl_as} and the function~$f_{\mathbb N\ell} (E)$ at low 
energies, it can only be used in a relatively small energy
interval defined by inequalities~\eqref{if1} and~\eqref{Enus1}.

Due to Eq.~\eqref{fnl_as}, equation~\eqref{tandelt-SS} at low energies can be reduced to
\begin{equation} 
\label{SSJM_phase_sc}
\tan{\delta_{\ell} (E_{\nu})}  = \frac{j_\ell\bigl(2\sqrt{E_\nu/s}\bigr)}{n_\ell\bigl(2\sqrt{E_\nu/s}\bigr)} .  
\end{equation}
This equation reveals the scaling at low energies:
the oscillator basis parameters~$\mathbb N$ and~$\hbar\Omega$ are not independent, they are entering  equations relating 
the $S$-matrix and phase shifts with the eigenenergies of the Hamiltonian matrix in the oscillator basis not separately but only through
 the scaling variable~$s$ combining them in a particular manner. The scaling is useful within our approach for selecting
 eigenenergies~$E_{\nu}$ obtained with different~$\mathbb N$ and~$\hbar\Omega$ for the further analysis of phase shifts and
$S$-matrix poles: the convergence of the results obtained by diagonalization of the Hamiltonian in oscillator basis is achieved
within some interval of~$\hbar\Omega$ values starting from some~$\mathbb N$; the converged results for~$E_{\nu}$ should describe the same phase
shifts with some accuracy, therefore, due to the scaling~\eqref{SSJM_phase_sc}, these converged~$E_{\nu}$ plotted as functions of the scaling 
parameter~$s$ should lie approximately on the same curve. By plotting~$E_{\nu}$~vs~$s$ we can pick up for further analysis only 
those~$E_{\nu}$ which form some curve as is illustrated later.

The scaling in variational oscillator-basis calculations of bound states was proposed in Refs.~\cite{Coon,CoonNTSE12}. We extend here
the scaling property of the oscillator-basis calculations to the continuum states. We prefer to use the scaling parameter~$s$ in energy units
rather than the scaling parameter~$\lambda_{sc}$ of  Refs.~\cite{Coon,CoonNTSE12, CoonNTSE13} in momentum units or the
scaling parameter
\begin{gather}
L=\sqrt{2(\mathbb N+7/2)} \,r_{0}
\label{LCoon}
\end{gather}
in the units of length suggested in Ref.~\cite{More}. The parameter~$L$ includes a small correction to the scaling proposed
in Refs.~\cite{Coon,CoonNTSE12} which was suggested in Ref.~\cite{More} based on numerical results. We obtain this correction
automatically in our approach. Having this correction in mind, we get
\begin{gather}
s\sim \lambda_{sc}^{2}\sim 1/L^{2};
\label{slambdaL}
\end{gather}
in other words, we propose generically the same scaling as discussed in 
Refs.~{\colour{red}\cite{Coon,CoonNTSE12, CoonNTSE13,Dick,More,IT-extrap, Forssen2014, Furnstahl-S-matrix, Furnstahl14, Furnstahl15, Sid16}} 
but using another scaling parameter and extending the scaling to continuum states.
 
We derive the scaling property in a very different approach than that utilized in Refs.~\cite{Coon,CoonNTSE12,Dick,More}. Therefore it is
interesting to compare these scalings in more detail. One can analytically continue the Eqs.~\eqref{tandelt-SS} 
and~\eqref{S-matr-SS} to the complex energy or complex momentum plane, in particular, one can use these expressions at negative
energies corresponding to bound states. Using asymptotic expressions of the functions~$C^{(+)}_{\mathbb N+2,\ell}(E)$
and~$C^{(-)}_{\mathbb N+2,\ell}(E)$ at large~$\mathbb N$ and negative energy~$E$ (see Refs.~\cite{true,trueZ}), we obtain from Eq.~\eqref{S-matr-SS}:
\begin{equation}
 S (E_{\nu})   =  (-1)^{\ell} \exp{\!\left(\!-4i\sqrt{\frac{E_{\nu}}{s}}\right)} , \qquad E_{\nu}<0.
\label{fnl_as-B}
\end{equation}
On the other hand, the $S$-matrix~$S (E_{\nu})$ at negative energies~$E_{\nu}$ in the vicinity of the pole associated with the bound
state at energy~$E_{b}<0$ can be expressed as~\cite{Baz}
 \begin{equation}
\label{bstate}
S (E_{\nu}) =\frac{D_{\ell}}{i\varkappa_{\nu}-ik_{b}},
\end{equation}
where $E_{\nu}=-\frac{\hbar^2\varkappa_{\nu}^2}{2m}$,  $E_b=-\frac{\hbar^2k_b^2}{2m}$, 
momenta~$\varkappa_{\nu}$ and~$k_b$ are supposed to be 
positive, and~$D_\ell$ can be expressed through the asymptotic normalization constant~$A_\ell$~\cite{Baz}:
\begin{equation}
\label{Dl}
D_\ell=(-1)^{\ell+1}\,i\,|A_\ell|^2 .
\end{equation}
Combining Eqs.~\eqref{fnl_as-B}--\eqref{Dl}, we obtain:
\begin{equation}
\label{ext_k}
\varkappa_{\nu}-k_b = - |A_\ell|^2\exp{\!\left(\! -\frac{4\varkappa_{\nu}\hbar}{\sqrt{2m s}} \right)} .
\end{equation}
This expression can be used for extrapolating the eigenenergies~$E_{\nu}$ (or respective momenta~$\varkappa_{\nu}$) obtained in a
finite oscillator basis to the infinite basis space supposing that~$E_{\nu}\to E_{b}$ as~$\mathbb N\to\infty$.

The respective expression for extrapolating the oscillator basis eigenenergies derived 
in Refs.~\cite{Coon,CoonNTSE12,Dick,More} rewritten in our notations, takes the form:
\begin{equation}
\label{ext_More}
E_{\nu}-E_b =C_\ell\exp{\!\left( \!-\frac{4k_{b}\hbar}{\sqrt{2m s}} \right)}  .
\end{equation}  
There is some similarity, however there is also an essential difference between Eqs.~\eqref{ext_k} and~\eqref{ext_More}. Both equations
have similar exponents in the right-hand-side, however the exponent in our Eq.~\eqref{ext_k} involves momentum~$\varkappa_{\nu}$
associated with the eigenenergy~$E_{\nu}$ while Eq.~\eqref{ext_More} involves momentum~$k_{b}$ associated with the converged
energy~$E_{b}$ in the limit~$\mathbb N\to\infty$.
In the vicinity of the $S$-matrix pole [see Eq.~\eqref{bstate}]~$ \varkappa_{\nu}$ should not differ much from~$k_{b}$;
we note however that~$k_{b}$ is conventionally treated as an additional fitting
parameter (see Refs.~\cite{Coon,CoonNTSE12, CoonNTSE13,Dick,More,IT-extrap, Forssen2014, Furnstahl-S-matrix, Furnstahl14, Furnstahl15}),
i.\,e., it is supposed that~$E_b\ne-\frac{\hbar^2k_b^2}{2m}$, and hence there may be an essential difference
between~$\varkappa_{\nu}$  and~$k_{b}$ in applications. 
Even more important is that the exponent in the right-hand-side controls the difference between the
energies~$E_{\nu}$ and~$E_{b}$ in Eq.~\eqref{ext_More} while in our Eq.~\eqref{ext_k} the exponent controls the difference between the
momenta~$\varkappa_{\nu}\sim\sqrt{|E_{\nu}|}$ and~$k_{b}\sim\sqrt{|E_{b}|}$. We plan to examine in detail in a separate publication
which of the Eqs.~\eqref{ext_k} 
and~\eqref{ext_More} describes better the results of diagonalizations of realistic Hamiltonians in the oscillator basis 
for negative eigenenergies~$E_{\nu}$  and which of them is
more accurate in extrapolating the results for bound states obtained in finite oscillator bases to the infinite basis space.

 Equations~\eqref{tandelt-SS} 
and~\eqref{S-matr-SS} can be used to obtain the phase shifts and $S$-matrix in some range of energies covered by eigenenergies~$E_{\nu}$
obtained with various~$\mathbb N$ and~$\hbar\Omega$. To interpolate the energy dependences of the phase shifts and $S$-matrix
within and to extrapolate them outside this interval, we need accurate formulas for the phase shifts and $S$-matrix as functions of energy
which we discuss in the next subsection.

\subsection{\boldmath Phase shifts and $S$-matrix at low energies}
The scattering $S$-matrix as a function of the complex momentum~$k$
is known~\cite{Baz, Taylor} to have the following symmetry properties:
\begin{subequations}
\label{Ssym}
\begin{align}
\label{S-k}
  S(-k)=\frac{1}{S(k)},  \\  
  \label{Sks}
  S(k^*)=\frac{1}{{S^*}(k)}, \\ 
  \label{S-ks}  S(-k^*)=S^*(k) ,
\end{align}
\end{subequations}
where star is used to denote the complex conjugation. The $S$-matrix can have poles either in the lower part of the complex momentum plane
or on the imaginary momentum axis~\cite{Baz, Taylor}. The poles in the lower part of the complex momentum plane 
at~$k=\kappa_r\equiv  k_r-i\gamma_r$ 
$(k_r,\gamma_r>0)$ due to the symmetry relations~\eqref{Ssym} are
accompanied by the poles at~$k=-\kappa^{*}_r\equiv -k_r-i\gamma_r$ and are
associated with resonances at the energy
\begin{gather}
\label{Ekappar}
  E_r=\frac{\hbar^2}{2m}(k_r^2 -\gamma_r^2)
  \intertext{and with the width}
  \Gamma= \frac{2\hbar^2}{m}k_r\gamma_r.
  \label{Gammakappar}
\end{gather}
Bound states at energy~$E_b=-\hbar^2k_b^2/2m$ are in correspondence with the poles on the positive imaginary momentum axis at~$k=ik_{b}$ 
($k_{b}>0$), however some positive imaginary momentum poles can appear to be the so-called false or redundant poles~\cite{Baz} which do not represent
any bound state. The poles at negative imaginary momentum at~$k=-ik_{v}$ ($k_{v}<0$) are associated with virtual states at
energy~$E_v=\hbar^2k_v^2/2m$.

If the $S$-matrix has a pole close to the origin either in the lower part of the complex momentum plane or on the imaginary momentum axis, it
can be expressed at low energies as
\begin{equation} 
\label{SA.Mstruct}
   S(k)=\Theta(k)\,S_p(k),
\end{equation}
where $\Theta(k)$ is a smooth function of~$k$ and the pole term~$S_p(k)$ in the case of a bound state or false pole ($p=b$),
virtual ($p=v$) or a resonant
state $(p={r})$ takes the form~\cite{Taylor}:
\begin{subequations}
\label{SSbvr}
\begin{align}
    S_b(k)&=-\frac{k+ik_b}{k-ik_b}, 
\label{SSb}    \\[1ex]  
    S_v(k)&=-\frac{k-ik_v}{k+ik_v}, 
\label{SSv}    \\[1ex] 
S_r(k)&=\frac{(k+\kappa_r)(k-\kappa^{*}_r)}{(k-\kappa_r)(k+\kappa^{*}_r)}.
    \label{SbSr}
\end{align}
\end{subequations}
The $S$-matrix is expressed through the phase shifts~$\delta_\ell(k)$ as
\begin{gather}
S(k)=e^{2i\delta_\ell(k)},
\label{Sdelta}
\end{gather}
hence the respective phase shifts
\begin{equation}
   \delta_\ell(k)=\phi(k)+\delta_p(k),
   \label{deltaphipamaz}
\end{equation}
where the pole contribution~$\delta_{p}(k)$ from the bound state
 takes the form 
 \begin{subequations}
\label{phasebfvr}
\begin{equation} 
\label{deltapEamaz}
\delta_b(E)=\pi-\arctan\sqrt{\frac{E}{|E_b|}},
\end{equation}
where $\pi$  appears 
due to the Levinson theorem~\cite{Taylor}. The contributions from the false, virtual and
resonant poles are
\begin{align} 
\label{deltaEf}
\delta_f(E)&=-\arctan\sqrt{\frac{E}{|E_f|}}, \\[.6ex]
\label{deltaEv}
\delta_v(E)&=\arctan\sqrt{\frac{E}{E_v}}, \\[.3ex]
\label{deltaEr}
 \delta_r(E) &= -\arctan
                     \frac{a\sqrt{E}}{E-b^2} ,
\end{align}
\end{subequations}
where the resonance energy~$E_r$ and width~$\Gamma$ can be expressed through the parameters~$a$ and~$b$ as\strut 
\begin{align}
\label{EamazGab}
E_r  
       &= b^2-a^2/2,   \\[.7ex]  
\Gamma&=a\sqrt{4b^2-a^2}.
\label{Gab}
\end{align}

Due to Eq.~\eqref{Sdelta}, the $S$-matrix symmetry~\eqref {S-k} require the phase shift~$\delta_{\ell}(E)$ to be an odd function of~$k$ and its expansion in
Taylor series of~$\sqrt{E}\sim k$ includes only odd powers of~$\sqrt{E}$:
\begin{gather}
\delta_{\ell}(E)=c\,\sqrt{E}+d\bigl(\sqrt{E}\bigr)^{3}+...
\label{Tayloramazdelta}
\end{gather}
More, since~$\delta_{\ell} \sim k^{2\ell+1}$ in the limit~$k\to0$, $c=0$ in the case of $p$-wave scattering, $c=d=0$ in the case of
$d$-wave scattering, etc.

In applications to the non-resonant $n\alpha$ scattering in the~$\frac{1}{2}^{+}$ state ($\ell=0$), we therefore are using the following
parametrization of the phase shifts:
\begin{equation} 
\label{S_phase_b}
   \delta_0(E)=\pi-\arctan\sqrt{\frac{E}{|E_b|}}+c\sqrt{E}+d\bigl(\sqrt{E}\bigr)^3 +  f \bigl(\sqrt{E}\bigr)^5 .
\end{equation} 
The bound state pole contribution here is associated with the so-called Pauli-forbidden state.
There are resonances in the $n\alpha$ scattering in the~$\frac{1}{2}^{-}$ and~$\frac{3}{2}^{-}$ states ($\ell=1$); hence we parametrize these
phase shifts as
\begin{equation} 
\label{S_phase_r}
   \delta_1(E)=-\arctan
                   \frac{a\sqrt{E}}{E-b^2}
                                         -\frac{a}{b^2}\sqrt{E} + d\bigl(\sqrt{E}\bigr)^3.
\end{equation}
This form guarantees that $ \delta_1\sim k^{3}$ in the limit of~$E\to0$. 

\section{Model problem\label{modelproblem}}
To test our SS-HORSE technique, we calculate the phase shifts and resonant parameters of $n\alpha$ scattering in a two-body approach
treating neutron and $\alpha$ as structureless particles whose interaction is described by the potential WSBG of 
 a Woods--Saxon type with parameters fitted by Bang and Gignoux~\cite{WSB}: 
\begin{multline}
\label{WSBpot}
V_{n\alpha} =
\frac{V_0}{1+ \exp\left[ (r - R_0) / \alpha _0 \right]} \\
\ +\ 
( \boldsymbol{l} \cdot \boldsymbol{s} ) \  \frac{1}{r} \; \frac{d\ }{d r} \; \frac{V_{ls}}{1 + \exp\left[ (r - R_1) / \alpha _1 \right]} ,
\end{multline}
$V_0=-43$~MeV, $V_{ls}=-40\rm~MeV\cdot fm^2$, $R_0$=2.0~fm, $\alpha _0$=0.70~fm, $R_1$=1.5~fm, $\alpha _1$=0.35~fm.  

{\colour{red}We study the  $n\alpha$ phase shifts both in the case of resonant scattering in the~$\frac32^{-}$ partial wave and in the case
of non-resonant scattering in the~$\frac12^{+}$ partial wave.}
The matrix in the oscillator basis of the relative motion Hamiltonian with 
{\colour{red}the  WSBG} interaction is
diagonalized using~$\hbar\Omega$ values ranging from~2.5 to 50~MeV in steps of 2.5~MeV and~$N_{\max}$ up to~20
for natural parity states~$\frac32^{-}$ 
and up to~19 for unnatural parity states~$\frac12^{+}.$

\subsection{\boldmath Partial wave $\frac32^{-}$}

\begin{figure}[t!]
\centerline{\includegraphics[width=\columnwidth]{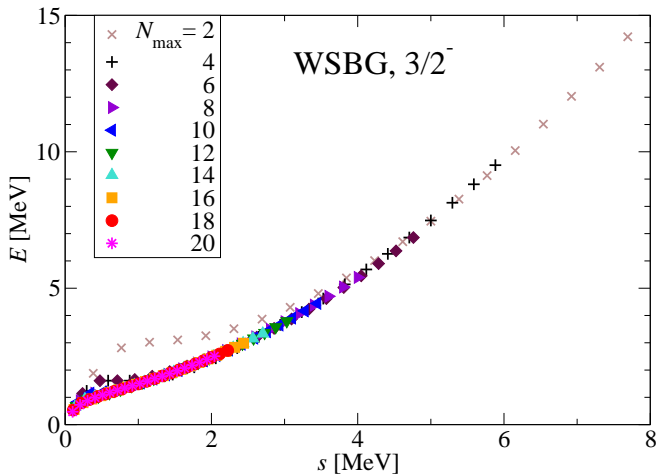}}
\caption{The lowest $\frac32^{-}$ eigenstates~$E_{0}$ of the model Hamiltonian with WSBG potential obtained with various~$N_{\max}$ and~$\hbar\Omega$
plotted as a function of the scaling parameter~$s$.}
\label{WSB32m_Es}
\end{figure}

\begin{figure}[b!]
\centerline{\includegraphics[width=\columnwidth]{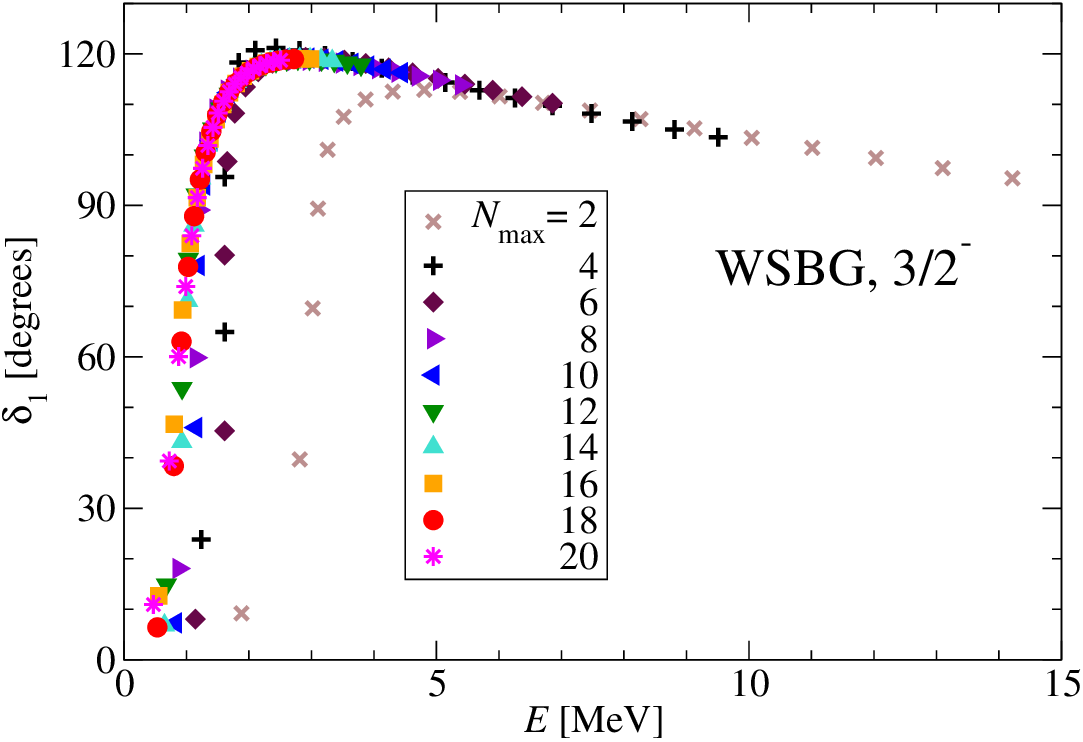}}
\caption{The $\frac32^{-}$ phase shifts obtained directly from  the WSBG eigenstates~$E_{0}$ using Eq.~\eqref{tandelt-SS}.}
\label{WSB32m_deltaE}
\end{figure}

The lowest eigenstates~$E_{0}$ obtained by diagonalization of the model Hamiltonian with the WSBG potential are presented in 
Fig.~\ref{WSB32m_Es} as a function of the scaling parameter~$s$. It is seen that the eigenstates obtained with large enough~$N_{\max}$
values form a single curve in Fig.~\ref{WSB32m_Es}; however the eigenstates obtained with smaller~$N_{\max}$ start deviating from
this curve at smaller~$\hbar\Omega$ which correspond to smaller~$s$ values reflecting the convergence patterns of calculations in
the finite oscillator basis. This feature is even more pronounced in the plot of the phase shifts obtained directly from 
eigenstates~$E_{0}$ using Eq.~\eqref{tandelt-SS} (see Fig.~\ref{WSB32m_deltaE}). We need to exclude from the further SS-HORSE analysis the
eigenstates deviating from the common curves in Figs.~\ref{WSB32m_Es} and~\ref{WSB32m_deltaE}.

\begin{figure}[t!]
\centerline{\includegraphics[width=\columnwidth]{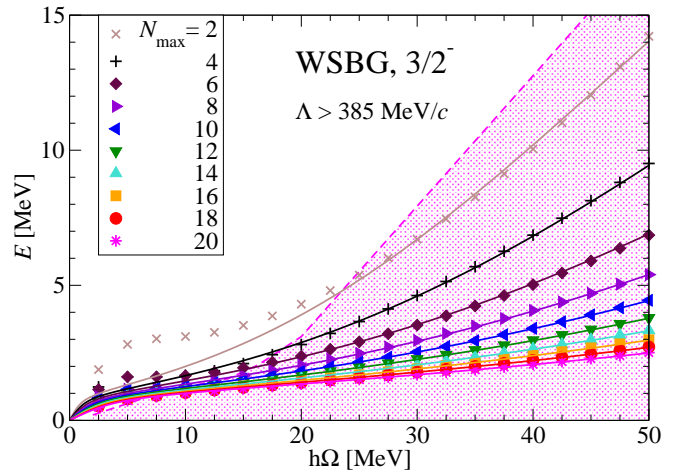}}
\caption{The lowest $\frac{3}{2\protect\vphantom{_{q}}}^{-}$ eigenenergies~$E_{0}$ of the model WSBG Hamiltonian obtained with various~$N_{\max}$ (symbols)
as  functions of~$\hbar\Omega$  and their selection 
for the SS-HORSE analysis according to inequality~$\Lambda>385$~MeV/$c$. 
The shaded area shows the selected~$E_{0}$ values. Solid lines are solutions of Eq.~\eqref{ImplicitENhw} for energies~$E_{0}$ with parameters~$a$,
$b$ and~$d$ obtained by the fit.}
\label{WSB32m_Ehw_Select0}
\end{figure}

\begin{figure}[b!]
\centerline{\includegraphics[width=\columnwidth]{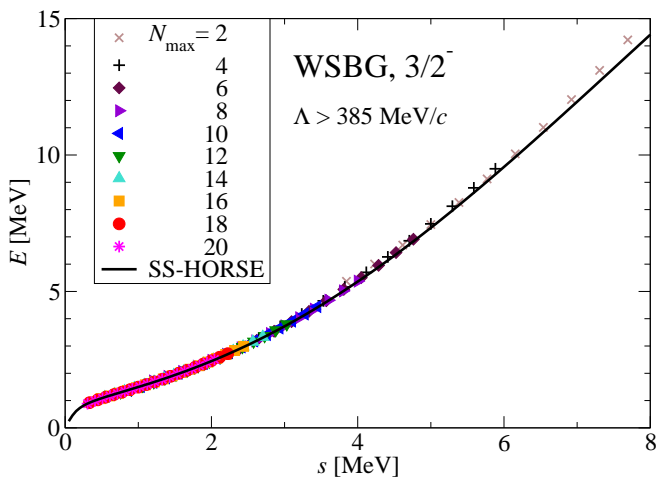}}
\caption{The  $\frac{3}{2\protect\vphantom{_{q}}}^{-}$ WSBG eigenstates~$E_{0}$ selected according to~$\Lambda>385$~MeV/$c$
plotted as a function of the scaling parameter~$s$. The solid curve depicts solutions of Eq.~\eqref{ImplicitENhw} for energies~$E_{0}$ 
with parameters~$a$, $b$ and~$d$ obtained by the fit with the respective selection of eigenstates.}
\label{WSB32m_Es_Select0}
\end{figure}

\begin{figure}[b!]
\centerline{\includegraphics[width=\columnwidth]{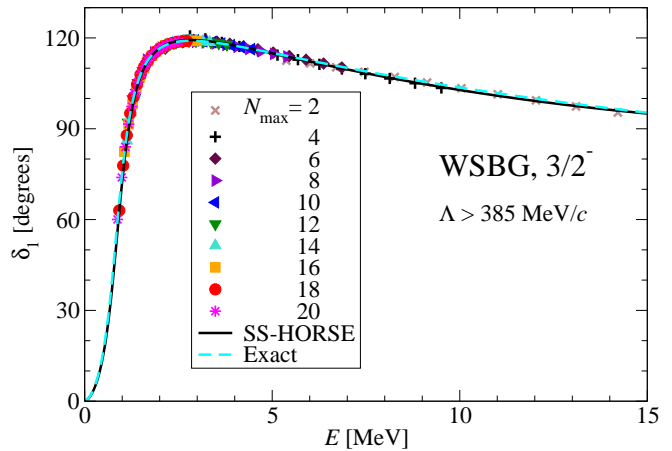}}
\caption{The $\frac32^{-}$ WSBG phase shifts obtained using Eq.~\eqref{tandelt-SS} 
directly from eigenstates~$E_{0}$ selected according to~$\Lambda>385$~MeV/$c$ (symbols).
The solid curve depicts the phase shifts of Eq.~\eqref{S_phase_r}  
with parameters~$a$, $b$ and~$d$ obtained by the fit with the respective selection of eigenstates;
the dashed curve is obtained by a numerical integration of the Schr\"odinger equation.}
\label{WSB32m_deltaE_Select0}
\end{figure}

As we already mentioned, the scaling property of our SS-HORSE formalism has much in common with those proposed in
Refs.~\cite{Coon,CoonNTSE12}. Using the nomenclature of Refs.~\cite{Coon,CoonNTSE12}, 
we should use only eigenenergies~$E_{0}$ which are not influenced 
by infra-red corrections. According to Refs.~\cite{Coon,CoonNTSE12}, these eigenenergies are obtained with~$N_{\max}$
and~$\hbar\Omega$ fitting inequality
\begin{equation}
\label{Lam}
\Lambda \equiv \sqrt{m \hbar\Omega( N_{\max}+\ell+3/2)} >\Lambda_{0},
\end{equation}
where~$\Lambda_{0}$ depends on the interaction between the particles. The value of~$\Lambda_0=385$~MeV/$c$ seems to be
adequate for the potential WSBG resulting in a reasonable selection of eigenenergies~$E_{0}$. The selection of eigenenergies
according to this  criterion is illustrated by the shaded area in Fig.~\ref{WSB32m_Ehw_Select0}
where we plot eigenenergies~$E_{0}$ obtained with various~$N_{\max}$ as functions of~$\hbar\Omega$. These selected
eigenstates plotted as a function of the scaling parameter~$s$ in Fig.~\ref{WSB32m_Es_Select0} and the respective SS-HORSE
phase shifts in Fig.~\ref{WSB32m_deltaE_Select0}
are seen to produce  smooth single curves.

 \begin{table*}[!] 
\caption{$\frac32^{-}$ resonance in $n\alpha$ scattering with model WSBG potential: fitting parameters~$a$, $b$, $d$ of Eq.~\eqref{ImplicitENhw},
resonance energy~$E_{r}$ and width~$\Gamma$, rms deviation of fitted energies~$\Xi$ and the number of these fitted energies~$D$
for different selections of eigenvalues in comparison with exact results for~$E_{r}$ and~$\Gamma$ obtained by numerical location of the
$S$-matrix pole. For the $N_{\max}\leq6$ selection, $\Xi$ and~$D$ for all energies from the previous selection are shown  within brackets.}
\label{WSBp32}  
\begin{ruledtabular}
\begin{tabular}{@{\hspace{6pt}}c@{\hspace{6pt}}c@{\hspace{6pt}}c@{\hspace{6pt}}c@{\hspace{6pt}}c@{\hspace{6pt}}c@{\hspace{6pt}}c@{\hspace{6pt}}c@{\hspace{6pt}}}
\raisebox{-1.8ex}[0pt][0pt]{Selection} & $a$ & $b^2$ & $d\cdot10^3$ & $E_r$ & $\Gamma$ & $\Xi$ & \raisebox{-1.8ex}[0pt][0pt]{$D$} \\
 & (MeV$^\frac12$) & (MeV) & (MeV$^{-\frac32}$) & (MeV) & (MeV) & (keV) &  \\
\noalign{\smallskip}\hline\noalign{\smallskip}
$\Lambda>385$~MeV/$c$ & 0.412 & 0.948 & 5.41 & 0.863 & 0.785 & 37 & 156 \\
$N_{\max}\leq6$ & 0.411 & 0.948 & 5.30 & 0.863 & 0.782 & 70(38) & 38(156) \\ 
Exact & & & & 0.836 & 0.780 & & \\
\end{tabular}
\end{ruledtabular} 
\end{table*}

The low-energy resonant $n\alpha$ scattering phase shifts in the~$\frac{3}{2}^{-}$  state are described by Eq.~\eqref{S_phase_r}. We
need to fit the parameters~$a$, $b$ and~$d$ of this equation.
Combining Eqs.~\eqref{tandelt-SS},
\eqref{bbN-Nmax} and~\eqref{S_phase_r} we derive  the following relation for resonant $n\alpha$ 
scattering in the~$\frac{3}{2}^{-}$  state 
{\colour{red}($\ell=1$)}:
\begin{multline} 
\label{ImplicitENhw}
   -\frac{S_{N_{\max}+3,\,1}(E_{0})}{C_{N_{\max}+3,\,1}(E_{0})} = \\
=   \tan\!\left(\!-\arctan\frac{a\sqrt{E_0}}{E_0-b^2}-\frac{a}{b^2}\sqrt{E_0}+d\bigl(\sqrt{E_0}\bigr)^3 \!\right)\!.
\end{multline}
We assign some values to the parameters~$a$, $b$ and~$d$ and solve this equation to find a set
of~$E_{0}$ values, $\mathcal{E}^{(i)}_{0}=E_{0}(N_{\max}^{i},\hbar\Omega^{i})$, $i=1,2,...,D$,
 for each combination
of~$N_{\max}$ and~$\hbar\Omega$ [note, $\hbar\Omega$ enters definitions of functions~$S_{N,\ell}(E)$ 
and~$C_{N,\ell}(E)$, see Eqs.~\mbox{\eqref{Snl}--\eqref{dimlessq}}].
The resulting set of~$\mathcal{E}^{(i)}_{0}$ is compared with the set of selected eigenvalues~$E_{0}^{(i)}$ obtained by the  
Hamiltonian diagonalization with respective~$N_{\max}$ and~$\hbar\Omega$ values, and we
minimize the rms deviation, 
\begin{equation}
 \label{ksi}
         \Xi = \sqrt{\frac{1}{D} \sum_{i=1}^{D}\Bigl(  E_0^{(i)}-{\cal E}_{0}^{(i)}   \Bigr)^2},
         \end{equation}
to find the optimal values of the parameters~$a$, $b$ and~$d$. The obtained parameters are listed in the first row of
Table~\ref{WSBp32}.  
The resonance energy~$E_{r}$ and width~$\Gamma$
obtained by Eqs.~\eqref{EamazGab} and~\eqref{Gab} are also presented in Table~\ref{WSBp32}. 
Note the accuracy of the fit: the rms deviation of 156~fitted energy eigenvalues is only 37~keV.

The behavior
of~$E_{0}$ as functions of~$\hbar\Omega$ dictated by Eq.~\eqref{ImplicitENhw} with the fitted optimal parameters 
for various~$N_{\max}$ values is depicted by solid curves in Figs.~\ref{WSB32m_Ehw_Select0} and~\ref{WSB32m_Es_Select0}. 
It is seen that these curves accurately
describe the selected eigenvalues~$E_{0}$ obtained by the Hamiltonian diagonalization. 
Note however a small deviation of the curve in Fig.~\ref{WSB32m_Es_Select0} from the  diagonalization results at large energies obtained 
with~$N_{\max}=2$ where the scaling become inaccurate, see Eq.~\eqref{if1}.
The phase shifts~$\delta_{1}(E)$ obtained by Eq.~\eqref{S_phase_r} with 
fitted parameters are shown in the Fig.~\ref{WSB32m_deltaE_Select0}. It is seen that the SS-HORSE phase shifts are in excellent
correspondence with the exact results obtained by numerical integration of the Schr\"odinger equation.
The $\frac32^{-}$ resonance energy and width are also well reproduced by our SS-{HORSE} technique (see Table~\ref{WSBp32}).
{\colour{red}  A small difference between the resonance energies~$E_{r}$ and widths~$\Gamma$ obtained by the SS-HORSE 
technique and by the numerical location of the respective $S$-matrix pole can be attributed to the fact that
the $\frac32^{-}$ resonance is wide enough and the respective $S$-matrix pole is located far enough from the
real energy axis; therefore the phase shifts even in the resonant region can be influenced by other $S$-matrix poles not accounted for
by our phase shift parametrization~\eqref{S_phase_r}.}

\begin{figure}[t!]
\centerline{\includegraphics[width=\columnwidth]{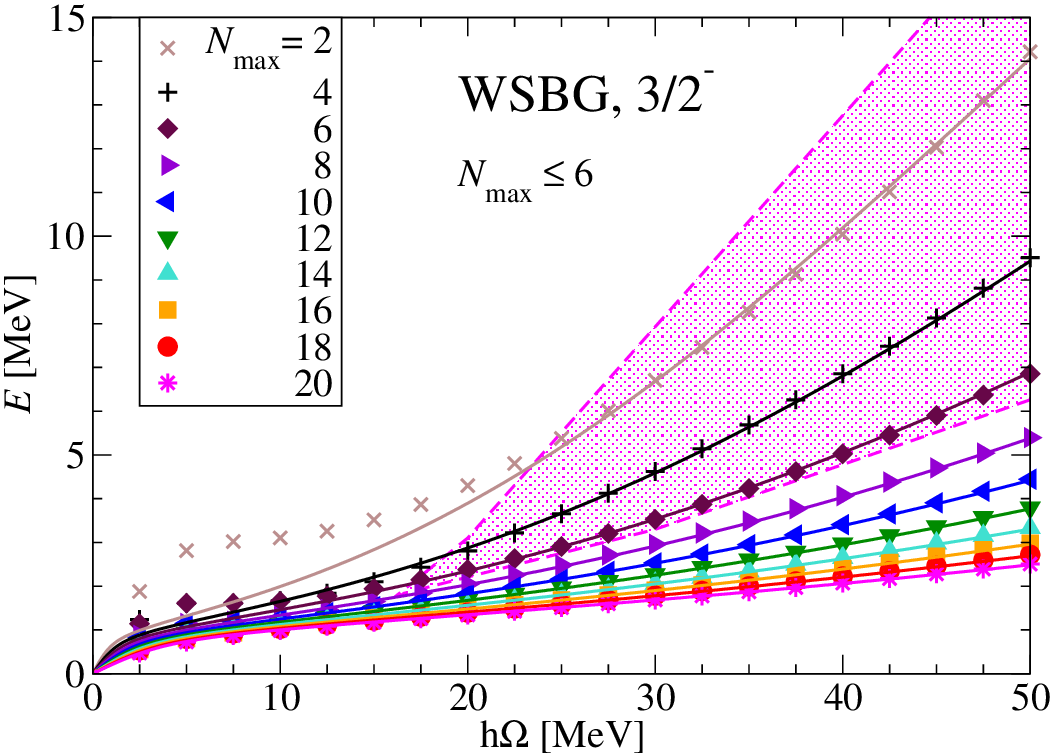}}
\caption{The  lowest $\frac32^{-}$ WSBG eigenstates~$E_{0}$ (symbols) and their ${N_{\max}\leq6}$ selection (shaded area).
See Fig.~\ref{WSB32m_Ehw_Select0}
for more details.}
\label{WSB32m_Ehw_Select2}
\end{figure}

\begin{figure}[b!]
\centerline{\includegraphics[width=\columnwidth]{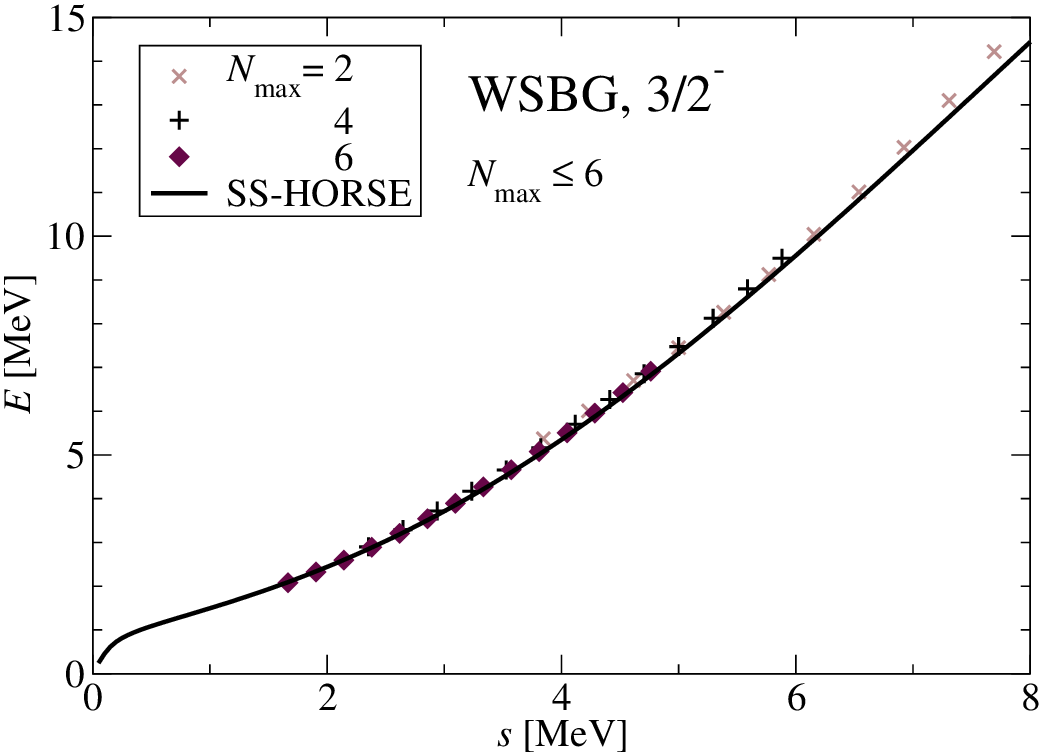}}
\caption{Selected lowest $\frac32^{-}$\strut\ WSBG eigenstates~$E_{0}$ obtained with ${N_{\max}\leq6}$ as a function of the scaling parameter~$s$. 
See Fig.~\ref{WSB32m_Es_Select0} for details.\strut}
\label{WSB32m_Es_Select2}
\end{figure}

\begin{figure}[!t]
\centerline{\includegraphics[width=\columnwidth]{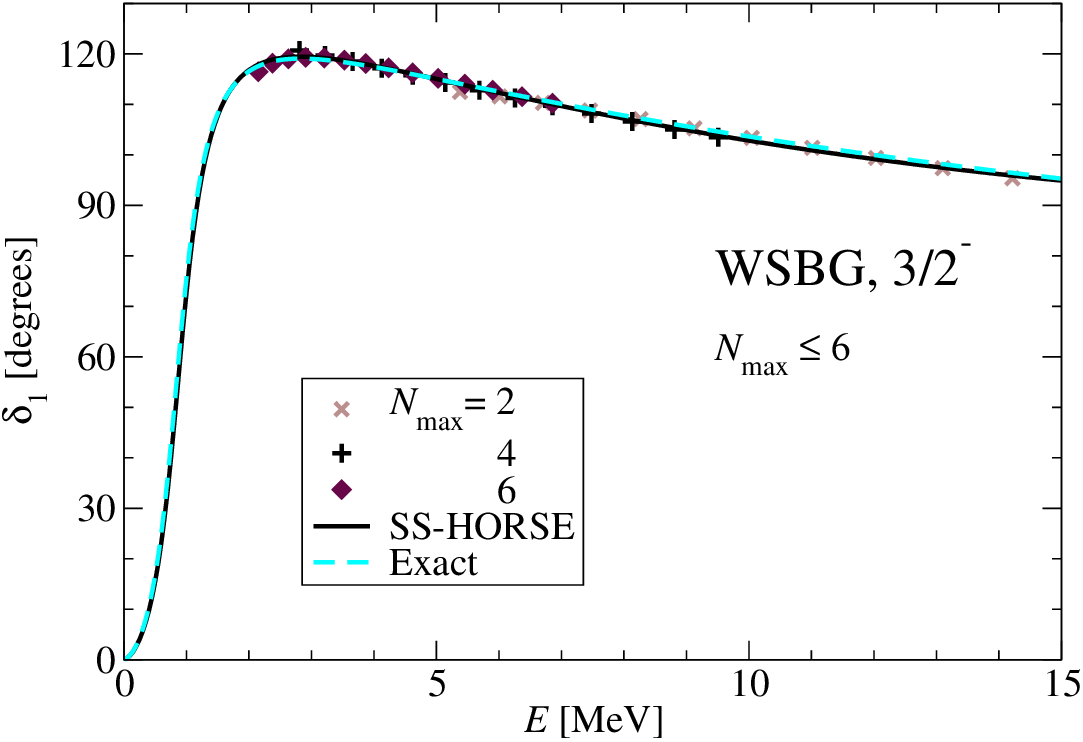}}
\caption{The $\frac32^{-}$ WSBG phase shifts generated by the selected eigenstates~$E_{0}$ obtained with ${N_{\max}\leq6}$. See
Fig.~\ref{WSB32m_deltaE_Select0} for details.}
\label{WSB32m_deltaE_Select2}
\end{figure}

 In the above analysis we used oscillator bases with~$N_{\max}$ values up to~${N_{\max}=20}$.  Such large~$N_{\max}$ are
accessible in two-body problems but are out of reach in modern many-body shell model applications.  Therefore it is very important to check
whether a reasonable accuracy of SS-HORSE phase shift and resonance parameter calculations can be achieved with significantly
smaller~$N_{\max}$.
 
 We remove from the set of selected $\frac32^{- }$ eigenstates~$E_{0}^{(i)}$ those obtained with~$N_{\max}>6$ and use this new selection illustrated 
by Figs.~\ref{WSB32m_Ehw_Select2} and~\ref{WSB32m_Es_Select2} to calculate phase shifts and resonant parameters. All
eigenenergies from this selection lie outside the resonance region as is seen in Fig.~\ref{WSB32m_deltaE_Select2} where we
plot the phase shifts as a function of energy. The SS-HORSE fit (see 
Table~\ref{WSBp32})
nevertheless accurately reproduces the exact phase shifts (see Fig.~\ref{WSB32m_deltaE_Select2}) even in the resonance region 
and the $\frac32^{- }$ resonance \mbox{energy}~$E_{r}$ and width~$\Gamma$ (see Table~\ref{WSBp32}). To get such accuracy, it is very important to use
the adequate phase shift
parametrization~\eqref{S_phase_r} which guarantees the low-energy phase shift behaviour~$\delta_{\ell}\sim k^{2\ell+1}$ and an
accurate description of the resonance region by the pole term~\eqref{deltaEr}: our previous study~\cite{RP2013} has clearly demonstrated that
it is impossible to reproduce the resonant parameters and phase shifts in a wide enough energy interval 
without paying special attention
to the low-energy phase shift description and by  using the less accurate Breit--Wigner resonant phase shifts instead of the pole term~\eqref{deltaEr} even
when  large~$N_{\max}$ eigenstates~$E_{0}$ are utilized to say nothing about the selection of  eigenstates obtained with
small~$N_{\max}$.

Solid lines in Figs.~\ref{WSB32m_Ehw_Select0} and~\ref{WSB32m_Ehw_Select2} present the eigenenergies~$E_{0}$ for various~$N_{\max}$ values as functions of~$\hbar\Omega$ 
obtained from 
the respective phase shift parametrization. It is seen that we accurately describe not only the eigenenergies from the shaded area
utilized in the fit but also those corresponding to a wider range of~$\hbar\Omega$ values.
It is even more interesting that in the case of Fig.~\ref{WSB32m_Ehw_Select2} 
where fitted are only the states with~$N_{\max}\le 6$, we also reproduce the eigenenergies obtained with  much larger~$N_{\max}$ 
values with nearly the same rms deviation as in the case of the
previous selection (see Table~\ref{WSBp32}) when those larger~$N_{\max}$ eigenenergies were included in the fit. 
In other words, our SS-HORSE fit to the diagonalization
results in small basis spaces makes it possible to `predict'  the diagonalization results obtained with much larger oscillator bases. 

{\colour{red}The predictive ability  of the SS-HORSE approach clearly demonstrates the  reliability of the potential truncation~\eqref{trunc}.
Note, the absolute values of discarded potential energy matrix elements~$ \tilde{V}_{NN'}^{\ell}$
with $N>\mathbb N$ or $N'>\mathbb N$ are not small in case of small enough~$N_{\max}$
values; however the contributions of these  discarded matrix elements are known to approximately cancel each other and their net contribution
appears to be small compared to the kinetic energy matrix elements~${T}_{NN}^{\ell}$ and~${T}_{N,N\pm 2}^{\ell}$ for~$N>\mathbb N$.
As  a result, in larger oscillator bases with the complete account of the potential energy we obtain nearly the same results as in the case
when the Hamiltonian matrix is extended to the same basis size by the kinetic energy matrix elements only. This feature is confirmed below
in our 5-body {NCSM-SS-HORSE} applications; it is very promising for shell-model applications to heavier nuclear systems}
and suggests a very efficient method of extrapolating the shell-model results 
to larger basis spaces.

\subsection{\boldmath Partial wave $\frac12^{+}$}

There are no resonances in the $n\alpha$ scattering in the~$\frac12^{+}$ partial wave. However, as it has been indicated in Refs.~\cite{PRC79, ApplMathInfSci},
the nuclear shell model should generate eigenstates in  non-resonant energy intervals in continuum to be consistent with
scattering observables. Therefore it is interesting to test with the WSBG potential the ability of the SS-HORSE approach to describe
the~$\frac12^{+}$  non-resonant $n\alpha$ scattering.

The low-energy $n\alpha$ scattering phase shifts in the~$\frac{1}{2}^{+}$  state are described by Eq.~\eqref{S_phase_b}. We
shall see that to get the same quality  fit as in the case of the odd-parity resonant scattering, we 
need in this case terms up
to the $\rm 5^{th}$ power of~$\sqrt{E}$ in the Taylor expansion of the background phase; therefore we preserve in Eq.~\eqref{S_phase_b} more
terms than in Eq.~\eqref{S_phase_r}.  $c$, $d$ and~$f$ are fitting parameters in Eq.~\eqref{S_phase_b}.  The WSBG potential supports a
bound state at energy~$E_{b}$ which mimics the Pauli-forbidden state in the $n\alpha$ scattering. We however treat~$E_{b}$ as an additional
fitting parameter as a preparation to many-body NCSM calculations where it is  impossible to obtain the energy of the Pauli-forbidden state.
This bound state appears as the lowest state with negative energy obtained by the Hamiltonian diagonalization and is unneeded for
our SS-HORSE analysis for which we use the first excited state~$E_{1}>0$ which is the lowest state in the continuum.

The excitation quanta~$N_{\max}$  is conventionally used to define the many-body 
NCSM basis space while the total oscillator quanta~$\mathbb N$ is entering our SS-HORSE equations. 
The $\frac12^{+}$ states in $^{5}$He are unnatural parity states, hence~$N_{\max}$ takes odd values within NCSM, the 
minimal oscillator quanta~$N_0=1$ in the five-body $n\alpha$ system, and
\begin{gather}
{\mathbb N=N_{\max}+N_0}
\label{NNmaxN0}
\end{gather}
is even.
To retain a correspondence with NCSM, we are using~$N_{\max}$ to define  the oscillator basis also in our model
two-body problem. We note that in this case the~$N_{\max}$ is formally related to~$\mathbb N$ according to Eq.~\eqref{bbN-Nmax} 
where~$\ell=0$, and~$N_{\max}$ should be even for even~$\mathbb N$. To have a closer correspondence with NCSM, we use
Eq.~\eqref{NNmaxN0} with~$N_{0}=1$ within our model two-body problem instead of Eq.~\eqref{bbN-Nmax} to relate~$N_{\max}$ 
to~$\mathbb N$, i.\,e., due to our NCSM-like definition, the $\frac{1}{2}^{+}$  eigenstates are labelled below by odd~$N_{\max}$ values.
Note, the definitions~\eqref{bbN-Nmax} and~\eqref{NNmaxN0} result in the same~$N_{\max}$ in the case of odd-parity $\frac{3}{2}^{-}$ 
and $\frac{1}{2}^{-}$ $n\alpha$ partial waves.

\begin{figure}[b!]
\centerline{\includegraphics[width=\columnwidth]{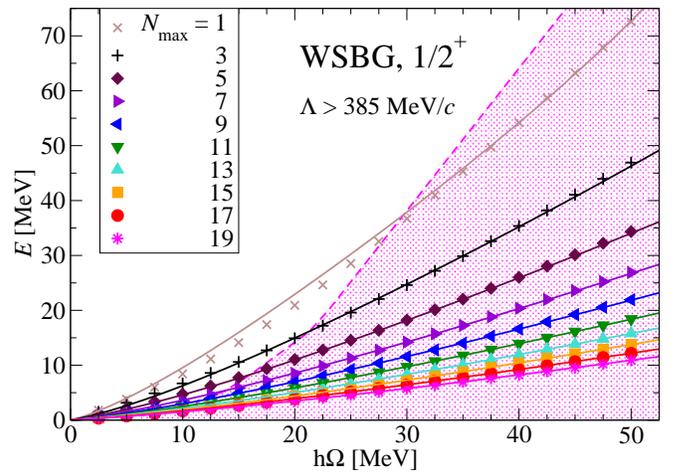}}
\caption{The lowest continuum $\frac12^{+}\protect\vphantom{_{q_{\int}}}$ WSBG eigenstates~$E_{1}$ (symbols) as  functions of~$\hbar\Omega$ and their~$\Lambda>385$~MeV/$c$ selection (shaded area). 
Solid lines are solutions of Eq.~\eqref{S_phase_implicit} for energies~$E_{1}$ with parameters~$E_{b}$, $c$, $d$ and~$f$
obtained by the fit with this selection of eigenstates.}
\label{WSB12p_Ehw_Select0}
\end{figure}

%

\begin{figure}[t!]
\centerline{\includegraphics[width=\columnwidth]{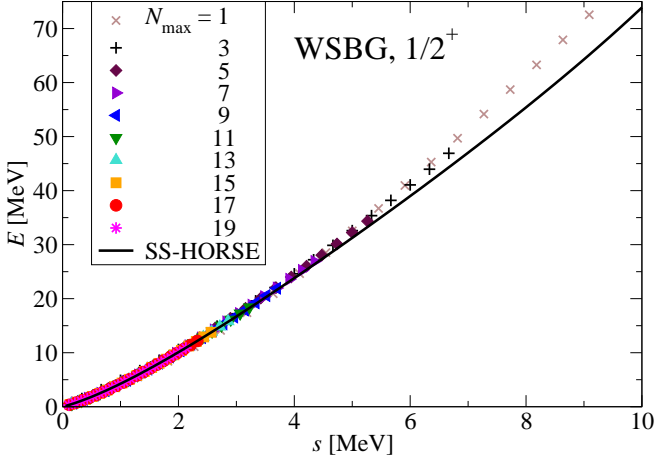}}
\caption{The  lowest continuum $\frac12^{+}$ WSBG eigenstates~$E_{1}$ as a function of the scaling parameter~$s$.
The solid curve depicts solutions of Eq.~\eqref{S_phase_implicit} for energies~$E_{1}$ 
with parameters~$E_{b}$, $c$, $d$ and~$f$  obtained by the fit with the $\Lambda>385$~MeV/$c$  selection of eigenstates.}
\label{WSB12p_Es0}
\end{figure}

\begin{figure}[b!]
\centerline{\includegraphics[width=\columnwidth]{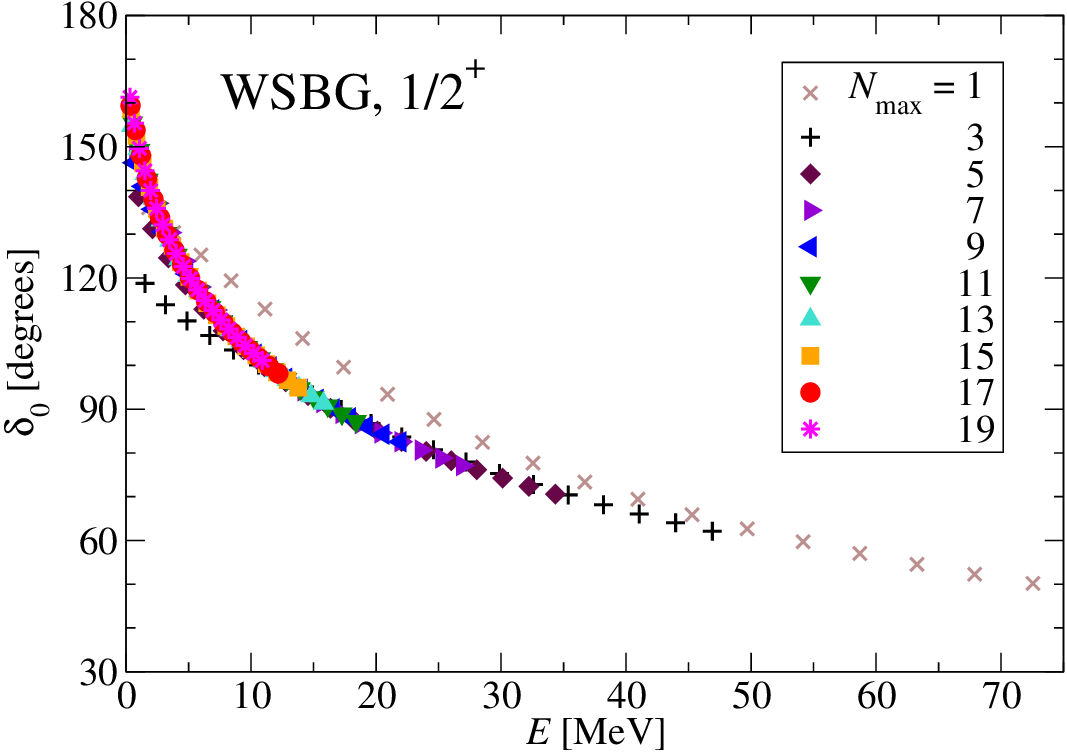}}
\caption{The  $\frac12^{+}$  phase shifts obtained directly from the WSBG eigenstates~$E_1$ using  Eq.~\eqref{tandelt-SS}.}
\label{WSB12p_deltaE}
\end{figure}

Combining Eqs.~\eqref{tandelt-SS}, \eqref{S_phase_b} and~\eqref{NNmaxN0}, we derive   for the $n\alpha$ 
scattering in the~$\frac{1}{2}^{+}$  partial wave:
\begin{multline} 
\label{S_phase_implicit}
   -\frac{S_{N_{\max}+3,\,0}(E_{\nu})}{C_{N_{\max}+3,\,0}(E_{\nu})} = 
   \tan \!\Biggl(\! \pi-\arctan\sqrt{\frac{E_{\nu}}{|E_b|}} \\ +c\sqrt{E_{\nu}}   
   +d\bigl(\sqrt{E_{\nu}}\bigr)^3 +  f \bigl(\sqrt{E_{\nu}}\bigr)^5 \!\Biggr)\! ,
\end{multline} 
where~$\nu=1$.
We assign some values to the fitting parameters~$E_{b}$, $c$, $d$ and~$f$ and solve Eq.~\eqref{S_phase_implicit} to find a set
of~$E_{1}$ values, $\mathcal{E}^{(i)}_{1}=E_{1}(N_{\max}^{i},\hbar\Omega^{i})$, $i=1,2,...,D$,
 for each combination
of~$N_{\max}$ and~$\hbar\Omega$ and minimize the rms deviation from the selected eigenvalues~$E_{1}^{(i)}$ obtained by the  
Hamiltonian diagonalization, see Eq.~\eqref{ksi} where the subindex~0 should be replaced by~1, to find the optimal values of the fitting parameters.

The lowest continuum $\frac12^{+}$ eigenstates~$E_{1}$ of the model WSBG Hamiltonian are shown as functions of~$\hbar\Omega$ 
for various~$N_{\max}$
in Fig.~\ref{WSB12p_Ehw_Select0} and as a function of the scaling parameter~$s$ in Fig.~\ref{WSB12p_Es0}. All  eigenenergies
in this case seem to lie approximately on the same curve in Fig.~\ref{WSB12p_Es0}; however,
as in the case of odd parity partial waves, the deviations from the common curve are much more pronounced in the plot of the SS-HORSE phase shifts 
corresponding to these eigenstates (see Fig.~\ref{WSB12p_deltaE}) which clearly indicates the need to select eigenstates for the 
SS-HORSE fitting.


 \begin{table*}[!t]
\caption{\strut$\frac12^{+}\protect\vphantom{_{q_{\int}}}$ $n\alpha$ scattering with model WSBG potential: fitting parameters~$E_{b}$, $c$, $d$ and~$f$ of 
Eq.~\eqref{S_phase_implicit}, rms deviation of fitted energies~$\Xi$ and the number of these fitted energies~$D$
for different selections of eigenvalues. For the $N_{\max}\leq5$ selection, $\Xi$ and~$D$ for all energies from the previous selection are shown  within brackets. }
\label{WSBs12}  
\begin{ruledtabular}
\begin{tabular}{ccccccc}
\raisebox{-1.8ex}[0pt][0pt]{Selection} & $E_{b}$ & $c$ & $d\cdot10^3$ &  $f\cdot10^5$ & $\Xi$ & \raisebox{-1.8ex}[0pt][0pt]{$D$} \\
 & (MeV) & (MeV$^{-\frac12}$) & (MeV$^{-\frac32}$) & (MeV$^{-\frac52}$) & (keV) &  \\
\noalign{\smallskip}
\hline\noalign{\smallskip}
$\Lambda>385$~MeV/$c$ &$-$6.841 &$-$0.157 & $+$1.19 & $-$0.888 & 163 & 151 \\
$N_{\max}\leq5$ & $-$6.853 &$-$0.156 & $+$1.19 & $-$0.888 & 332(163) & 35(151) \\ 
Exact  &   $-$9.85 &  &  &  &  &  \\ 
\end{tabular}
\end{ruledtabular} 
\end{table*}

%
%

\begin{figure}[t!]
\centerline{\includegraphics[width=\columnwidth]{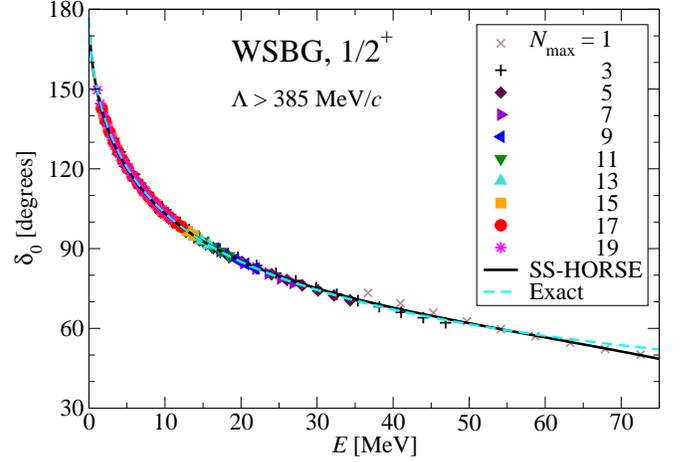}}
\caption{The $\frac12^{+}$ WSBG phase shifts generated by the~$\Lambda>385$~MeV/$c$ selected eigenstates~$E_1$ (symbols).
The solid curve depicts the phase shifts of Eq.~\eqref{S_phase_b}  
with parameters~$E_{b}$, $c$, $d$ and~$f$  obtained by the fit with this selection of eigenstates;
the dashed curve is obtained by a numerical integration of the Schr\"odinger equation.}
\label{WSB12p_deltaE_Select0}
\end{figure}

As in the case of the odd parity~$\frac32^{-}$ state,
 we use the $\Lambda>385$~MeV/$c$ selection of eigenenergies as is
illustrated by Fig.~\ref{WSB12p_deltaE_Select0} and 
by the shaded area in Fig.~\ref{WSB12p_Ehw_Select0}. 
The obtained fitting parameters of Eq.~\eqref{S_phase_implicit} are presented in Table~\ref{WSBs12}. It is interesting that the
fitted energy~$E_{b}$ differs essentially from the exact value which is the energy of the  bound state in the WSBG potential. The SS-HORSE
$\frac12^{+}$ phase shifts nevertheless are seen in Fig.~\ref {WSB12p_deltaE_Select0} to be  nearly indistinguishable from the exact ones
up to the energies of about 70~MeV where the SS-HORSE phase shifts governed by~$N_{\max}=1$ eigenstates slightly differ from exact.
We note that the WSBG bound state has a large binding energy, the respective $S$-matrix pole is far enough from the real
momentum axis and hence has a minor influence on the phase shifts. This result indicates that one should not take seriously  the 
energies of bound states obtained by the fit to the scattering data only, at least for well-bound states.
{\colour{red}We note however that the energy of this Pauli-forbidden state can be accurately calculated within the SS-HORSE approach by
including the lowest WSBG negative energy eigenstate~$E_{0}$ in the fit. Nevertheless, such a fit is of no interest for many-body NCSM applications
which do not generate  Pauli-forbidden states.}

\begin{figure}[t!]
\centerline{\includegraphics[width=\columnwidth]{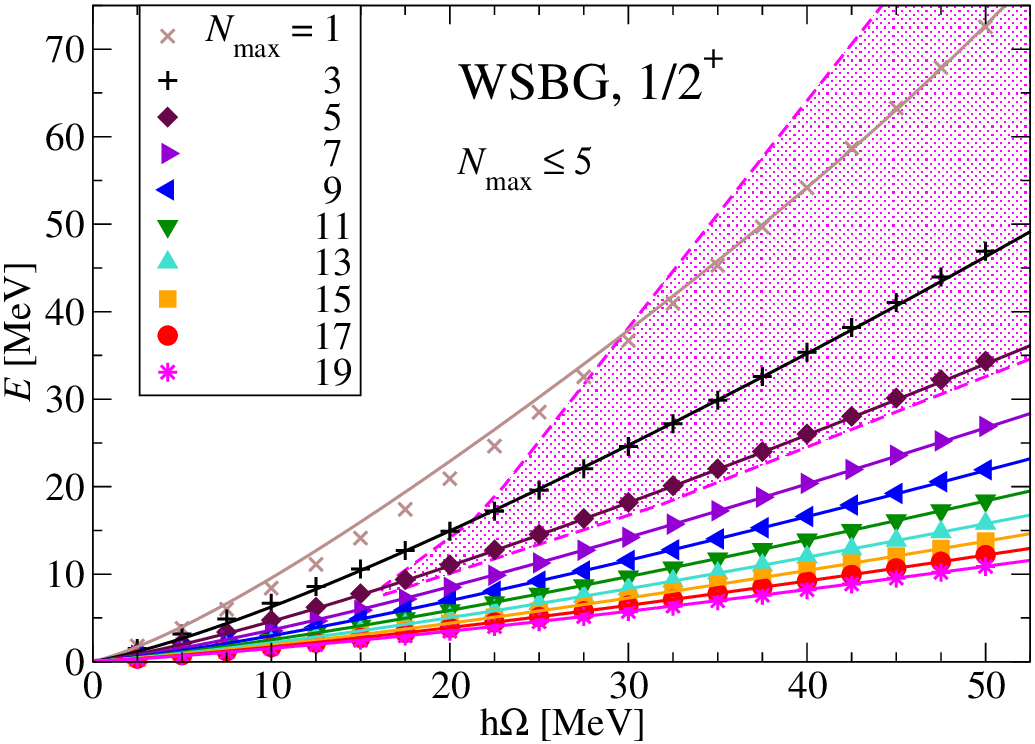}}
\caption{The lowest continuum $\frac12^{+}$ WSBG eigenstates~$E_{1}$ (symbols) and their  ${N_{\max}\leq5}$ selection (shaded area)
See Fig.~\ref{WSB32m_Ehw_Select0}
for more details.}
\label{WSB12p_Ehw_Select2}
\end{figure}

%

\begin{figure}[b!]
\centerline{\includegraphics[width=\columnwidth]{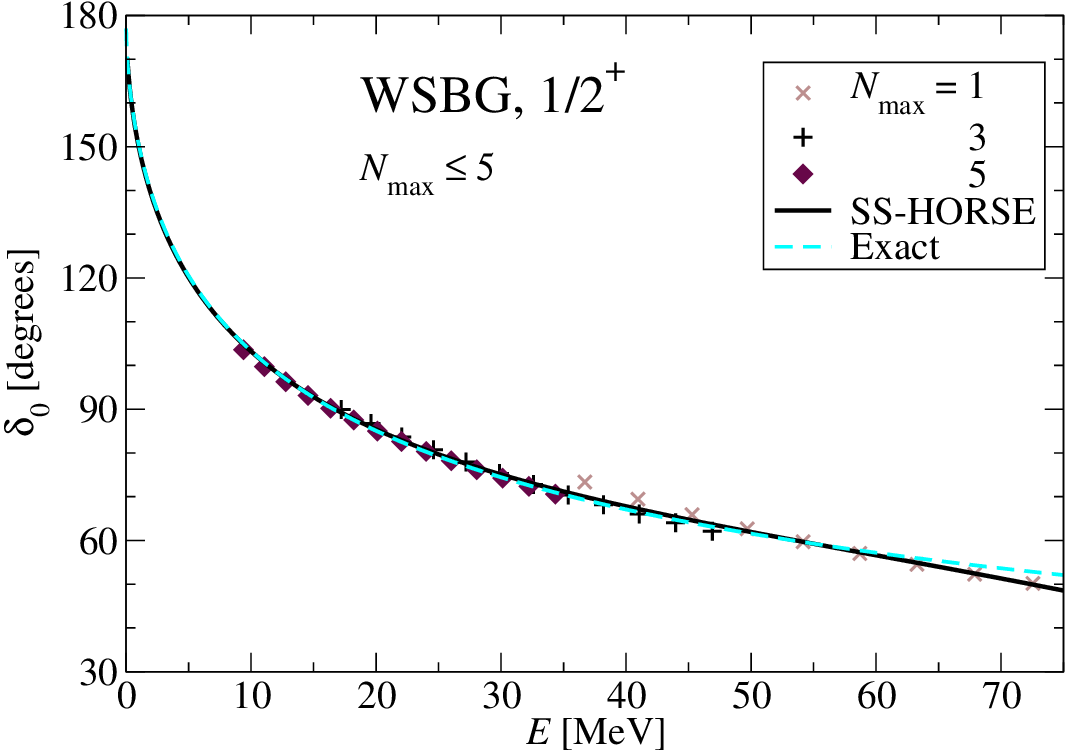}}
\caption{The $\frac12^{+}$ WSBG phase shifts generated by the selected eigenstates~$E_{1}$ obtained with ${N_{\max}\leq6}$. See
Fig.~\ref{WSB12p_deltaE_Select0} for details.}
\label{WSB12p_deltaE_Select2}
\end{figure}

To examine a possibility of  describing the low-energy $\frac12^{+}$ phase shifts using only the diagonalization results in small basis spaces,
we remove from the previous selection the eigenenergies~$E_{1}$ obtained with~$N_{\max}>5$ as is illustrated by
Figs.~\ref{WSB12p_Ehw_Select2} 
and~\ref{WSB12p_deltaE_Select2}. We obtain nearly the same values of the fitting parameters as is
seen from Table~\ref{WSBs12}. The largest though still small enough difference is obtained for the fitted~$E_{b}$ values which, as has been
already noted, does not play an essential role in the phase shifts. Therefore it is not surprising that we get an excellent description of the 
exact phase shifts presented in Fig.~\ref{WSB12p_deltaE_Select2}. Figure~\ref{WSB12p_Ehw_Select2} demonstrates that we describe
accurately not only the eigenstates~$E_{1}$ involved in the fitting procedure but also those obtained in much larger basis spaces
which were not fitted. The rms deviation in the description of energies of all~$\Lambda>385$~MeV/$c$ selected eigenstates is exactly the same
as in the case when all these eigenstates were included in the fit.

{\colour{red}\subsection{Scaling and convergence trends}}
As we  already noted, the scaling of the eigenstates of finite Hamiltonian matrices in oscillator basis has been proposed 
by S.~Coon and collaborators in Refs.~\cite{Coon, CoonNTSE12} who studied the convergence patterns of the bound states. They have
demonstrated that the eigenenergies~$E_{\nu}$ as functions of the scaling parameter~$\lambda_{sc}\sim \sqrt{s}$ tend to a constant 
as~$\lambda_{sc}$ approaches~0; this constant is the convergence limit of the respective eigenenergy in the infinite basis. Our study
extends the scaling patterns of the harmonic oscillator eigenstates to the case of states in the continuum. In this case the eigenenergies should 
approach~0 as the basis is expanded infinitely. The solid 
line in Fig.~\ref{WSB12p_Es0}
demonstrates the behaviour of eigenenergies in the continuum~$E_{1}$ as a function of the scaling parameter~$s$ in the case of a system
which has a bound state and does not have resonances in the low-energy region; the respective low-energy 
phase shifts are described by Eq.~\eqref{S_phase_b}, a general
formula for this case. The eigenstates are seen to be a smooth monotonic function of~$s$ (or~$\lambda_{sc}$) which tends, as
expected, to zero as~$s\to0$.

{\colour{red}It is seen in Fig.~\ref{WSB12p_Es0} that in the high-energy region the WSBG eigenstates deviate from the solid curve 
presenting the solutions of Eq.~\eqref{S_phase_implicit} for the respective~$N_{\max}$ and~$\hbar\Omega$ values. Note, these 
eigenstates correspond to small~$N_{\max}$ values for which the scaling condition~\eqref{if1} at large energies is not fulfilled and hence the 
scaling properties~\eqref{fnl_as} and~\eqref{SSJM_phase_sc} become inaccurate. As a result, the solutions of Eq.~\eqref{S_phase_implicit}
plotted as a function of the scaling parameter~$s$ deviate from the  WSBG eigenstates while the same WSBG eigenstates are perfectly 
described by the Eq.~\eqref{S_phase_implicit} solutions when plotted as functions of~$\hbar\Omega$ for each~$N_{\max}$ in
Fig.~\ref{WSB12p_Ehw_Select0}. The inaccuracy of the scaling is much less pronounced in Fig.~\ref{WSB32m_Es_Select0} where
the energies are much smaller.}

\begin{figure}[t!]
\centerline{\includegraphics[width=\columnwidth]{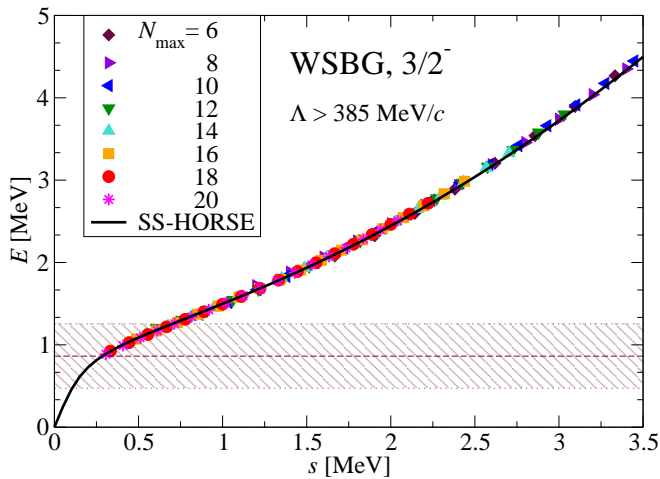}}
\caption{The same as Fig.~\ref{WSB32m_Es_Select0} but in a larger scale. The dashed line corresponds to the
resonance energy~$E_{r}$, the shaded area shows the resonance width.}
\label{WSB32m_res}
\end{figure}

The solid lines in 
Figs.~\ref{WSB32m_Es_Select0} and~\ref{WSB32m_Es_Select2}
demonstrate the behaviour of the eigenstates~$E_{0}$  as a function of the scaling parameter~$s$ when the low-energy 
phase shifts are given by Eq.~\eqref{S_phase_r} which is a general
formula describing a system which does not have a bound state but has a low-energy resonance. We see again a smooth monotonically
increasing function of~$s$ with a large enough derivative at large~$s$. At smaller~$s$ when the energy approaches the resonant region,
the derivative of~$E_{0}(s)$ decreases; this decrease of the derivative is more pronounced for narrow resonances.
Figure~\ref{WSB32m_res} where the function~$E_{0}(s)$ from
Fig.~\ref{WSB32m_Es_Select0} is shown in a larger scale together with the resonant region, demonstrates that the further decrease
of~$s$ strongly enhances the derivative of this function at the energies below the resonance energy~$E_{r}$. When the 
function~$E_{0}(s)$ leaves the resonant region at smaller~$s$ values, the next eigenstate~$E_{1}(s)$ (not shown in the figure)
approaches  the resonant region from above.

These are the general convergence trends of the  positive energy eigenstates obtained in the oscillator basis.  

Concluding this section, we have demonstrated using the WSBG potential as an example that the proposed SS-HORSE technique is
adequate for the description of low-energy scattering phase shifts and 
resonance energies~$E_{r}$ and
widths~$\Gamma$. A very encouraging sign for many-body shell-model applications is that 
the resonance parameters and phase shifts can be obtained nearly without loosing the accuracy by 
using within the SS-HORSE approach only
the Hamiltonian eigenstates obtained in small basis spaces; more, having the low-lying energies from small basis spaces 
we are able to `predict'  accurately the values of 
eigenenergies in much larger
oscillator bases.\\

\section{\boldmath SS-HORSE NCSM calculation of resonances in $n\alpha$ scattering\label{sec:5He}}

We discuss here the application of our SS-HORSE technique to  $n\alpha$ scattering phase shifts and resonance parameters based 
on {\em ab initio} many-body calculations of $^{5}$He 
within the NCSM with the realistic JISP16 $NN$ interaction.
The NCSM calculations are performed using the code MFDn~\cite{mamazfdn1,mamazfdn2} with~$2\leq N_{\max}\leq18$ for both parities and 
with~$\hbar\Omega$ values ranging from 10 to 40~MeV in steps of 2.5~MeV.

As it has been already noted above, for the SS-HORSE analysis we need the $^{5}$He energies relative to the $n+\alpha$
threshold. Therefore from each of the  $^{5}$He NCSM odd (even) parity
eigenenergies we subtract the $^{4}$He ground state energy
obtained by the NCSM with the same~$\hbar\Omega$    and the same~$N_{\max}$ (with~$N_{\max}-1$) excitation quanta, 
and in what follows  these subtracted energies are called NCSM
eigenenergies~$E_{\nu}$. Only these $^{5}$He NCSM eigenenergies relatively to  the $n+\alpha$ threshold
are discussed below.

{\colour{red}We note here that the NCSM utilizes the truncation based on the many-body oscillator quanta~$N_{\max}$ while the SS-HORSE
requires the oscillator quanta truncation of the interaction describing the relative motion of neutron and $\alpha$ particle. A justification
of using~$N_{\max}$ for the SS-HORSE analysis is obvious if the $\alpha$ particle is described by the simplest four-nucleon oscillator function
with excitation quanta~$N^{\alpha}_{\max}=0$.  Physically it is clear that the use of~$N_{\max}$ within the SS-HORSE should work well also
in a more general case when the $\alpha$ particle is presented by the wave function with~$N^{\alpha}_{\max}>0$ due to the
dominant role of the zero-quanta component in the $\alpha$ particle wave function. Instead of trying to rigorously justify the use 
of~$N_{\max}$ within the SS-HORSE by lengthy algebraic manipulations,
we suggest an \mbox{\em a posteriori} justification: 
we demonstrate below that we obtain $n\alpha$ phase shift parametrizations consistent with the NCSM results obtained with 
very different~$N_{\max}$
and~$\hbar\Omega$ values; more, we are able to `predict' the NCSM results with large~$N_{\max}$ using the phase shift parametrizations
based on the NCSM calculations with much smaller model spaces. It will be clearly impossible if the use of~$N_{\max}$ truncation for
the SS-HORSE analysis will not work properly.

}

\begin{figure}[t!]
\centerline{\includegraphics[width=\columnwidth]{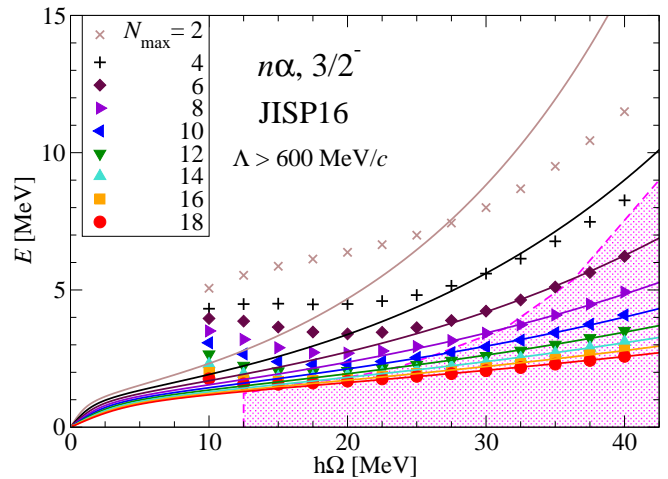}}
\caption{The lowest $^{5}$He $\frac32^{-}$ eigenstates~$E_{0}$  obtained by the NCSM with various~$N_{\max}$ (symbols)
as  functions of~$\hbar\Omega$. The shaded area shows the~$E_{0}$ values selected for the SS-HORSE analysis according
to inequality~$\Lambda>600$~MeV/$c$. Solid lines are solutions of Eq.~\eqref{ImplicitENhw} for energies~$E_{0}$ with parameters~$a$,
$b$ and~$d$ obtained by the fit.}
\label{n-4He32m_Ehw_Select0}
\end{figure}

\subsection{\boldmath Partial wave $\frac32^{-}$}
\begin{figure}[b!]
\centerline{\includegraphics[width=\columnwidth]{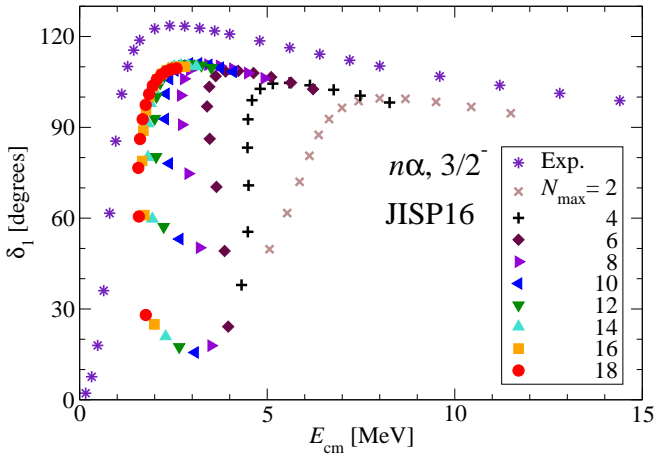}}
\caption{The $\frac32^{-}$ $n\alpha$ phase shifts obtained directly from  the $^{5}$He eigenstates~$E_{0}$ using Eq.~\eqref{tandelt-SS} and the
phase shift analysis of experimental data of Refs.~\cite{naExpB} (stars).}
\label{n-4He32m_deltaE}
\end{figure}

We utilize the same Eq.~\eqref{ImplicitENhw} to fit the parameters describing the low-energy~$\frac32^{-}$ and~$\frac12^{-}$ phase
shifts as in the model problem; the only difference is that the lowest energy eigenstates~$E_{0}$ are obtained now from the 
many-body NCSM
calculations. These lowest $\frac32^{-}\vphantom{1_{q_{\int}}}$ NCSM eigenstates are shown in Fig.~\ref{n-4He32m_Ehw_Select0} as functions
of~$\hbar\Omega$ for various~$N_{\max}$ values. Figure~\ref{n-4He32m_Es}\strut\ presents these eigenstates~$E_{0}$ as a function of the
scaling parameter~$s$ while Fig.~\ref{n-4He32m_deltaE} presents the  $\frac32^{-}$ phase shifts obtained directly from  them
using Eq.~\eqref{tandelt-SS}. Figures~\ref{n-4He32m_Es} and~\ref{n-4He32m_deltaE} clearly demonstrate the  need of the
eigenstate selection since many  points in these figures deviate strongly from the common curves formed by other points. On the other hand,
these figures demonstrate the convergence achieved in large~$N_{\max}$ calculations: the deviation from the common curves occurs  
at smaller~$\hbar\Omega$ values as~$N_{\max}$ increases and all results from the largest available NCSM basis spaces seem
to lie on the single common curves with the exception of only very few eigenenergies obtained with~$\hbar\Omega<15$~MeV.

\begin{figure}[t!]
\centerline{\includegraphics[width=\columnwidth]{n-4He32m_Es.eps}}
\caption{The lowest  $^{5}$He $\frac32^{-}$ eigenstates~$E_{0}$  
as a function of the scaling parameter~$s$.}
\label{n-4He32m_Es}
\end{figure}

\begin{figure}[t!]
\centerline{\includegraphics[width=\columnwidth]{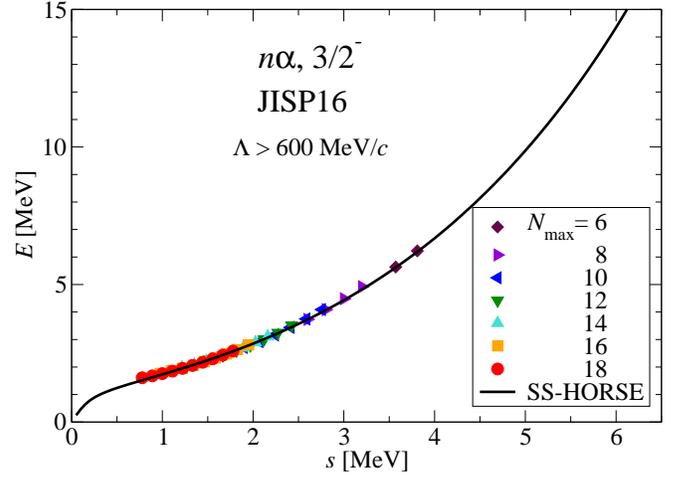}}
\caption{The $^{5}$He  $\frac32^{-}$ eigenstates~$E_{0}$ selected according to~$\Lambda>600$~MeV/$c$
plotted as a function of the scaling parameter~$s$ (symbols). See Fig.~\ref{WSB32m_Es_Select0} for other details.}
\label{n-4He32m_Es_Select0}
\end{figure}

\begin{figure}[b!]
\centerline{\includegraphics[width=\columnwidth]{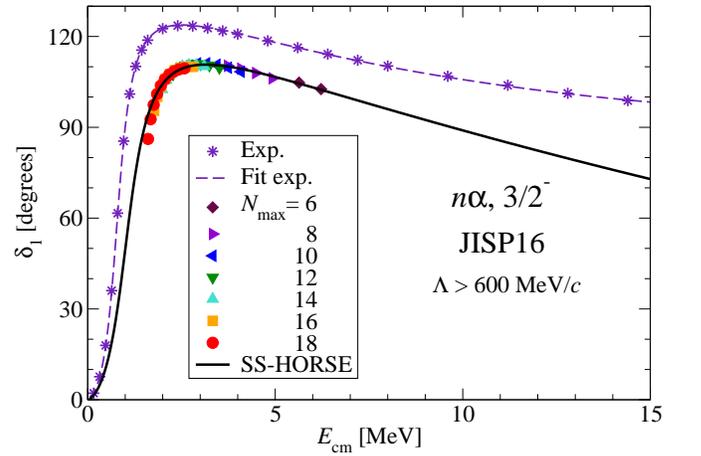}}
\caption{The $\frac32^{-}$ $n\alpha$ phase shifts obtained using Eq.~\eqref{tandelt-SS} 
directly from  $^{5}$He  eigenstates~$E_{0}$ selected according to~$\Lambda>600$~MeV/$c$ (symbols).
The solid curve depicts the phase shifts of Eq.~\eqref{S_phase_r}  
with parameters~$a$, $b$ and~$d$ obtained by the fit with the respective selection of eigenstates; stars and the dashed curve depict the
phase shift analysis of experimental data of Refs.~\cite{naExpB} and the fit by Eq.~\eqref{S_phase_r}.}
\label{n-4He32m_deltaE_Select0}
\end{figure}

Our first selection is the eigenstates fitting inequality~$\Lambda>600$~MeV/$c$, the value recommended in Refs.~\cite{Coon, CoonNTSE12} for the JISP16
$NN$ interaction. This selection is illustrated by the shaded area in Fig.~\ref{n-4He32m_Ehw_Select0}; 
common curves are formed
by the selected eigenenergies~$E_{0}$ plotted as a function of the scaling parameter~$s$ in Fig.~\ref{n-4He32m_Es_Select0} and by
the phase shifts obtained directly from these eigenenergies with the help of Eq.~\eqref{S_phase_r} in Fig.~\ref{n-4He32m_deltaE_Select0}.
We  get an accurate fit of the selected NCSM eigenenergies with the rms deviation of 31~keV, the obtained values of the fitting 
parameters~$a$, $b$, $d$ of Eq.~\eqref{ImplicitENhw} and the $\frac32^{-}$ resonance
energy~$E_{r}$ and width~$\Gamma$ are presented in Table~\ref{n-4Hep32}. The fit accuracy is also illustrated by solid lines 
in Figs.~\ref{n-4He32m_Ehw_Select0},
\ref{n-4He32m_Es_Select0} and~\ref{n-4He32m_deltaE_Select0} obtained using our fitting parameters: these curves are seen to
reproduce the selected NCSM energies~$E_{0}$ in Figs.~\ref{n-4He32m_Ehw_Select0} and~\ref{n-4He32m_Es_Select0} and the 
corresponding phase shifts in Fig.~\ref{n-4He32m_deltaE_Select0}.

 \begin{table*}[t!]
\caption{$\frac32^{-}$ resonance in $n\alpha$ scattering from the  $^{5}$He NCSM calculations with JISP16 $NN$ interaction: fitting parameters~$a$, $b$, $d$ of Eq.~\eqref{ImplicitENhw},
resonance energy~$E_{r}$ and width~$\Gamma$, rms deviation of fitted energies~$\Xi$ and the number of these fitted energies~$D$
for different selections of eigenvalues in comparison with the analysis of experimental data in various approaches
of Refs.~\cite{Hale} and~\cite{PRC79} and with the
fit  by Eq.~\eqref{S_phase_r} of the phase shifts~$\delta_{1}$ extracted from experimental data in Ref.~\cite{naExpB}. 
For the~${N_{\max}\leq4}$ selection, $\Xi$ and~$D$ for all energies from the manual selection are shown  within brackets.}
\label{n-4Hep32}  
\begin{ruledtabular}
\begin{tabular}{@{\hspace{6pt}}c@{\hspace{6pt}}c@{\hspace{6pt}}c@{\hspace{6pt}}c@{\hspace{6pt}}c@{\hspace{6pt}}c@{\hspace{6pt}}c@{\hspace{6pt}}c@{\hspace{6pt}}}
\raisebox{-1.8ex}[0pt][0pt]{Selection} & $a$ & $b^2$ & $d\cdot10^4$ & $E_r$ & $\Gamma$ & $\Xi$ & \raisebox{-1.8ex}[0pt][0pt]{$D$} \\
 & (MeV$^\frac12$) & (MeV) & (MeV$^{-\frac32}$) & (MeV) & (MeV) & (keV) &  \\
\noalign{\smallskip}\hline\noalign{\smallskip}
$\Lambda>600$~MeV/$c$ & 0.505 & 1.135 & $-$0.9 & 1.008 & 1.046 & 31 & 46 \\
Manual & 0.506 & 1.019 & $+$93.2 & 0.891 & 0.989 & 70 & 68 \\
$N_{\max}\leq4$ & 0.515 & 1.025 & $+$101 & 0.892 & 1.008 & 106(81) & 11(68) \\[1ex] 
\multicolumn{3}{c}{Nature:} \\
$R$-matrix \cite{Hale} & & & & 0.80 & 0.65 & & \\
$J$-matrix \cite{PRC79} & & & & 0.772 & 0.644 & & \\
Fit $\delta_{1}$ of Ref.~\cite{naExpB}& 0.358 & 0.839 & $+$55.9 & 0.774 & 0.643 & 0.21$^\circ$ & 26\\
\end{tabular}
\end{ruledtabular}
\end{table*}

The JISP16 $NN$ interaction generates the $\frac32^{-}$ phase shifts reproducing qualitatively but not quantitively the results of phase
shift analysis of Refs.~\cite{naExpB}  of $n\alpha$ scattering data as is
seen in  Fig.~\ref{n-4He32m_deltaE_Select0}. We obtain the resonance energy~$E_{r}$ slightly above the experimental value, the difference
is about 0.2~MeV (see Table~\ref{n-4Hep32}). The resonance width~$\Gamma$ is also overestimated by JISP16, the difference between the
JISP16 prediction and experiment is about 0.4~MeV. We present in Fig.~\ref{n-4He32m_deltaE_Select0} and in the last row
of Table~\ref{n-4Hep32} also the fit by Eq.~\eqref{S_phase_r}
of the phase shift analysis of experimental data of Refs.~\cite{naExpB} obtained by minimizing the rms deviation of the phase shifts (column~$\Xi$ 
in the Table). The  fit parameters derived from the experimental data are seen to be markedly different from those derived from JISP16
 by the NCSM-SS-HORSE approach.

Returning to the $\frac32^{-}$ $^{5}$He eigenstates depicted in Fig.~\ref{n-4He32m_Ehw_Select0}, we see that the solid curves 
presenting our fit in this figure describe not only the selected eigenstates from the shaded area but also many other eigenstates not involved in the
fit. This signals that the~$\Lambda>600$~MeV/$c$ selection is too restrictive and we can use for the SS-HORSE analysis and fits
many more NCSM eigenstates. We can use within the SS-HORSE approach all eigenstates forming with the others a common curve in Fig.~\ref{n-4He32m_Es}
and especially in Fig.~\ref{n-4He32m_deltaE} which is, as we have noted, more sensitive to convergence patterns. There is however a restriction:
unacceptable for the SS-HORSE are  eigenstates~$E_{\nu}$ obtained with any given~$N_{\max}$ from the range of~$\hbar\Omega$
values where their energy decreases with~$\hbar\Omega$, i.\,e., we can select only those eigenstates with a  given fixed~$N_{\max}$ which 
derivative~$\frac{E_{\nu}}{\hbar\Omega}>0$~--- Eqs.~\eqref{ImplicitENhw} and~\eqref{S_phase_implicit} do not exclude mathematically
the possibility of having~$\frac{E_{\nu}\vphantom{^{T}}}{\hbar\Omega}<0$ but such solutions can arise only with unphysical
parameters of these equations.

We would like to use within the SS-HORSE as many NCSM eigenstates as possible in order to enlarge the energy interval where the phase
shifts are fitted and to improve the accuracy of 
the fit parameters. From this point of view, the selection according to
inequality~$\Lambda>\Lambda_{0}$ is not favorable. The~$\Lambda>\Lambda_{0}$ rule excludes states with~$\hbar\Omega<\hbar\Omega_{0}$ 
where~$\hbar\Omega_{0}$  depends on~$N_{\max}$ and decreases as~$N_{\max}$ increases.
As is seen from our study of the model problem, in particular, from
Figs.~\ref{WSB32m_deltaE}, \ref{WSB32m_Ehw_Select0}, 
\ref{WSB12p_Ehw_Select0},
\ref{WSB12p_deltaE}, we can utilize for the SS-HORSE the eigenstates obtained with sufficiently large~$N_{\max}$ and with very 
small~$\hbar\Omega$; the same conclusion follows from our {\em ab initio} many-body study of the system of four neutrons (tetraneutron)
in the continuum~\cite{4n}.
According to the~$\Lambda>\Lambda_{0}$ rule we either exclude these large~$N_{\max}$\,--\,small~$\hbar\Omega$
eigenstates or include in the fit some small~$N_{\max}$ states which strongly deviate from common curves on the plots
of~$E_{\nu}$ vs~$s$ or~$\delta_{\ell}$ vs~$E$.

The ultraviolet cutoff~$\Lambda_{0}$ was introduced in Refs.~\cite{Coon, CoonNTSE12}
with an idea that the oscillator basis should be able to describe in the many-body system  the short-range (high-momentum)
behaviour of the two-nucleon
interaction employed in the calculations; thus the~$\hbar\Omega$ cannot be too small since oscillator functions with small~$\hbar\Omega$
have a large radius  (corresponding to small momentum) and are not able to catch the short-range (high-momentum) peculiarities of a particular $NN$ potential.
We imagine this concept to be insufficient  at least in some cases. In light nuclei where binding energies are not large, the
structure of the wave function can be insensitive to the short-range $NN$ potential behaviour associated with high relative momentum. Much more
important is the radius of the state under consideration, e.\,g., we can expect an adequate description of the ground state only if 
the highest oscillator function in the basis has at least one node 
within the radius of this state, two nodes are required within the radius of the first excited state, etc. Therefore the minimal 
acceptable~$\hbar\Omega$ value depends strongly on the state under consideration and may be insensitive to the inter-nucleon 
interaction. This is particularly important for loosely-bound nuclear states or for low-energy scattering states. In the case of scattering, the
wave function at low energies can have a very distant first node and not only permits but just requires the use of oscillator functions
with very small~$\hbar\Omega$ values and large radius.

\begin{figure}[t!]
\centerline{\includegraphics[width=\columnwidth]{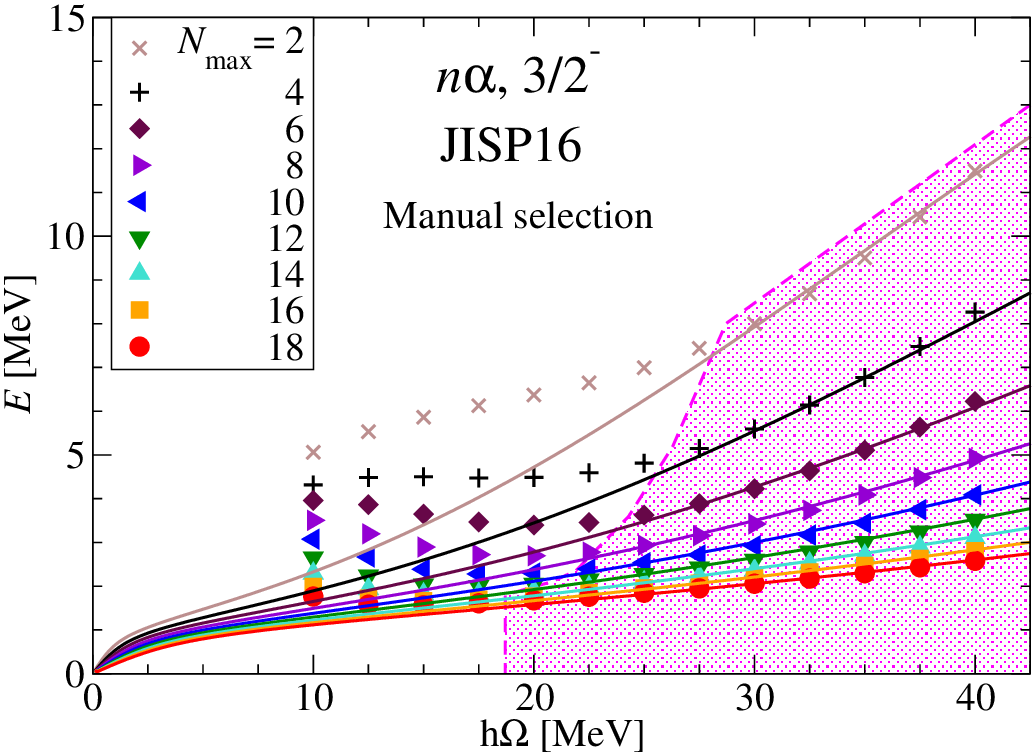}}
\caption{The lowest  $^{5}$He $\frac{3}{2}^{-}$ eigenstates~$E_{0}$ (symbols) and their manual selection (shaded area).
See Fig.~\ref{n-4He32m_Ehw_Select0} for  more details.}
\label{n-4He32m_Ehw_Select1}
\end{figure}
\begin{figure}[b!]
\centerline{\includegraphics[width=\columnwidth]{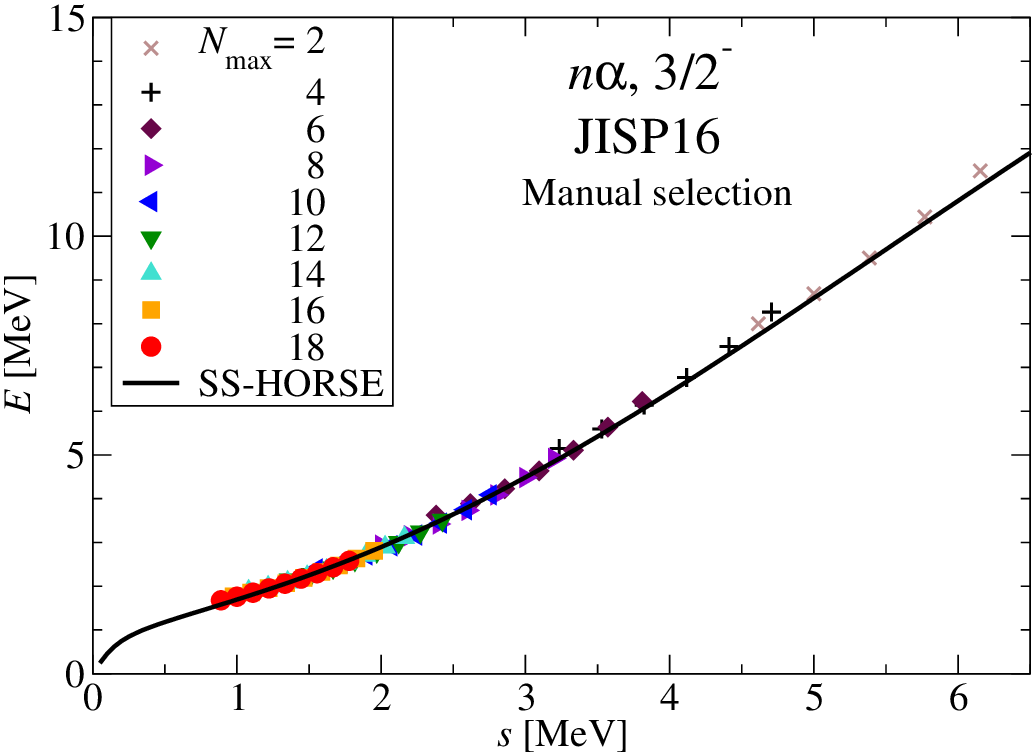}}
\caption{Manually selected $^{5}$He  $\frac{3}{2\protect\vphantom{_{q}}}^{-}$ eigenstates~$E_{0}$ plotted as a function of the scaling parameter~$s$ (symbols). 
See Fig.~\ref{WSB32m_Es_Select0} for other details.}
\label{n-4He32m_Es_Select1}
\end{figure}
\begin{figure}[t!]
\centerline{\includegraphics[width=\columnwidth]{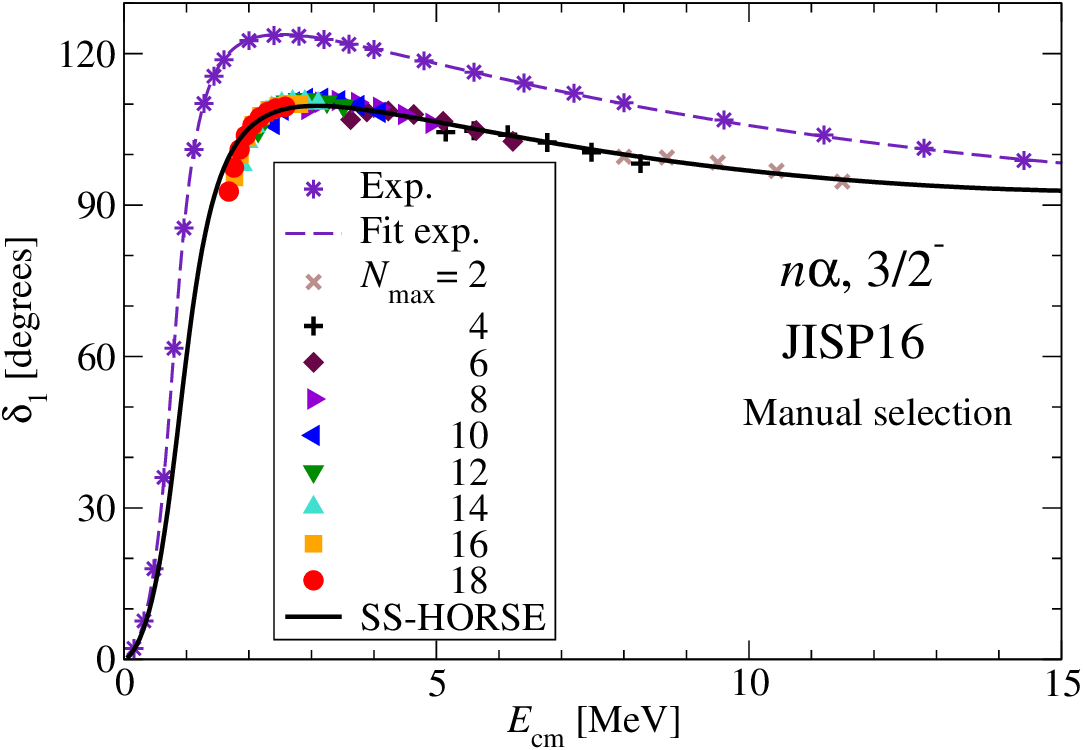}}
\caption{The $\frac{3}{2\protect\vphantom{_{q}}}^{-}$ $n\alpha$ phase shifts generated by the manually selected  $^{5}$He eigenstates~$E_0$. 
See Fig.~\ref{n-4He32m_deltaE_Select0} for details.}
\label{n-4He32m_deltaE_Select1}
\end{figure}

\begin{figure}[b!]
\centerline{\includegraphics[width=\columnwidth]{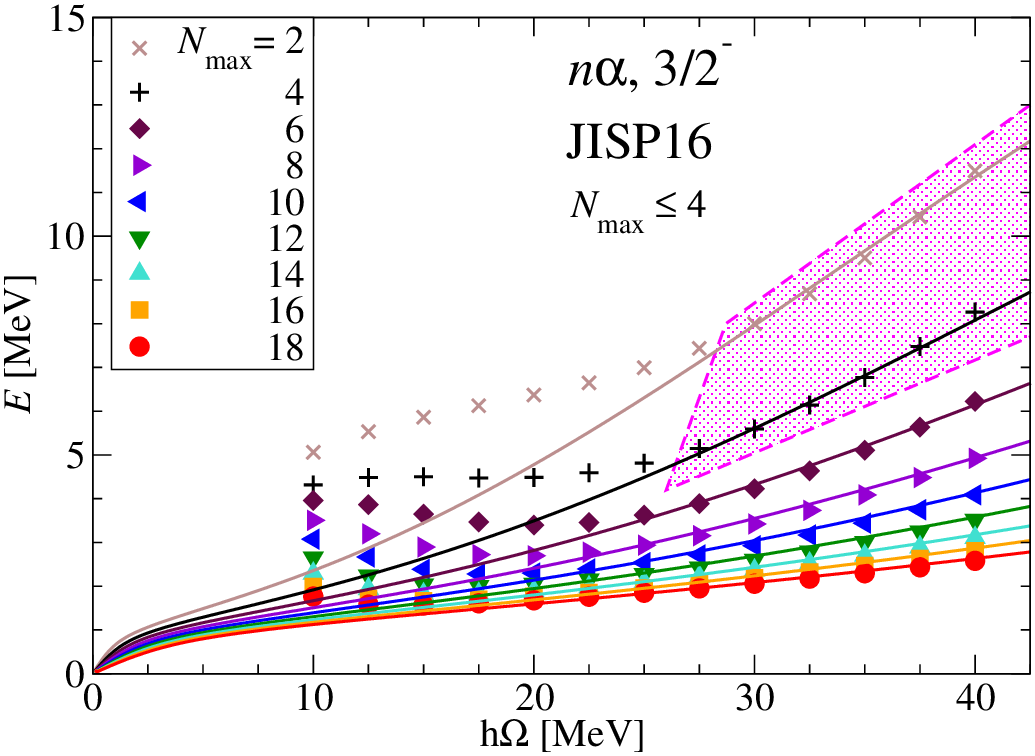}}
\caption{The lowest $^{5}$He $\frac32^{-}$ eigenstates~$E_{0}$ (symbols) and their ${N_{\max}\leq4}$ selection (shaded area).
See Fig.~\ref{n-4He32m_Ehw_Select0}
for more details.}
\label{n-4He32m_Ehw_Select2}
\end{figure}

We cannot formulate a simple rule or formula for selecting eigenstates acceptable for the SS-HORSE analysis, instead we pick up
manually individual states with eigenenergies~$E_{0}$ lying to the right from the minimum of the~$\hbar\Omega$ dependence for
each~$N_{\max}$ in Fig.~\ref{n-4He32m_Ehw_Select0} and lying on or close to the common curve in Figs.~\ref{n-4He32m_Es}
and~\ref{n-4He32m_deltaE}. These manually selected eigenstates and the respective phase shifts
are presented in Figs.~\ref{n-4He32m_Ehw_Select1}, \ref{n-4He32m_Es_Select1}
and~\ref{n-4He32m_deltaE_Select1}. The results of the fit with this selection of eigenstates are presented in the second line of Table~\ref{n-4Hep32}. 
We obtain an accurate fit with the rms deviation of eigenenergies of 70~keV; this number however depends on the selection
criteria like the acceptable distance from the common curve formed by other points in Figs.~\ref{n-4He32m_Es_Select1}
and~\ref{n-4He32m_deltaE_Select1}. Comparing Figs.~\ref{n-4He32m_Ehw_Select0} and~\ref{n-4He32m_Ehw_Select1} we see that
our manual selection makes it possible to describe eigenenergies with small~$N_{\max}$ which were far from theoretical curves
in Fig.~\ref{n-4He32m_Ehw_Select0}. These small~$N_{\max}$ states have large energies, and their inclusion in the SS-HORSE
analysis extends the description of the phase shifts in the high-energy region in Fig.~\ref{n-4He32m_deltaE_Select1} pushing them
closer to the phase shift analysis of the experimental $n\alpha$ scattering data in this region as compared 
with Fig.~\ref{n-4He32m_deltaE_Select0}. These changes in the phase shift behavior at larger energies correspond to a drastic change
of the fitting parameter~$d$ which is the coefficient of the highest power term in the expansion~\eqref{S_phase_r}. At smaller energies
including the resonance region, the phase shifts obtained from the fits with the manual and with the~~$\Lambda>600$~MeV/$c$ 
selections are
nearly the same, and we get close values of the respective fitting parameters~$a$ and~$b$ and hence small changes of the 
resonance energy~$E_{r}$ and width~$\Gamma$ due to the switch from one selection to the other.

\begin{figure}[!t]
\centerline{\includegraphics[width=\columnwidth]{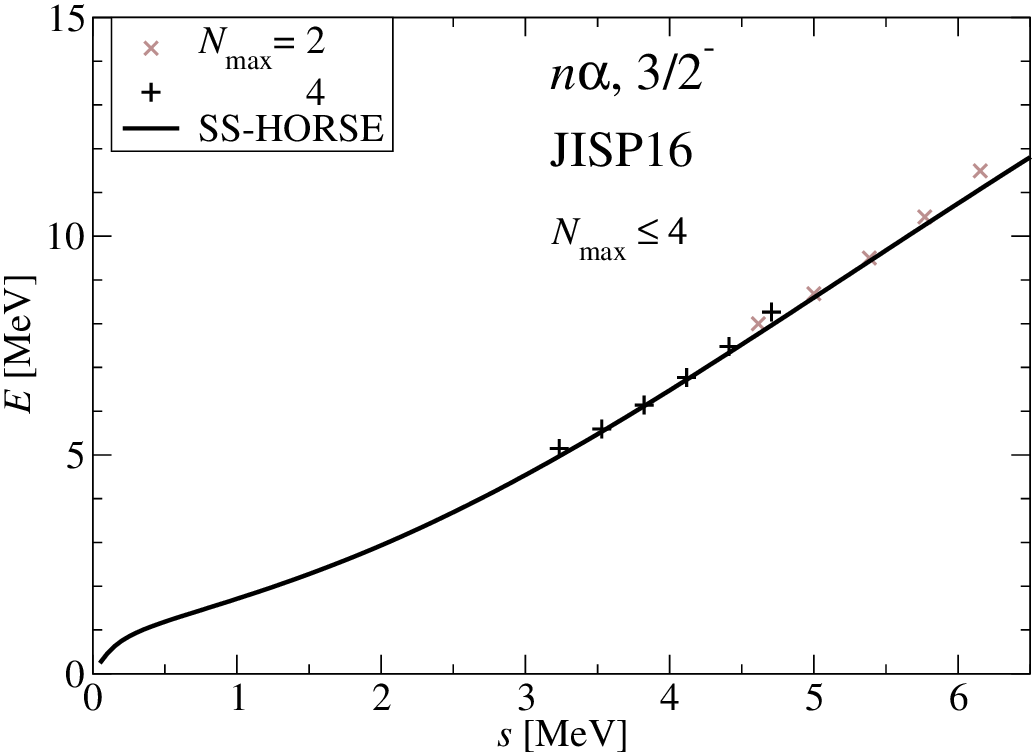}}
\caption{Selected lowest $^{5}$He $\frac32^{-}$ eigenstates~$E_{0}$ obtained in NCSM  
with ${N_{\max}\leq4}$ as a function of the scaling parameter~$s$. 
See Fig.~\ref{WSB32m_Es_Select0} for other details.\strut}
\label{n-4He32m_Es_Select2}
\end{figure}

\begin{figure}[!b]
\centerline{\includegraphics[width=\columnwidth]{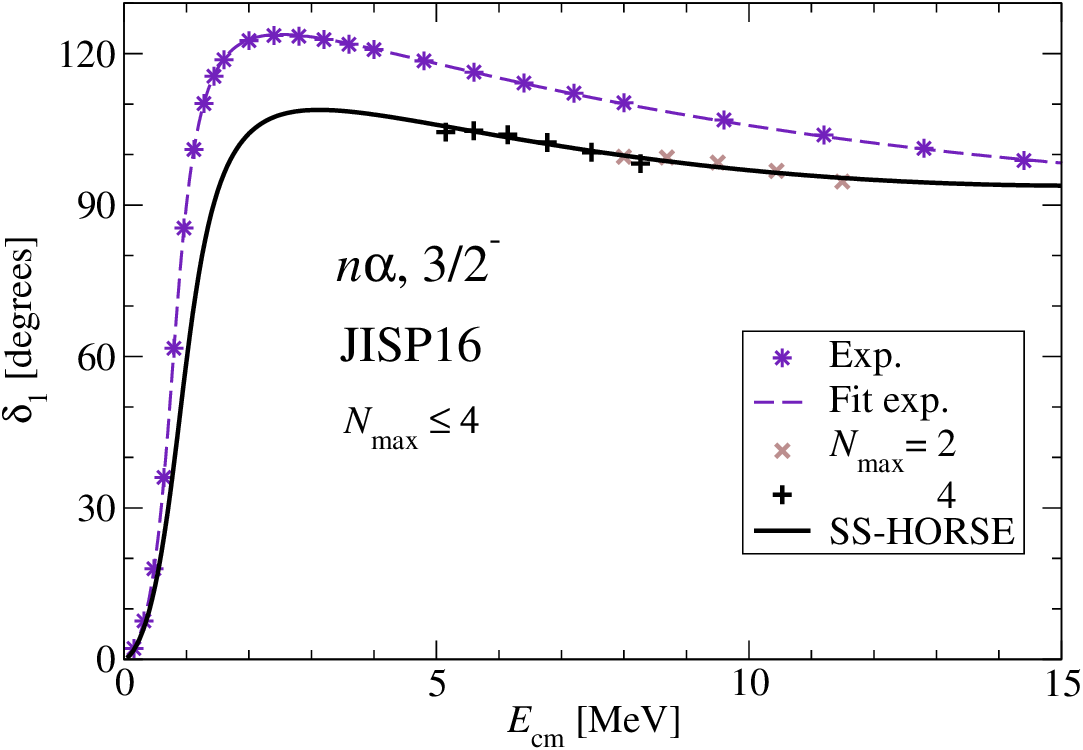}}
\caption{The $\frac{3}{2\protect\vphantom{_{q}}}^{-}$ $n\alpha$ phase shifts generated by the selected $^{5}$He eigenstates~$E_{0}$ obtained in NCSM 
with ${N_{\max}\leq4}$. See
Fig.~\ref{n-4He32m_deltaE_Select0} for details.}
\label{n-4He32m_deltaE_Select2}
\end{figure}

It is very interesting to investigate whether  we can get reasonable phase shifts and resonance parameters using only the NCSM eigenstates from
small basis spaces. From our manually selected eigenstates we select only those obtained with~$N_{\max}=2$ and~4. This selection
and the results obtained by the fit are depicted in Figs.~\ref{n-4He32m_Ehw_Select2}, \ref{n-4He32m_Es_Select2} 
and~\ref{n-4He32m_deltaE_Select2}. All eigenenergies~$E_{0}$ involved in this fit are significantly above the resonant region (see
Fig.~\ref{n-4He32m_deltaE_Select2}). Nevertheless we obtain from these 11~small-$N_{\max}$ eigenstates nearly the same phase shifts
as those from all 68~manually selected eigenstates  and very close values of fit parameters and of the resonance energy and width presented in
Table~\ref{n-4Hep32}. Figure~\ref{n-4He32m_Ehw_Select2} demonstrates that, as in the case of the model problem, with these
eigenstates~$E_{0}$ from many-body NCSM calculations with~$N_{\max}\leq4$ we can accurately `predict' the $^{5}$He eigenstates
obtained in much larger basis spaces and in  a wider range of~$\hbar\Omega$. The rms deviation~$\Xi$ of all manually selected eigenstates by
this~${N_{\max}\leq4}$ fit is only 81~keV as compared with 70~keV from the fit to all those  eigenstates.

\subsection{\boldmath Partial wave $\frac12^{-}$}

\begin{figure}[t!]
\centerline{\includegraphics[width=\columnwidth]{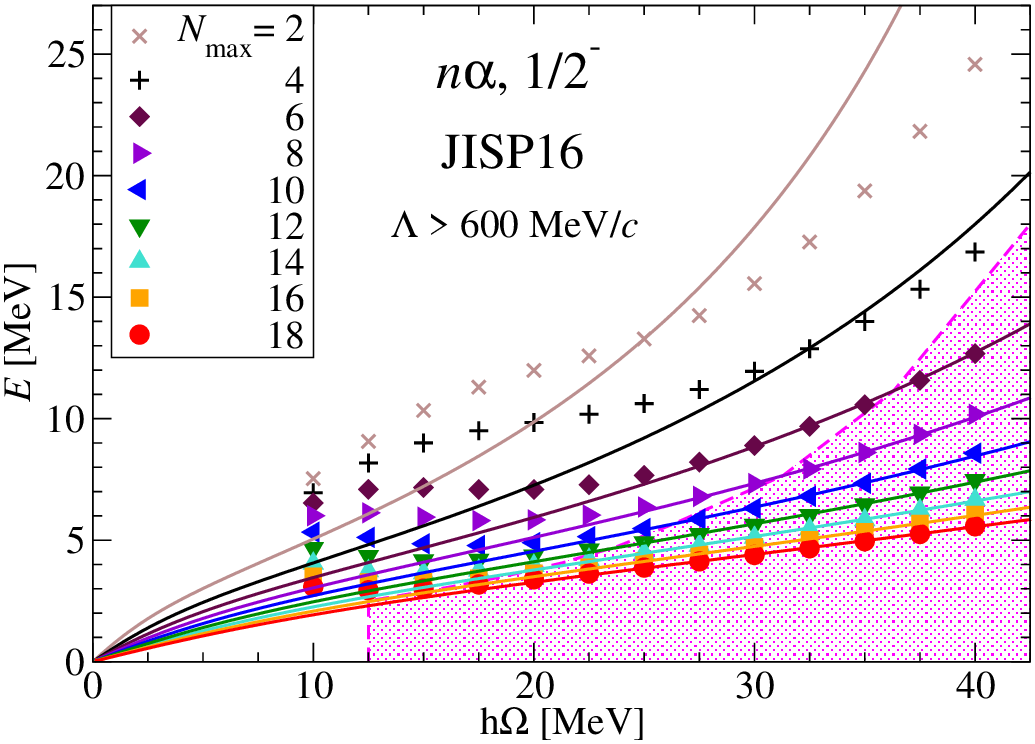}}%
\caption{The lowest  $^{5}$He $\frac12^{-}$ eigenstates~$E_{0}$ (symbols) 
and their~$\Lambda>600$~MeV/$c$ selection (shaded area). 
See Fig.~\ref{n-4He32m_Ehw_Select0}  for more details.}
\label{n-4He12m_Ehw_Select0}
\end{figure}

\begin{figure}[!b]
\centerline{\includegraphics[width=\columnwidth]{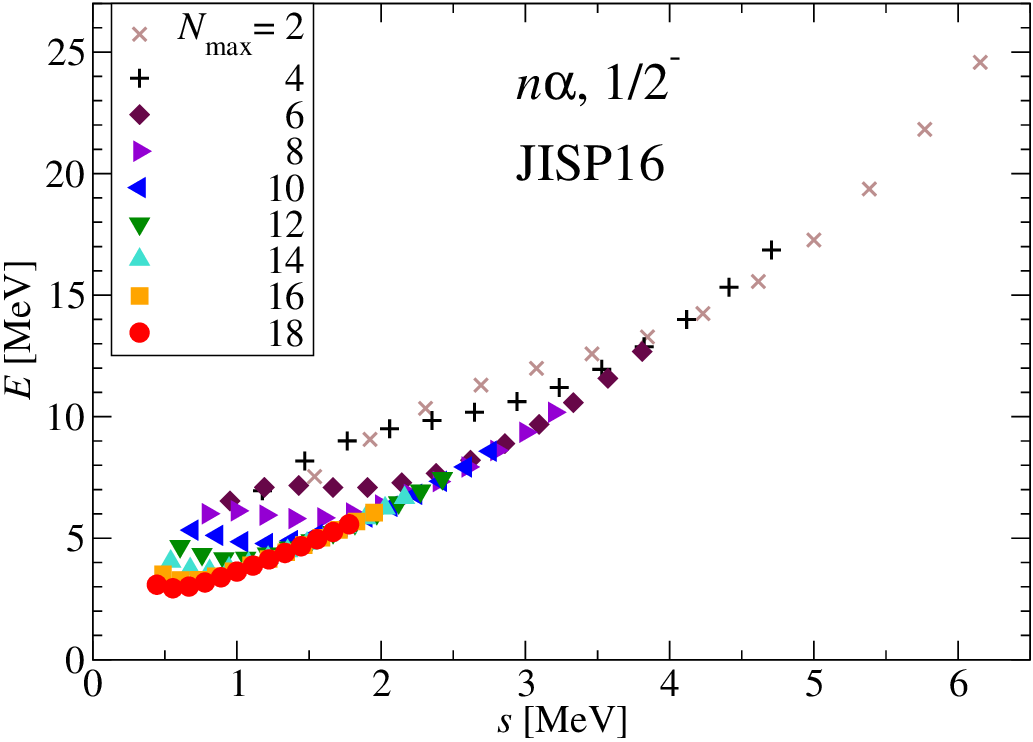}}
\caption{The  lowest  $^{5}$He $\frac12^{-}$ eigenstates~$E_{0}$ as a function of the scaling parameter~$s$.} 
\label{n-4He12m_Es}
\end{figure}

\begin{figure}[t!]
\centerline{\includegraphics[width=\columnwidth]{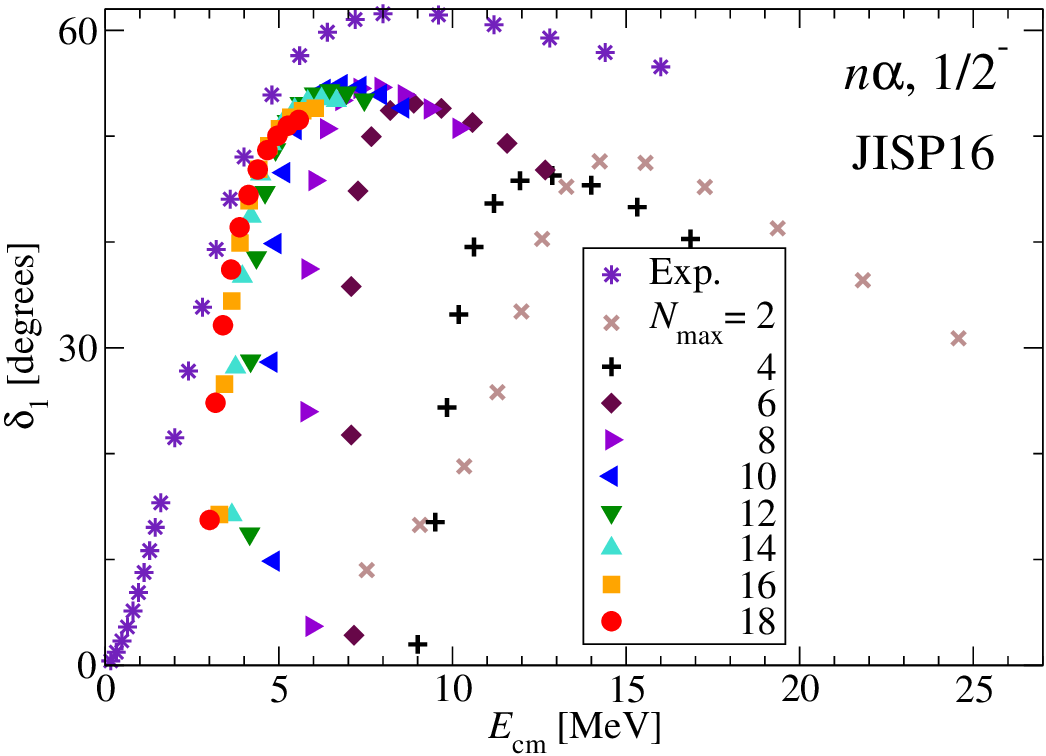}}
\vspace{1ex}
\caption{The  $\frac{1}{2\protect\vphantom{_{q}}}^{-}$   $n\alpha$ phase shifts obtained directly from  
the $^{5}$He eigenstates~$E_{0}$ using Eq.~\eqref{tandelt-SS}.
See Fig.~\ref{n-4He32m_deltaE} for other details.}
\label{n-4He12m_deltaE}
\vspace{5ex}
%
%
%
\centerline{\includegraphics[width=\columnwidth]{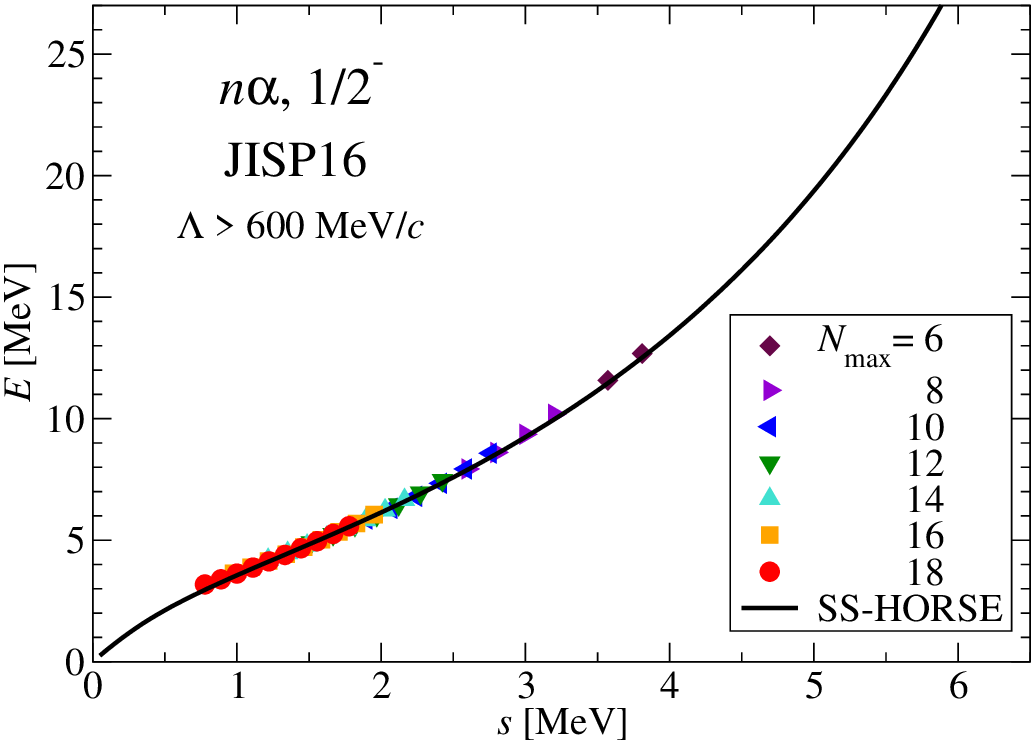}}
\caption{The $^{5}$He  $\frac12^{-}$ eigenstates~$E_{0}$ selected according to~$\Lambda>600$~MeV/$c$ as a function of the scaling parameter~$s$.
See  Fig.~\ref{WSB32m_Es_Select0} for other details.}
\label{n-4He12m_Es_Select0}
\vspace{-6ex}
\end{figure}

The lowest  $\frac12^{-}$ eigenstates of $^{5}$He from the NCSM calculations with JISP16 $NN$ interaction are presented in
Fig.~\ref{n-4He12m_Ehw_Select0} as functions of~$\hbar\Omega$ and in Fig.~\ref{n-4He12m_Es} as a function of the scaling
parameter~$s$, Fig.~\ref{n-4He12m_deltaE} presents the respective $n\alpha$ phase shifts. The eigenenergies in
 Figs.~\ref{n-4He12m_Es} and~\ref{n-4He12m_deltaE} tend to form single common curves demonstrating  the convergence
of many-body NCSM calculations, however we see that many eigenstates diverge from the common curves and lie far from them
thus demonstrating the need to select the states for the SS-HORSE analysis.

 \begin{table*}[t!] 
\caption{Parameters of the $\frac{1}{2\protect\vphantom{_{q}}}^{-}$  resonance in $n\alpha$ scattering from the  $^{5}$He 
NCSM calculations with JISP16 $NN$ interaction. See Table~\ref{n-4Hep32} for details.}
\label{n-4Hep12}  
\begin{ruledtabular}
\begin{tabular}{@{\hspace{6pt}}c@{\hspace{6pt}}c@{\hspace{6pt}}c@{\hspace{6pt}}c@{\hspace{6pt}}c@{\hspace{6pt}}c@{\hspace{6pt}}c@{\hspace{6pt}}c@{\hspace{6pt}}}
\raisebox{-1.8ex}[0pt][0pt]{Selection} & $a$ & $b^2$ & $d\cdot10^4$ & $E_r$ & $\Gamma$ & $\Xi$ & \raisebox{-1.8ex}[0pt][0pt]{$D$} \\
 & (MeV$^\frac12$) & (MeV) & (MeV$^{-\frac32}$) & (MeV) & (MeV) & (keV) &  \\
\noalign{\smallskip}\hline\noalign{\smallskip}
$\Lambda>600$~MeV/$c$ & 1.680 & 3.443 & $-$3.6 & 2.031 & 5.559 & 61 & 46 \\
Manual & 1.699 & 3.299 & 21.3 & 1.856 & 5.456 & 11 & 60 \\
$N_{\max}\leq4$& 2.460 & 6.734 & $-$0.15 & 3.710 & 11.24 & 109(893) & 9(60) \\
$4\leq N_{\max}\leq6$   & 1.718 & 3.310 & 25.0 & 1.834 & 5.511 & 25(92)  & 10(60)  \\[1ex] 
\multicolumn{3}{c}{Nature:} \\
$R$-matrix \cite{Hale} & & & & 2.07 & 5.57 & & \\
$J$-matrix \cite{PRC79} & & & & 1.97 & 5.20 & & \\
Fit $\delta_{1}$ of Ref.~\cite{naExpB}& 1.622 & 3.276 &$+$46.3 & 1.960 & 5.249 & 0.038$^\circ$ & 26 \\
\end{tabular}
\end{ruledtabular} 
\end{table*}

\begin{figure}[b!]
\centerline{\includegraphics[width=\columnwidth]{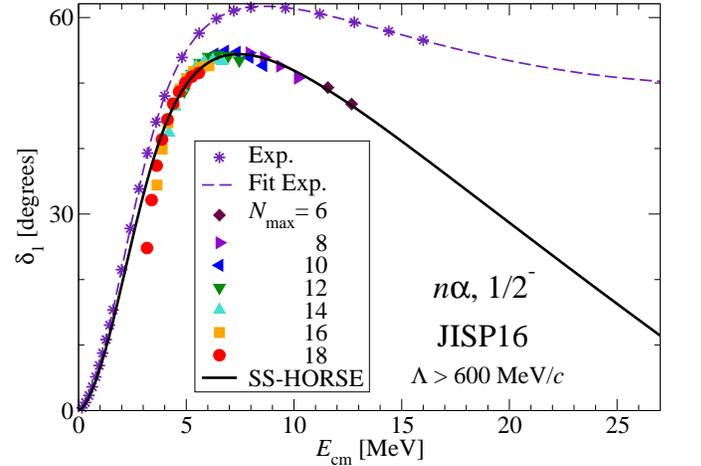}}
\caption{The $\frac{1}{2\protect\vphantom{_{q}}}^{-}$  $n\alpha$ phase shifts generated by the~$\Lambda>600$~MeV/$c$ selected $^{5}$He eigenstates~$E_0$.
See Fig.~\ref{n-4He32m_deltaE_Select0} for details.}
\label{n-4He12m_deltaE_Select0}
\end{figure}

As in the case of the $\frac32^{-}$ partial wave, we start from the~$\Lambda>600$~MeV/$c$ eigenstate selection recommended for the
JISP16 $NN$ interaction in Refs.~\cite{Coon, CoonNTSE12} which is
 illustrated in Figs.~\ref{n-4He12m_Ehw_Select0}  and~\ref{n-4He12m_Es_Select0},
the respective phase shifts are shown in Fig.~\ref{n-4He12m_deltaE_Select0}.
The selected states form reasonably smooth common curves in Figs.~\ref{n-4He12m_Es_Select0} and~\ref{n-4He12m_deltaE_Select0} 
making possible an accurate fit of parameters in Eq.~\eqref{S_phase_r};
the obtained fitted parameters can be found in 
Table~\ref{n-4Hep12}. We get a good description of the $\frac12^{-}$ resonance energy and width however the phase shift behaviour 
extracted from the experimental $n\alpha$ scattering data is
reproduced qualitatively but not quantitatively (see Fig.~\ref{n-4He12m_deltaE_Select0}). Note however that the fit parameters derived from  the
experimental data and JISP16 results (Table~\ref{n-4Hep12}) are close with the exception of the parameter~$d$ which contribution is very 
small at energies below 20~MeV.
Figure~\ref{n-4He12m_Ehw_Select0}  shows that we reproduce not only the eigenstate energies from the shaded area that were fitted
but also many other eigenstates not included in the fit, especially small-$N_{\max}$ eigenstates, thus suggesting to perform a
manual eigenstate selection which will involve many more eigenenergies in the SS-HORSE analysis.

\begin{figure}[t!]
\centerline{\includegraphics[width=\columnwidth]{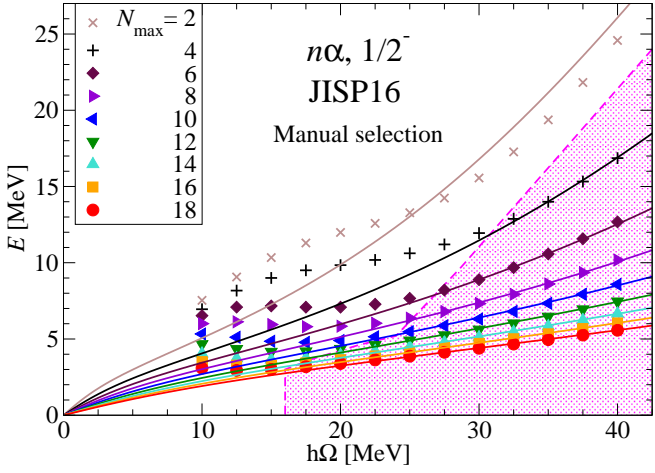}}
\caption{The lowest  $^{5}$He $\frac{1}{2}^{-}$ eigenstates~$E_{0}$ (symbols) and their manual selection (shaded area).
See Fig.~\ref{n-4He32m_Ehw_Select0} for more details.}
\label{n-4He12m_Ehw_Select1}
\vspace{-2ex}
\end{figure}

\begin{figure}[b!]
\centerline{\includegraphics[width=\columnwidth]{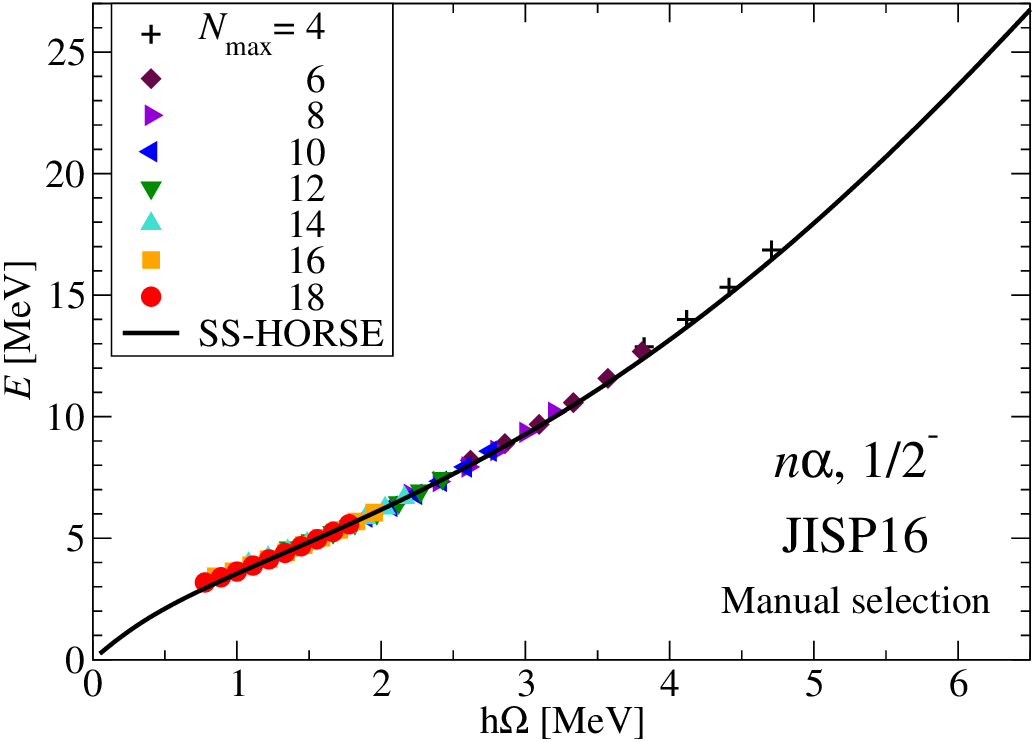}}
\caption{Manually selected $^{5}$He  $\frac{1}{2\protect\vphantom{_{q}}}^{-}$ eigenstates~$E_{0}$ plotted as a function of the scaling parameter~$s$ (symbols). 
See Fig.~\ref{WSB32m_Es_Select0} for other details.}
\label{n-4He12m_Es_Select1}
\end{figure}

\begin{figure}[t!]
\centerline{\includegraphics[width=\columnwidth]{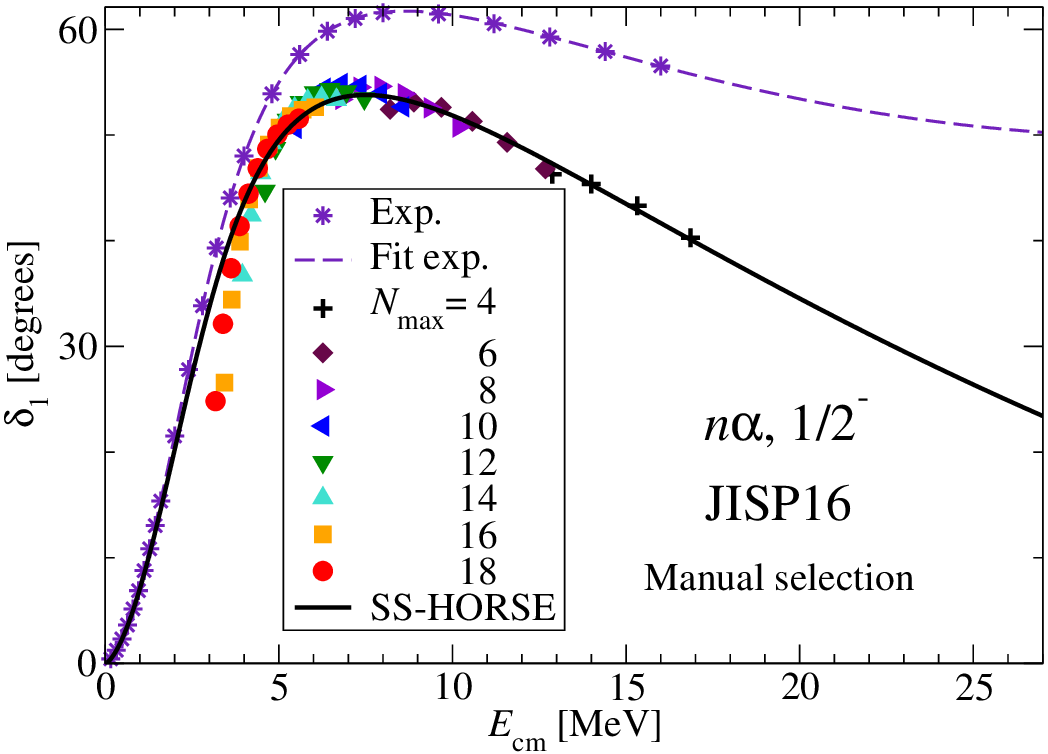}}
\caption{The $\frac{1}{2\protect\vphantom{_{q}}}^{-}$ $n\alpha$ phase shifts generated by the manually selected  $^{5}$He eigenstates~$E_0$. 
See Fig.~\ref{n-4He32m_deltaE_Select0} for details.}
\label{n-4He12m_deltaE_Select1}
\end{figure}

Our manual selection of the lowest $\frac12^{-}$ eigenstates in $^{5}$He is shown in Figs.~\ref{n-4He12m_Ehw_Select1}
and~\ref{n-4He12m_Es_Select1} while the respective $n\alpha$ phase shifts are presented in Fig.~\ref{n-4He12m_deltaE_Select1},
the results of the fit are given in Table~\ref{n-4Hep12}. As in the case of the $\frac32^{-}$ $n\alpha$ partial wave, the inclusion of the
additional eigenstates in the fit does not affect the phase shifts at smaller energies including the resonance region.  
However, including the additional eigenstates pushes the
phase shifts up in the direction of the phase shift analysis at larger energies. The  $\frac12^{-}$ resonance energy and width and the
parameters of the phase shift fit by Eq.~\eqref{S_phase_r} are seen from Table~\ref{n-4Hep12} to change only slightly with the exception
of the parameter~$d$ responsible for the phase shift behaviour at higher energies.

\begin{figure}[b!]
\centerline{\includegraphics[width=\columnwidth]{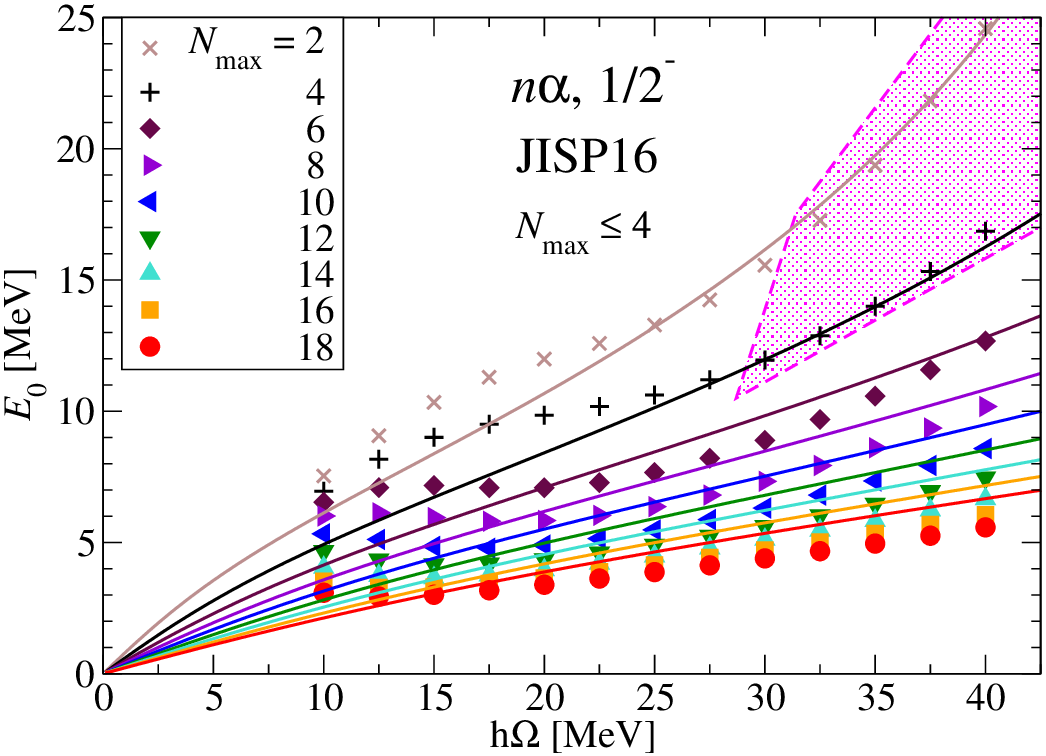}}
\caption{The lowest $^{5}$He $\frac12^{-}$ eigenstates~$E_{0}$ (symbols) and their ${N_{\max}\leq4}$ selection (shaded area).
See Fig.~\ref{n-4He32m_Ehw_Select0}
for more details.}
\label{n-4He12m_Ehw_Select3}
\end{figure}

\begin{figure}[t!]
\centerline{\includegraphics[width=\columnwidth]{n-4He12m_Es_Select3.eps}}
\caption{Selected lowest $^{5}$He $\frac12^{-}$ eigenstates~$E_{0}$ obtained in NCSM  
with ${N_{\max}\leq4}$ as a function of the scaling parameter~$s$. 
See Fig.~\ref{WSB32m_Es_Select0} for other details.\strut}
\label{n-4He12m_Es_Select3}
\end{figure}
\begin{figure}[!b]
\centerline{\includegraphics[width=\columnwidth]{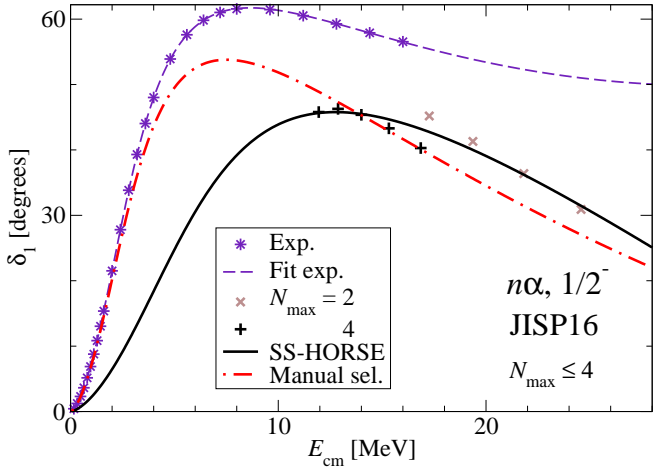}}
\caption{The $\frac{1}{2\protect\vphantom{_{q}}}^{-}$ $n\alpha$ phase shifts generated by the selected $^{5}$He eigenstates~$E_{0}$ obtained in NCSM 
with ${N_{\max}\leq4}$. The dash-dotted curve depicts the phase shifts obtained by the fit to all manually selected eigenstates.
See Fig.~\ref{n-4He32m_deltaE_Select0} for other details.}
\label{n-4He12m_deltaE_Select3}
\vspace{-1.4ex}
\end{figure}

It is very interesting and important to examine whether it is possible to get a reasonable description of the resonance and
phase shifts in the~$\frac{1}{2\vphantom{_{q}}}^{-}$ $n\alpha$ scattering using only eigenstates obtained in many-body NCSM calculations 
in small bases. In the case of the~$\frac{3}{2_{\vphantom{q}}}^{-}$ $n\alpha$ scattering we manage to derive very good phase shifts from
the~$N_{\max}\leq4$ NCSM eigenstates. Therefore we try the~$N_{\max}\leq4$ eigenstate selection also in the~$\frac12^{-}$ 
partial wave, 
see Figs.~\ref{n-4He12m_Ehw_Select3}, \ref{n-4He12m_Es_Select3} and~\ref{n-4He12m_deltaE_Select3}.
This selection clearly fails to reproduce the phase shifts and resonance \mbox{parameters} which differ essentially from the converged results obtained
with the manual selection of the~$\frac12^{-}$ $^{5}$He eigenstates (see Fig.~\ref{n-4He32m_deltaE_Select0} and Table~\ref{n-4Hep12}); we
see also in Fig.~\ref{n-4He12m_Ehw_Select3} that the fit to the~${N_{\max}\leq4}$ eigenstates from the shaded area fails to `predict'
the eigenenergies~$E_{0}$ obtained with larger~$N_{\max}$ values. That is not surprising because the plots of the~$N_{\max}\leq4$
eigenenergies as a function of the scaling parameter~$s$ (Fig.~\ref{n-4He12m_Es_Select3}) and of the respective phase shifts as
a function of energy (Fig.~\ref{n-4He12m_deltaE_Select3}) do not form smooth common curves.

\begin{figure}[!]
\centerline{\includegraphics[width=\columnwidth]{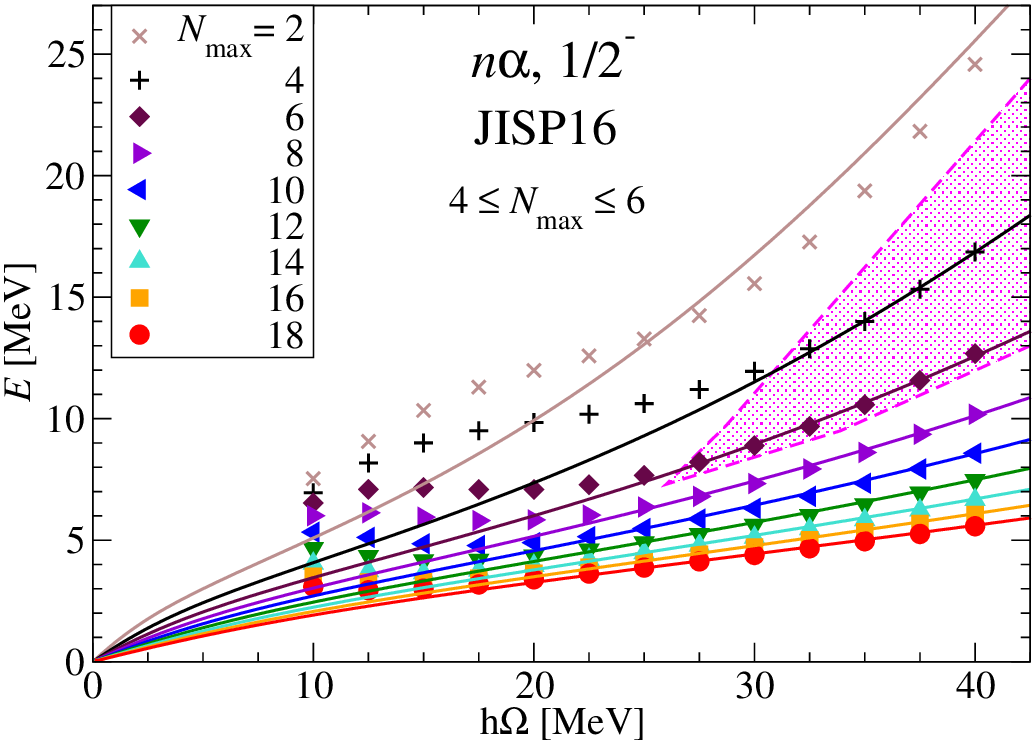}}
\vspace{-1.8ex}
\caption{The lowest $^{5}$He $\frac12^{-}$ eigenstates~$E_{0}$ (symbols) and their ${4\leq N_{\max}\leq6}$ selection (shaded area).
See Fig.~\ref{n-4He32m_Ehw_Select0}
for more details.}
\label{n-4He12m_Ehw_Select2}
\vspace{4ex}
%
\centerline{\includegraphics[width=\columnwidth]{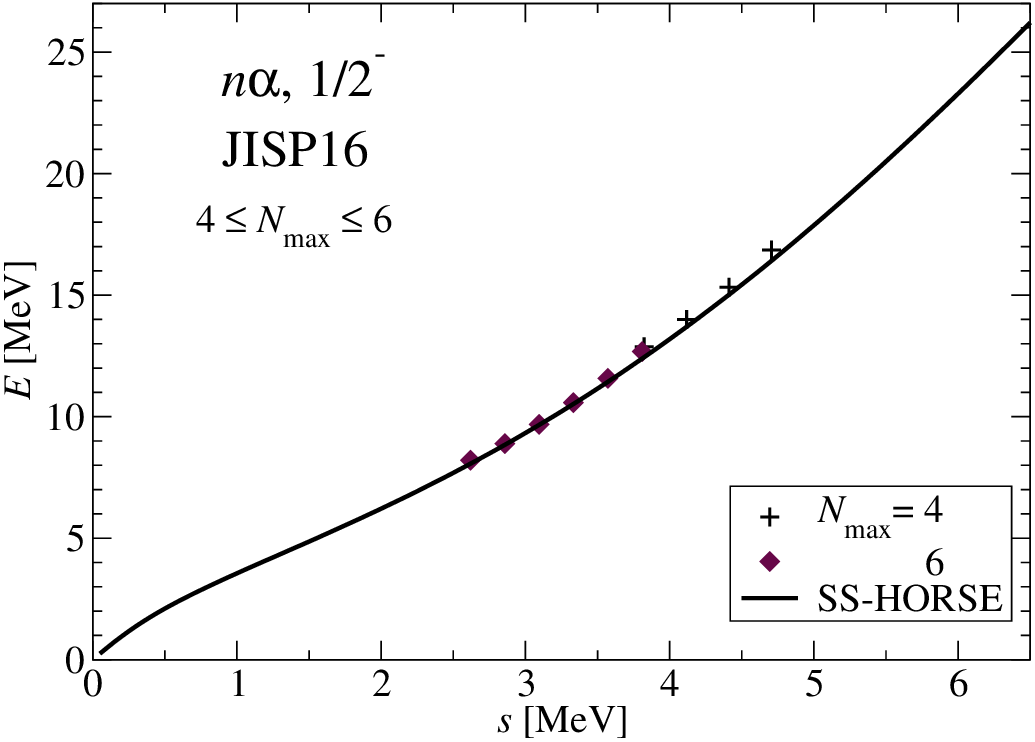}}
\vspace{-1.5ex}
\caption{Selected lowest $^{5}$He $\frac12^{-}$ eigenstates~$E_{0}$ obtained in NCSM  
with ${4\leq N_{\max}\leq6}$ as a function of the scaling parameter~$s$. 
See Fig.~\ref{WSB32m_Es_Select0} for other details.\strut}
\label{n-4He12m_Es_Select2}
\vspace{3ex}
%
\centerline{\includegraphics[width=\columnwidth]{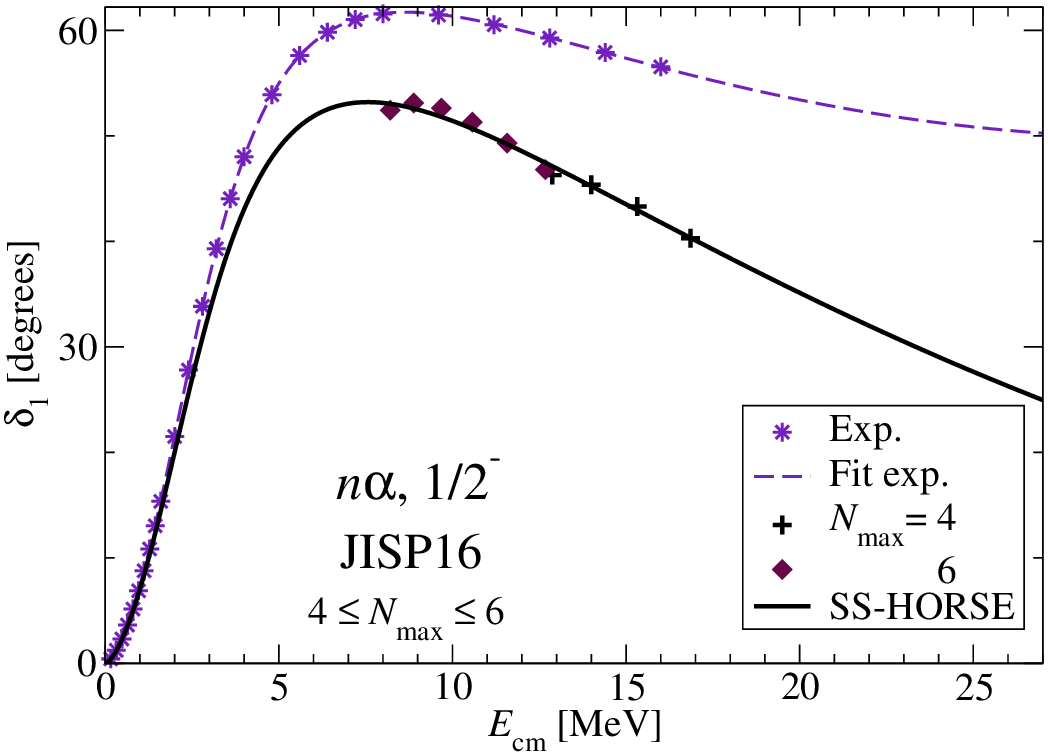}}
\vspace{-1.5ex}
\caption{The 
$\frac12^{-}$ $n\alpha$ phase shifts generated by the selected $^{5}$He eigenstates~$E_{0}$ obtained in NCSM 
with ${4\leq N_{\max}\leq6}$.
See Fig.~\ref{n-4He32m_deltaE_Select0} for  details.}
\label{n-4He12m_deltaE_Select2}
\vspace{-1.5ex}
\end{figure}

However an entirely different result is obtained by selecting for the SS-HORSE analysis the $^{5}$He $\frac12^{-}$ NCSM results
from basis spaces with~${4\leq N_{\max}\leq6}$. For the ${4\leq N_{\max}\leq6}$ selection we pick up 10~eigenstates with the
smallest~$N_{\max}$ values out of 60~manually selected before $\frac{1}{2\vphantom{_{q}}}^{-}$ eigenstates. This eigenstate selection and the respective
results are illustrated by Figs.~\ref{n-4He12m_Ehw_Select2}, \ref{n-4He12m_Es_Select2} and~\ref{n-4He12m_deltaE_Select2}. 
The selected eigenenergies are seen to form sufficiently smooth curves in Figs.~\ref{n-4He12m_Es_Select2} and~\ref{n-4He12m_deltaE_Select2}. 
The parameter fit results in nearly the same phase shifts  (Fig.~\ref{n-4He12m_deltaE_Select2}) as in the case of the manual eigenstate selection,
we get also very close values of the resonance energy and width and fitting parameters listed in Table~\ref{n-4Hep12}.
Figure~\ref{n-4He12m_Ehw_Select2} demonstrates that  by using only 10 \mbox{small-$N_{\max}$} eigenstates from the shaded area we accurately
`predict' the energies of many higher-$N_{\max}$ eigenstates: the rms deviation~$\Xi$ of energies of all 60~manually selected 
eigenstates is 92~keV (see Table~\ref{n-4Hep12}).
Of course, 92~keV is much larger than the~$\Xi$ value of~11~keV obtained in the full fit to all
these 60~eigenenergies, but it is still an indication of a good quality `prediction' of many-body eigenenergies~$E_{0}$ obtained with
much larger bases in a wide range of~$\hbar\Omega$ values.

\subsection{\boldmath Partial wave $\frac12^{+}$}

 \begin{table*}[t!]
\caption{\strut$\frac12^{+}\protect\vphantom{_{q_{\int}}}$ $n\alpha$ scattering from the  $^{5}$He NCSM calculations with JISP16 $NN$ interaction: 
fitting parameters~$E_{b}$, $c$, $d$ and~$f$ of 
Eq.~\eqref{S_phase_implicit}, rms deviation of fitted energies~$\Xi$ and the number of these fitted energies~$D$
for different selections of eigenvalues in comparison with the fit  by Eq.~\eqref{S_phase_b} of the phase shifts~$\delta_{0}$ 
extracted from experimental data in Ref.~\cite{naExpB}. 
For the~${5\leq N_{\max}\leq7}$ selection, $\Xi$ and~$D$ for all energies from the manual selection are shown  within brackets.}
\label{n-4Hes12}  
\begin{ruledtabular}
\begin{tabular}{@{\hspace{1pt}}ccccccc@{\hspace{1pt}}}
\raisebox{-1.8ex}[0pt][0pt]{Selection} & $E_{b}$ & $c$ & $d\cdot10^3$ &  $f\cdot10^5$ & $\Xi$ & \raisebox{-1.8ex}[0pt][0pt]{$D$} \\
 & (MeV) & (MeV$^{-\frac12}$) & (MeV$^{-\frac32}$) & (MeV$^{-\frac52}$) & (keV) &  \\
\noalign{\smallskip}\hline\noalign{\smallskip}
$\Lambda>600$~MeV/$c$ & $-$5.996 &$-$0.171 & $-$8.02 & 6.48 & 85 & 41 \\
Manual & $-$6.733 &$-$0.183 & $-$13.0 & 30.8 & 120 & 53 \\
$5\leq N_{\max}\leq7$ &  $-$3.347 &$-$0.151 & 63.0 & $-$86.7 & 168(259) & 13(53) \\[1.5ex]
Fit $\delta_{0}$ of Ref.~\cite{naExpB}  & $-$13.75 &$-$0.156 & $-$429 & 220 & 0.018$^\circ$ & 26 \\ 
\end{tabular}
\end{ruledtabular} 
\end{table*}

In this subsection we examine a possibility to describe neutron-nucleus non-resonant scattering using as input for the SS-HORSE
analysis the results of many-body shell model calculations. The SS-HORSE fit is done in the same manner as in the case of resonant
scattering. The difference is that the non-resonant low-energy $n\alpha$ scattering phase shifts 
in the~$\frac{1}{2}^{+}$  state are described by Eq.~\eqref{S_phase_b} instead of Eq.~\eqref{S_phase_r} which parameters are
fitted using Eq.~\eqref{S_phase_implicit} instead of Eq.~\eqref{ImplicitENhw}. The parameter~$E_{b}$ of this equation mimics the
Pauli-forbidden state in the $n\alpha$ scattering. As compared with the discussion of the~$\frac{1}{2}^{+}$ scattering by the model
WSBG potential which supports the  Pauli-forbidden state, this bound state does not appear as a result of the NCSM $^{5}$He calculations.
Therefore we should use for the SS-HORSE fit the lowest~$\frac{1}{2}^{+}$ state obtained by the NCSM with the eigenenergy~$E_{0}$
and set~$\nu=0$ in Eq.~\eqref{S_phase_implicit}.

\begin{figure}[t!]
\centerline{\includegraphics[width=\columnwidth]{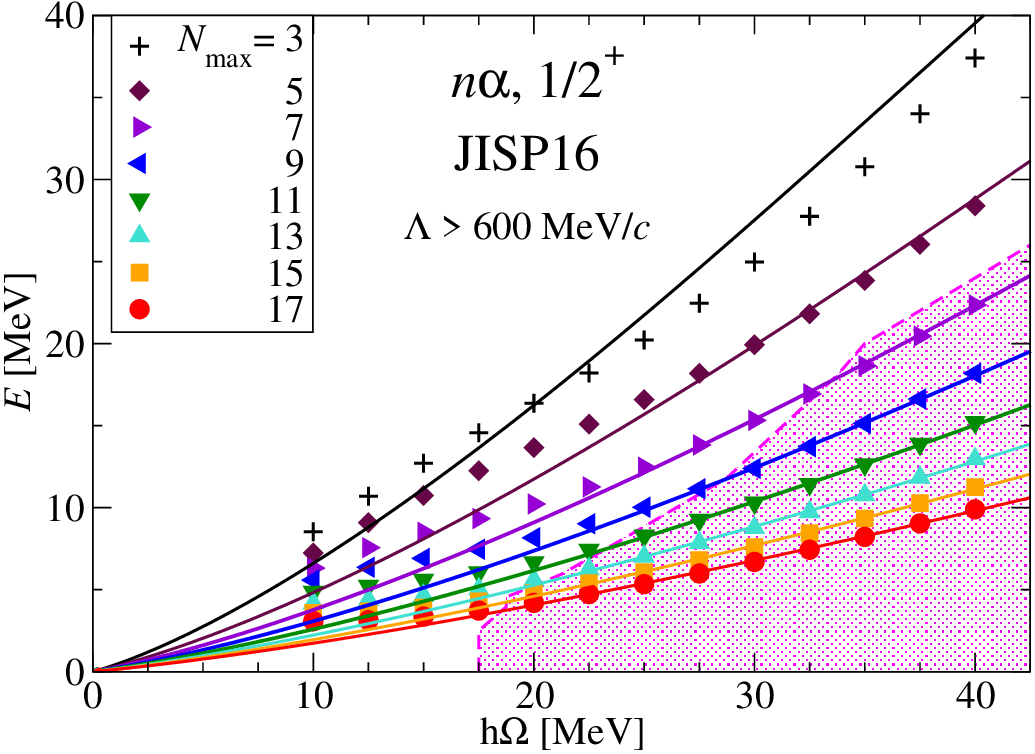}}
\vspace{-1ex}
\caption{The lowest $^{5}$He $\frac12^{+}\protect\vphantom{_{q_{\int}}}$ eigenstates~$E_{0}$ (symbols) 
and their~$\Lambda>600$~MeV/$c$ selection (shaded area). See Fig.~\ref{WSB12p_Ehw_Select0} for more details.}
\label{n-4He12p_Ehw_Select0}
\end{figure}

\begin{figure}[t!]
\centerline{\includegraphics[width=\columnwidth]{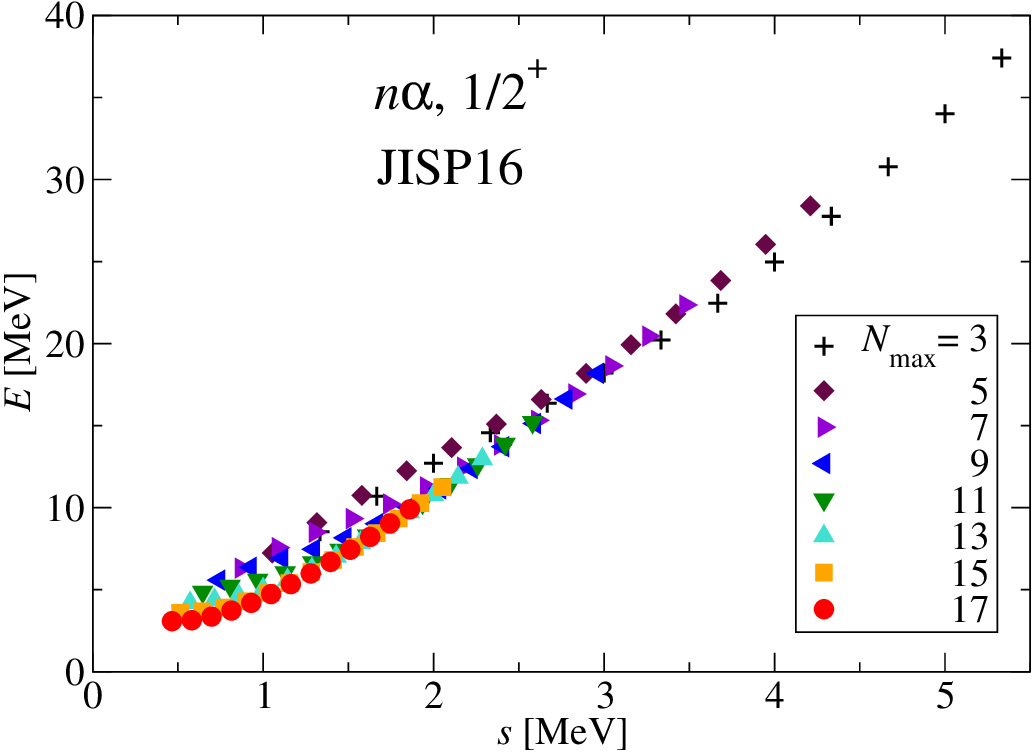}}
\vspace{-1ex}
\caption{The  lowest $^{5}$He $\frac12^{+}$ eigenstates~$E_{0}$ as a function of the scaling parameter~$s$.} 
\label{n-4He12p_Es}
\end{figure}

\begin{figure}[b!]
\centerline{\includegraphics[width=\columnwidth]{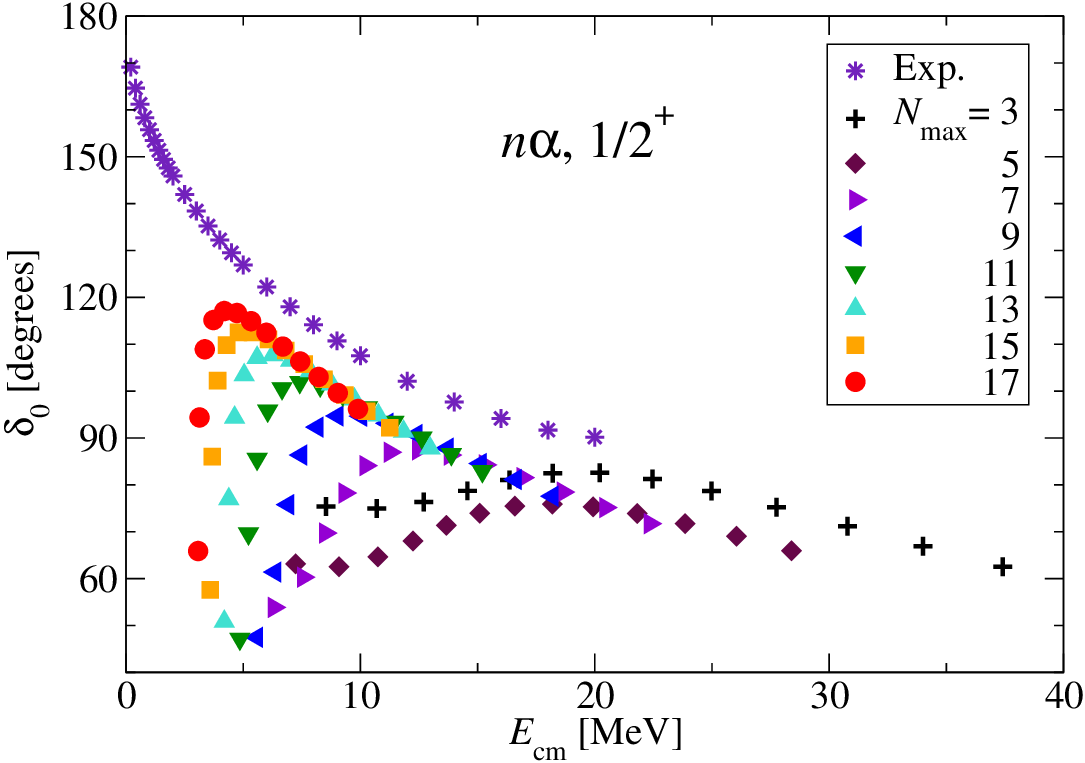}}
\caption{The  $\frac{1}{2\protect\vphantom{_{q}}}^{+}$ $n\alpha$ phase shifts obtained directly from the $^{5}$He eigenstates~$E_0$ using  Eq.~\eqref{tandelt-SS}.
See Fig.~\ref{n-4He32m_deltaE} for other details.}
\label{n-4He12p_deltaE}
\end{figure}

These lowest  $\frac12^{+}$ $^{5}$He eigenstates~$E_{0}$ are shown as functions of~$\hbar\Omega$ 
for various~$N_{\max}$
in Fig.~\ref{n-4He12p_Ehw_Select0} and as a function of the scaling parameter~$s$ in Fig.~\ref{n-4He12p_Es}. 
We see a tendency of eigenstates to approach the common curve at smaller~$\hbar\Omega$ values with increasing~$N_{\max}$ which
signals that the convergence is achieved at smaller energies in larger basis spaces. This tendency is 
much more pronounced in the plot of the SS-HORSE phase shifts 
corresponding to the NCSM eigenstates in Fig.~\ref{n-4He12p_deltaE}. This figure however also clearly indicates the need 
to select eigenstates for the 
SS-HORSE fitting.

\begin{figure}[t!]
\centerline{\includegraphics[width=\columnwidth]{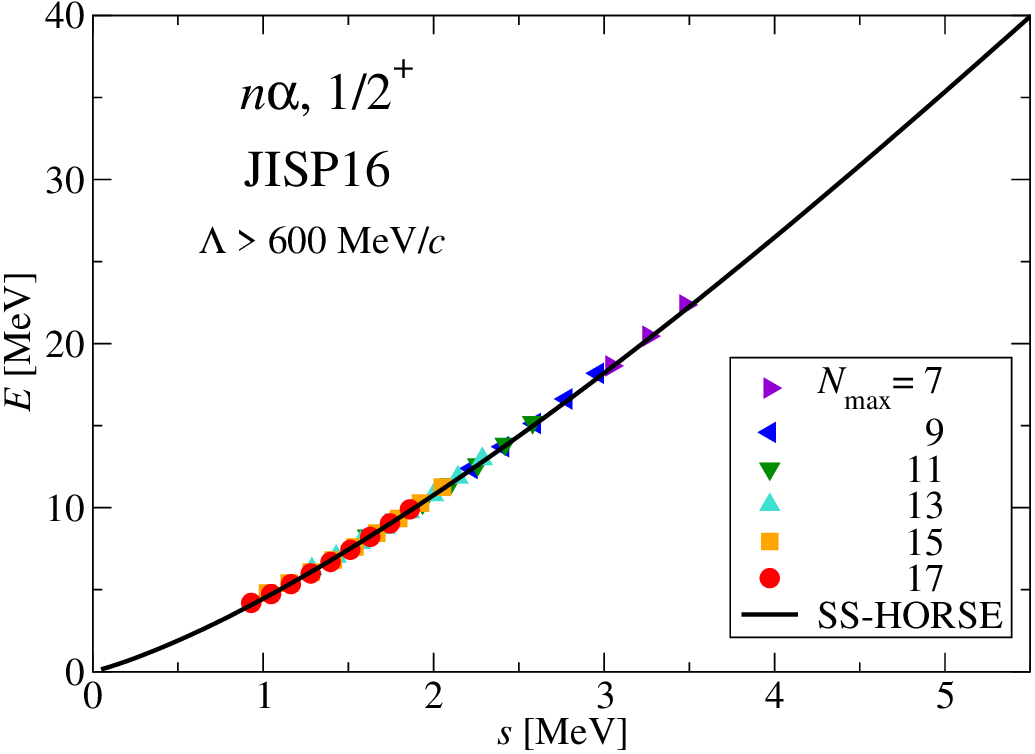}}
\caption{The $^{5}$He  $\frac12^{+}$ eigenstates~$E_{0}$ selected according to~$\Lambda>600$~MeV/$c$ 
as a function of the scaling parameter~$s$. See Fig.~\ref{WSB12p_Es0} for other details.}
\label{n-4He12p_Es_Select0}
\vspace{5.5ex}
%
%
%
\centerline{\includegraphics[width=\columnwidth]{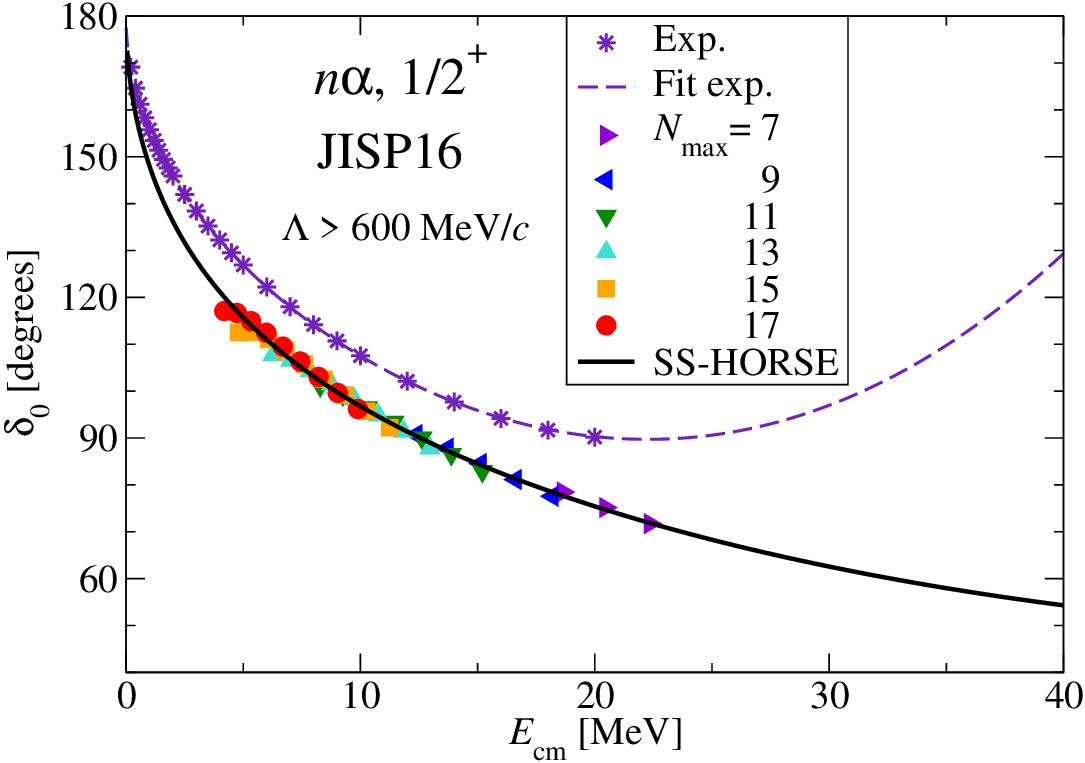}}%
\caption{The $\frac12^{+}$ $n\alpha$ phase shifts generated by the~$\Lambda>600$~MeV/$c$ selected $^{5}$He eigenstates~$E_0$ (symbols).
The solid curve depicts the phase shifts of Eq.~\eqref{S_phase_b}  
with parameters~$E_{b}$, $c$, $d$ and~$f$  obtained by the fit with this selection of eigenstates;
stars and the dashed curve depict the
phase shift analysis of experimental data of Refs.~\cite{naExpB} and the fit by Eq.~\eqref{S_phase_b}.}
\label{n-4He12p_deltaE_Select0}
\vspace{-6ex}
\end{figure}

\begin{figure}[b!]
\centerline{\includegraphics[width=\columnwidth]{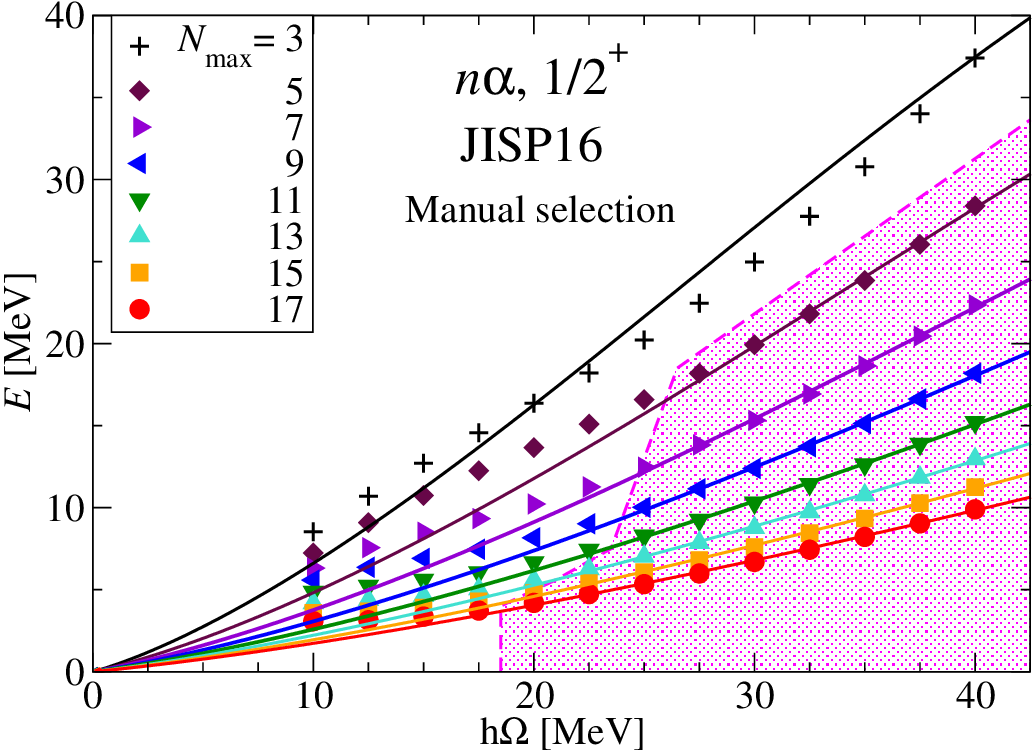}}
\caption{The lowest  $^{5}$He $\frac{1}{2}^{+}$ eigenstates~$E_{0}$ (symbols) and their manual selection (shaded area).
See Fig.~\ref{WSB12p_Ehw_Select0} for more details.}
\label{n-4He12p_Ehw_Select1}
\end{figure}

\begin{figure}[!]
\centerline{\includegraphics[width=\columnwidth]{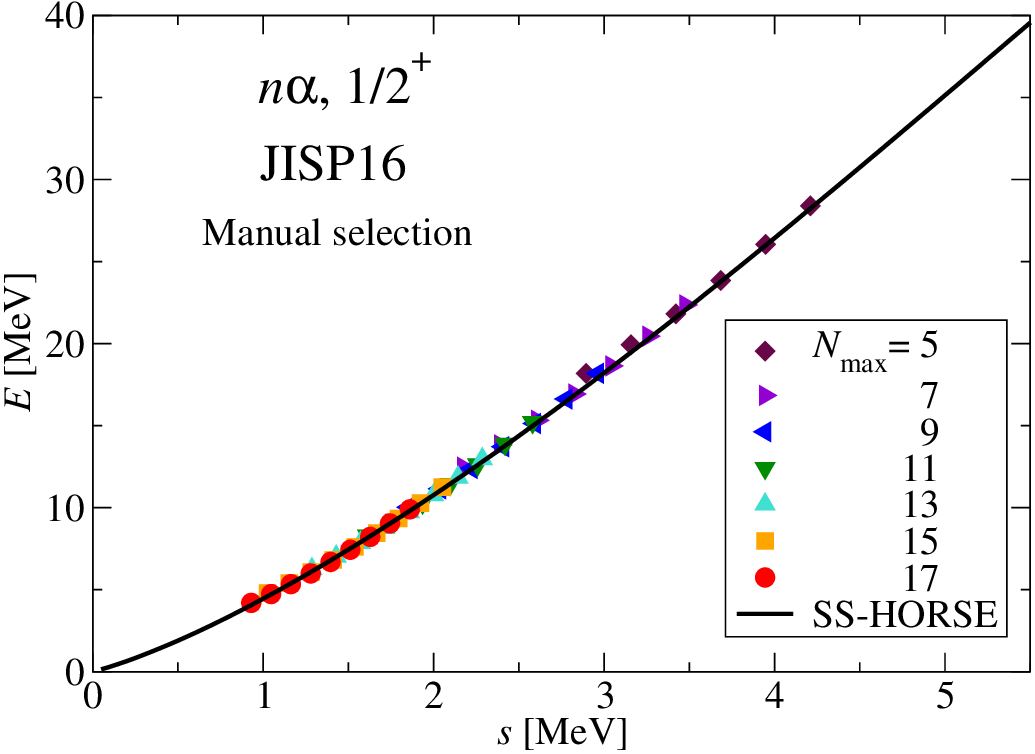}}
\vspace{-1.5ex}
\caption{Manually selected $^{5}$He  $\frac{1}{2\protect\vphantom{_{q}}}^{+}$ eigenstates~$E_{0}$ plotted as a function of the scaling parameter~$s$ (symbols). 
See Fig.~\ref{WSB12p_Es0} for details.}
\label{n-4He12p_Es_Select1}
\vspace{4ex}
%
%
\centerline{\includegraphics[width=\columnwidth]{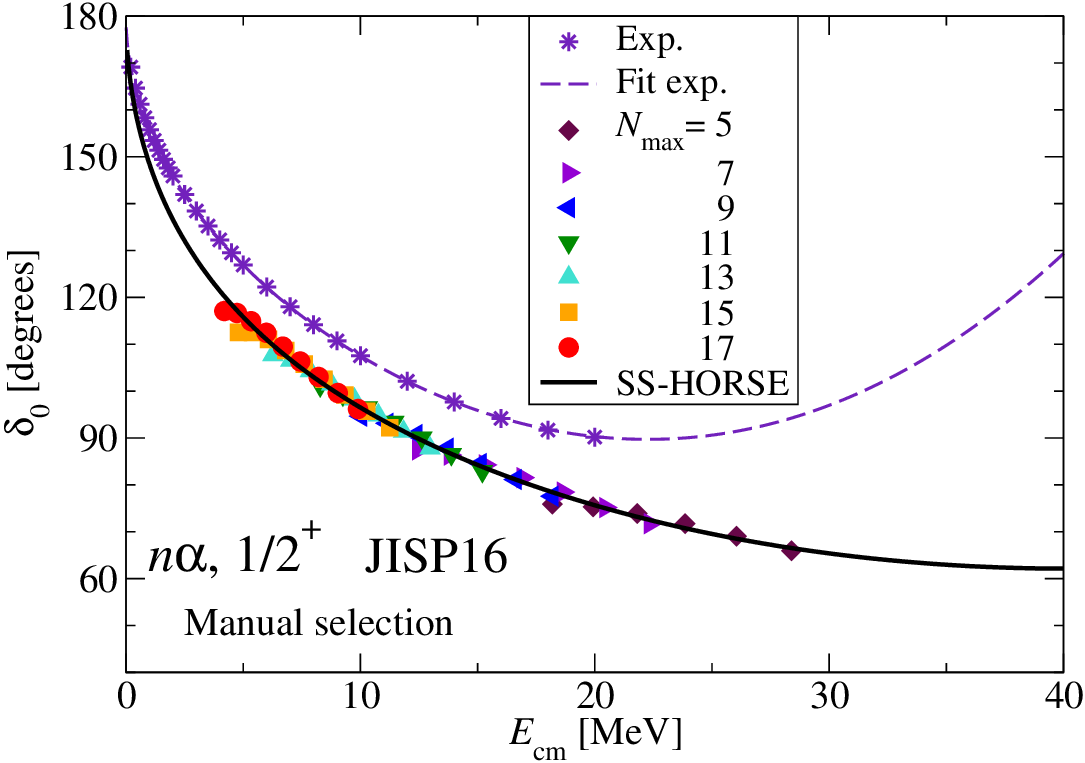}}
\vspace{-1.5ex}
\caption{The $\frac{1}{2\protect\vphantom{_{q}}}^{+}$ $n\alpha$ phase shifts generated by the manually selected  $^{5}$He eigenstates~$E_0$. 
See Fig.~\ref{n-4He12p_deltaE_Select0} for details.}
\label{n-4He12p_deltaE_Select1}
\vspace{4ex}
%
%
\centerline{\includegraphics[width=\columnwidth]{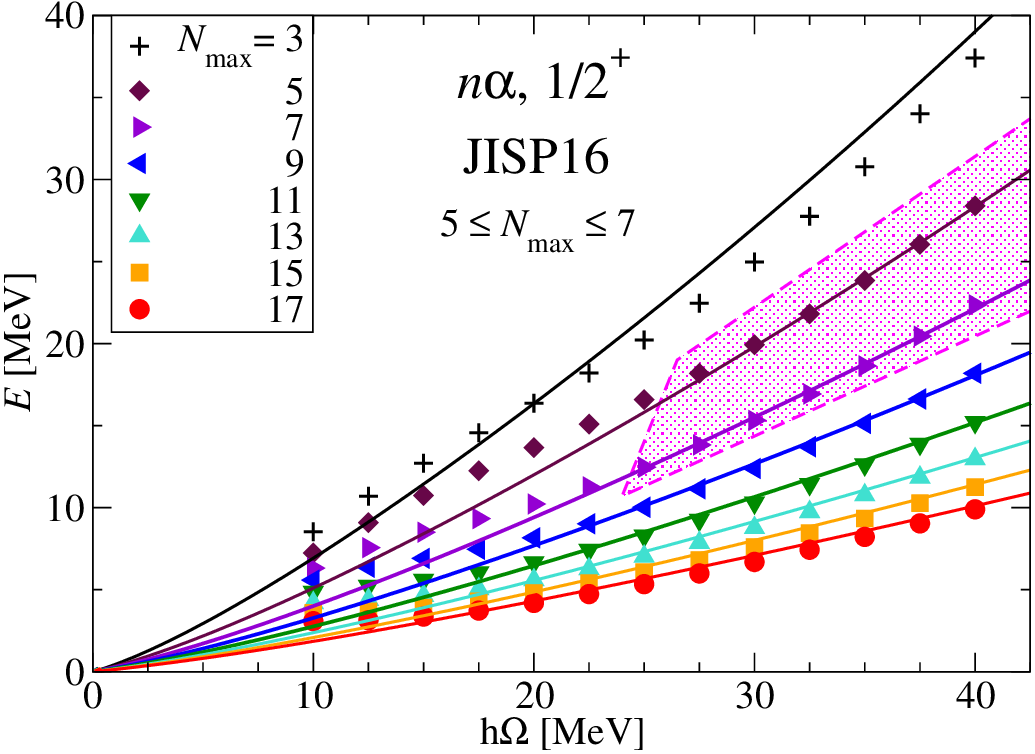}}
\vspace{-1.5ex}
\caption{The lowest $^{5}$He $\frac12^{+}$ eigenstates~$E_{0}$ (symbols) and their ${5\leq N_{\max}\leq7}$ selection (shaded area).
See Fig.~\ref{WSB12p_Ehw_Select0} for more details.}
\label{n-4He12p_Ehw_Select2}
\end{figure}

We start with selecting eigenstates according to the inequality~$\Lambda>600$~MeV/$c$ as is illustrated by 
Figs.~\ref{n-4He12p_Ehw_Select0} and~\ref{n-4He12p_Es_Select0}, the respective phase shifts are shown in
Fig.~\ref{n-4He12p_deltaE_Select0}, and the obtained fitting parameters are presented in Table~\ref{n-4Hes12}. We obtain a
reasonable accuracy of the fit with the rms deviation of the fitted energies of~85~keV.
We reproduce reasonably the phase shift behaviour by the JISP16 $NN$ interaction. We note that at energies~$E_{\rm cm}>25$~MeV
the fit by Eq.~\eqref{S_phase_b} of the results of the phase shift analysis 
start going up with
the energy. This seems unphysical, however the $\frac12^{+}$ phases extracted from the $n\alpha$ scattering data are available only
up to~$E_{\rm cm}=20$~MeV; the phase shift analysis at higher energies is needed to obtain a more realistic fit in this energy interval where
the NCSM-SS-HORSE phase shifts look more realistic.

Figure~\ref{n-4He12p_Ehw_Select0} demonstrates that it would be reasonable to perform a manual selection and to
include in the fit more eigenstates thus extending the energy interval of the fitted phase shifts. Our manual selection of the lowest
$\frac12^{+}$ $^{5}$He eigenstates and the respective phase shifts are presented in Figs.~\ref{n-4He12p_Ehw_Select1},
\ref{n-4He12p_Es_Select1}, \ref{n-4He12p_deltaE_Select1} and Table~\ref{n-4Hes12}. Some of the fitting parameters are  profoundly altered
due to the inclusion of additional eigenstates in the fit, however the resulting phase shifts are nearly the same with an exception of the
energies~$E_{\rm cm}>30$~MeV where these additional eigenstates push the phase shifts slightly up. The phase shift analysis is
unavailable at these energies, therefore it is impossible to judge whether this adjustment of the phase shifts improves the description of the
experiment.

\begin{figure}[!t]
\centerline{\includegraphics[width=\columnwidth]{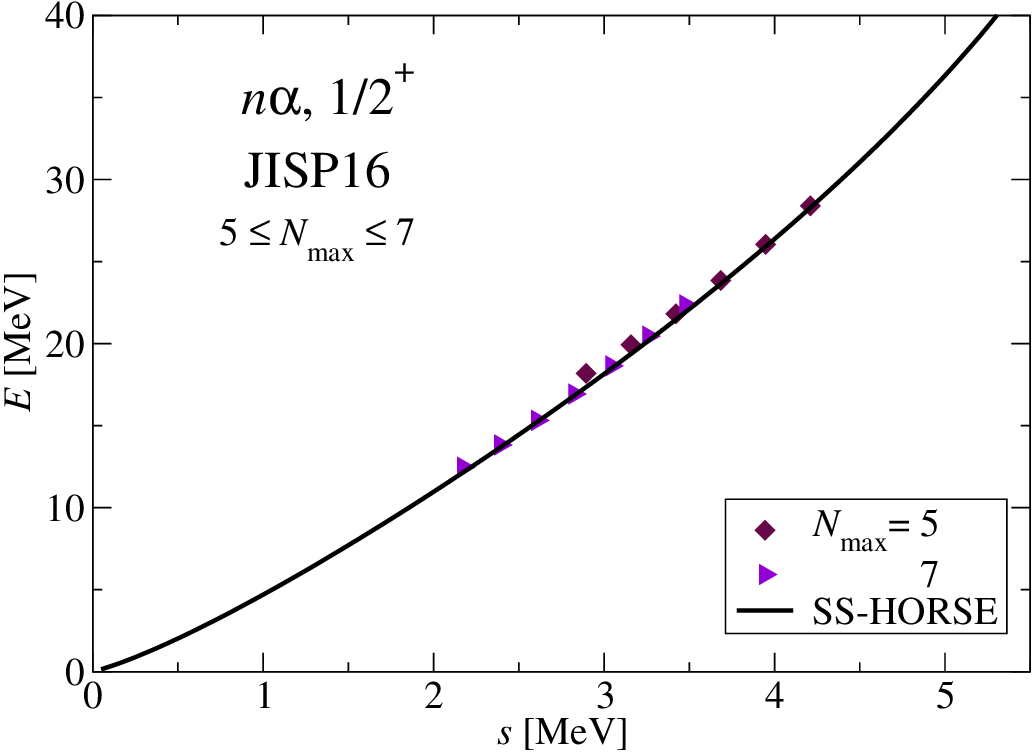}}
\caption{Selected lowest $^{5}$He $\frac12^{+}$ eigenstates~$E_{0}$ obtained in NCSM  
with ${5\leq N_{\max}\leq7}$ as a function of the scaling parameter~$s$. 
See Fig.~\ref{WSB12p_Es0} for details.}
\label{n-4He12p_Es_Select2}
\end{figure}

\begin{figure}[!b]
\centerline{\includegraphics[width=\columnwidth]{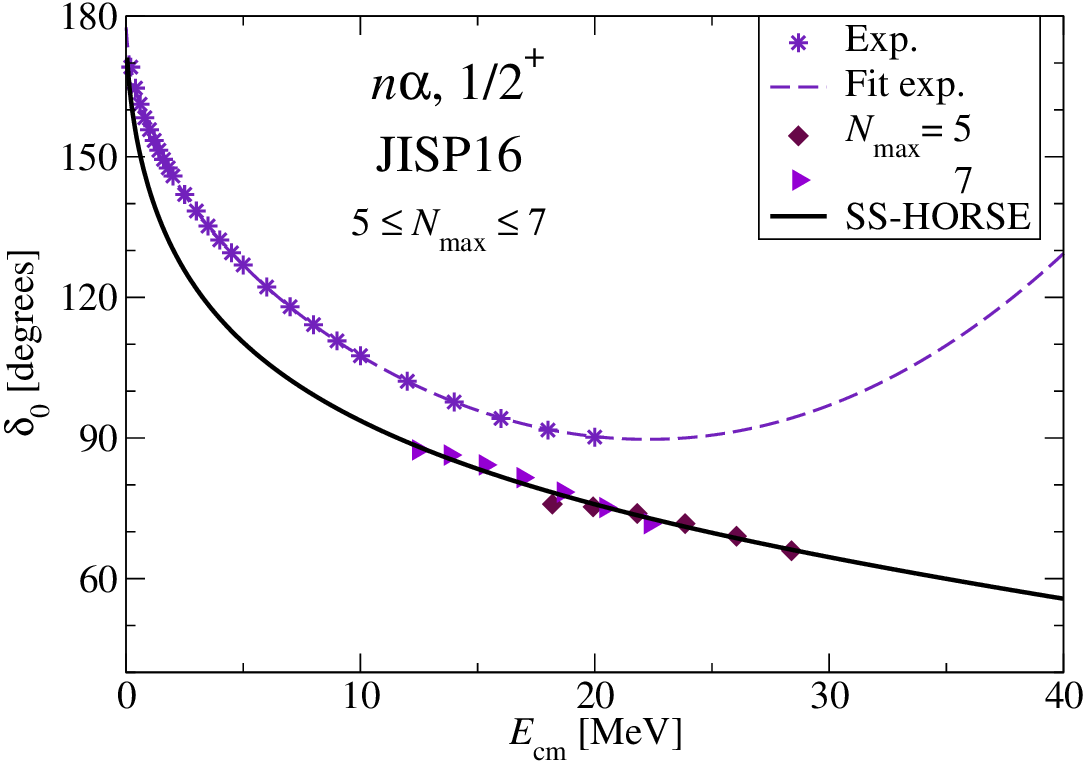}}
\caption{The $\frac{1}{2\protect\vphantom{_{q}}}^{+}$ $n\alpha$ phase shifts generated by the selected $^{5}$He eigenstates~$E_{0}$ obtained in NCSM 
with ${5\leq N_{\max}\leq7}$. See Fig.~\ref{n-4He12p_deltaE_Select0} for details.}
\label{n-4He12p_deltaE_Select2}
\end{figure}

It is interesting and important to examine the possibility of describing the eigenenergies and non-resonant phase shifts obtained in many-body calculations in
large basis spaces by SS-HORSE fits based on results in much smaller basis spaces. As in the case of $\frac{1}{2\protect\vphantom{_{q}}}^{-}$ states, we
do not succeed by choosing the eigenstates from the smallest available NCSM basis spaces with~$N_{\max}=3$ and~5: note, in both
cases the results from the smallest basis space with~$N_{\max}=2$ for $\frac12^{-}$ and~$N_{\max}=3$ for $\frac12^{+}$ states are not
included in our respective manual selections. \mbox{However} picking up eigenstates obtained with~$5\leq N_{\max}\leq7$ from the manual
selection of the $^{5}$He $\frac12^{+}$ eigenstates, we obtain reasonable phase shifts and `predictions' for the eigenstates
with larger~$N_{\max}$, see Figs.~\ref{n-4He12p_Ehw_Select2}, \ref{n-4He12p_Es_Select2} and~\ref{n-4He12p_deltaE_Select2}.
It is interesting that we get  similar phase shifts with three different selections of the $\frac12^{+}$ eigenstates while the respective
fitting parameters shown in  Table~\ref{n-4Hes12} differ essentially. The rms deviation of all 53~manually selected $\frac12^{+}$ eigenstates 
resulting from the fit to 13~eigenstates from the $5\leq N_{\max}\leq7$ selection is 259~keV that is much worse than the `predictions' of
the odd parity eigenstates. We suppose that this is related to the fact that the $\frac12^{+}$ eigenstates lie higher in energy than the
$\frac32^{-}$ and $\frac12^{-}$ eigenstates and the SS-HORSE fits, especially those to the small-$N_{\max}$ eigenstates, involve the
phase shifts at higher energies where our low-energy phase shift expansions become less accurate and require higher order
terms in Taylor series and more fitting parameters.

\section{Conclusions\label{sec:Conclusions}}
We  develop a SS-HORSE approach allowing us to obtain  low-energy  scattering phase shifts and resonance energy and width
in variational calculations with the oscillator basis, in the nuclear shell model in particular. 
The SS-HORSE technique is based on the general properties of the oscillator basis and on
the HORSE ($J$-matrix) formalism in scattering theory, it utilizes general low-energy expansions of the $S$-matrix including the poles
associated with the bound and resonant states.

The SS-HORSE approach is carefully verified using a model two-body problem with a Woods--Saxon type potential and is shown to be able to obtain
accurate scattering phase shifts and resonance energy and width even with small oscillator bases. Next the SS-HORSE method is successfully applied
to the study of the $n\alpha$ scattering phases and resonance based on the NCSM calculations of $^{5}$He with the realistic JISP16
$NN$ interaction.

Within the SS-HORSE approach we obtain and generalize to the states lying above nuclear disintegration thresholds the scaling property
of variational calculations with oscillator basis suggested in Refs.~\cite{Coon, CoonNTSE12} which states that the eigenenergies do not
depend separately on~$\hbar\Omega$ and the maximal oscillator quanta~$\mathbb N$ of the states included in the basis but
only on their combination~$s$ (or the scaling parameter~$\lambda_{sc}$ as suggested in Refs.~\cite{Coon, CoonNTSE12},
$s\sim\lambda_{sc}^{2}$). We demonstrate a typical behavior of eigenstates in the continuum as functions of~$s$ in cases when the system
has or does not have a low-energy resonance. The scaling property is useful for \mbox{extrapolating} the \mbox{results} obtained in smaller basis spaces
to larger bases, and we demonstrate using both the model problem and many-body NCSM calculations that we are able to `predict'
accurately the eigenenergies obtained in large bases using the results from much smaller calculations.

We anticipate that the suggested SS-HORSE method will be useful in numerous shell model studies of low-energy nuclear resonances.

We plan to extend 
 the SS-HORSE approach to the case of scattering of charged particles in  future publications. We intend also
to examine an application of the SS-HORSE method to the study of $S$-matrix poles corresponding to bound states and to develop
the SS-HORSE extrapolation of the variational bound state energies to the infinite basis space.\\

\subsection*{Acknowledgements}

We are thankful to L.~D.~Blokhintsev and Pieter Maris for valuable discussions. 
This work was supported in part
by the U.S. Department of Energy
 under grants No.~DESC0008485 (SciDAC/NUCLEI) and DE-FG02-87ER40371.
The development  and application of the SS-HORSE approach was supported by the Russian Science Foundation under
project No.~16-12-10048. Computational resources
were provided by the National Energy Research Scientific Computing Center (NERSC)  
which is supported by the U.S. Department of Energy under Contract No.~DE-AC02-05CH11231.


\begin{thebibliography}{33}


\bibitem{VMS} P. Maris, J. P. Vary and A. M. Shirokov, 
Phys. Rev. C {\bf 79}, 014308 (2009).

\bibitem{Coon} S. A. Coon, M. I. Avetian, M. K. G. Kruse, U. van Kolck, P. Maris and J.~P.~Vary,
Phys. Rev. C {\bf 86}, 
054002 (2012).

\bibitem{CoonNTSE12}  S. A. Coon,
  in {\em Proc. Int. Workshop Nucl. Theor. Supercomputing Era (NTSE-2012), Khabarovsk, Russia, June 18--22, 2012}, eds.
  A. M. Shirokov and A.~I.~Mazur. Pacific National University, Khabarovsk, 2013,
 p.~171, 
\url{http://www.ntse-2012.khb.ru/Proc/S_Coon.pdf}.

\bibitem{Dick} R. J. Furnstahl, G. Hagen and T. Papenbrock, Phys. Rev. C {\bf 86}, 031301 
 (2012).

\bibitem{More} S. N. More, A. Ekstr\"om, R. J. Furnstahl, G. Hagen and T. Papenbrock,
Phys. Rev. C {\bf 87}, 044326 (2013).

\bibitem{CoonNTSE13} S. A. Coon and M. K. G. Kruse, 
 in {\em Proc. Int. Conf. Nucl. Theor. Supercomputing Era (NTSE-2013), Ames, IA, USA, May 13--17, 2013}, eds.
  A. M. Shirokov and A.~I.~Mazur. Pacific National University, Khabarovsk, 2014,
 p.~314, \url{http://www.ntse-2013.khb.ru/Proc/Coon.pdf}.

\bibitem{IT-extrap} M. K. G. Kruse, E. D. Jurgenson, P. Navr\'atil, B.~R.~Barrett and W. E. Ormand, Phys. Rev. C {\bf 87},  044301 (2013).

\bibitem{Forssen2014}  D. S\"a\"af and C. Forss\'en, Phys. Rev. C {\bf 89}, 011303(R) (2014).

\bibitem{Furnstahl-S-matrix} R. J. Furnstahl, S. N. More and T. Papenbrock, Phys. Rev. C {\bf 89},  044301 (2014).

\bibitem{Furnstahl14} S. K\"onig, S. K. Bogner, R. J. Furnstahl, S. N. More and T. Papenbrock, Phys. Rev. C {\bf 90}, 064007 (2014).

\bibitem{Furnstahl15} R. J. Furnstahl, G. Hagen, T. Papenbrock and
K.~A.~Wendt, J. Phys. G {\bf 42},  034032 (2015).

\bibitem{Furnstahl-HH} K.	 A. Wendt, C. Forss\'en, T. Papenbrock and D. S\"a\"af, Phys. Rev. C {\bf 91}, 061301(R)  (2015). 

{\colour{red}\bibitem{Sid16} S. A. Coon and M. K. G. Kruse, Int. J. Mod. Phys. E {\bf 25}, 1641011 (2016).}

\bibitem{PRC79} A. M. Shirokov, A. I. Mazur, J. P. Vary and E. A. Mazur, 
        Phys. Rev. C {\bf 79}, 014610 (2009).

\bibitem{ApplMathInfSci} A. M. Shirokov, A. I. Mazur, E. A. Mazur and J. P. Vary, Appl. Math. Inf. Sci. {\bf 3}, 245 (2009).

\bibitem{JimmyNTSE13} J. Rotureau, in {\em Proc. Int. Conf. Nucl. Theor. Supercomputing Era (NTSE-2013), Ames, IA, USA, May 13--17, 2013}.  eds.
  A. M. Shirokov and A.~I.~Mazur. Pacific National University, Khabarovsk, 2014,
 p.~236, 
 \url{http://www.ntse-2013.khb.ru/Proc/Rotureau.pdf}.
 
\bibitem{Jimmy} G. Papadimitriou, J. Rotureau, N.~Michel, M.~P{\l}oszajczak and B. R. Barrett, 
Phys. Rev. C {\bf 88}, 044318 (2013).

\bibitem{NavrJPG09} P. Navr\'atil, S. Quaglioni, I. Stetcu and B.~R.~Barrett, 
J. Phys. G {\bf 36},  083101 (2009). 

\bibitem{NavratilNTSE13}  P. Navr\'atil, in {\em Proc. Int. Conf. Nucl. Theor. Supercomputing Era (NTSE-2013), Ames, IA, USA, May 13--17, 2013},  eds.
  A. M. Shirokov and A.~I.~Mazur. Pacific National University, Khabarovsk, 2014, p.~211, 
  \url{http://www.ntse-2013.khb.ru/Proc/Navratil.pdf}.


\bibitem{QuagNav1} S. Quaglioni, P. Navr\'atil, G. Hupin, J.~Langhammer, C.~Romero-Redondo and  R. Roth, 
Few-body Syst. {\bf 54}, 877 (2013).

\bibitem{Vary2013} B.~R.~Barrett, P.~Navr\'atil and J.~P.~Vary, 
Progr. Part. Nucl. Phys {\bf 69}, 131 (2013). 


\bibitem{QuagNav2} G. Hupin, S. Quaglioni and P. Navr\'atil, 
Phys. Rev. C {\bf 90}, 061601(R) (2014).

{\colour{red}\bibitem{PhysScr} P.~Navr\'ati, S.~Quaglioni, G.~Hupin, C.~Romero-Redondo and A.~Calci, Phys. Scr. {\bf 91},
053002 (2016).}

\bibitem{Hell} E. J. Heller and H. A. Yamany, Phys. Rev. A {\bf 9}, 1201 (1974); Phys. Rev. A {\bf 9},
 1209 (1974).

\bibitem{YaFi} H. A. Yamany and L. Fishman.  J. Math. Phys. {\bf 16}, 410 (1975).


\bibitem{Fill} G. F.  Filippov and I. P. Okhrimenko, Yad. Fiz. {\bf 32}, 932 (1980) [Sov. J. Nucl. Phys. {\bf 32}, 480 (1980)]; 
G. F. Filippov, Yad. Fiz. {\bf 33}, 928 (1981) [Sov. J. Nucl. Phys. {\bf 33}, 488 (1981)]. 

\bibitem{NeSm}
Yu.~F.~Smirnov and Yu.~I.~Nechaev, Kinam {\bf 4}, 445 (1982); 
Yu.~I.~Nechaev and Yu.~F.~Smirnov, Yad. Fiz. {\bf 35}, 1385 (1982) [Sov. J.
Nucl. Phys. {\bf 35}, 808 (1982)].

\bibitem{M1} V.~A.~Knyr, A.~I.~Mazur and Yu.~F.~Smirnov, Yad. Fiz. {\bf 52},
754 (1990) [Sov. J. Nucl. Phys. {\bf 52}, 483 (1990)];
Yad. Fiz. {\bf 54}, 1518 (1991) [Sov. J. Nucl. Phys. {\bf 54}, 927 (1991)];
Yad. Fiz. {\bf 56}(10), 72 (1993) [Phys. At. Nucl. {\bf 56}, 1342 (1993)]; 
A.~I.~Mazur and S.~A.~Zaytsev, Yad. Fiz. {\bf 62}, 656 (1999) [Phys. At. Nucl. {\bf 62}, 
608 
(1999)]. 

\bibitem{Bang}
J. M. Bang, A.~I.~Mazur, A.~M.~Shirokov, Yu.~F.~Smirnov and
S.~A.~Zaytsev, Ann. Phys. (NY) {\bf 280}, 299 (2000).

\bibitem{Rub1}
V. I. Kukulin, V. N. Pomerantsev and O.~A.~Rubtsova, JETP Lett. {\bf 90}, 402 (2009);
O. A. Rubtsova, V.~I.~Kukulin, V. N. Pomerantsev and  A. Faessler, Phys. Rev. C {\bf 81}, 064003 (2010).


\bibitem{Lif}  I. M. Lifshitz, Zh. Eksp. Teor. Fiz. {\bf 17}, 1017 (1947) ({\em in Russian}); {\em ibid.}
 {\bf 17}, 1076 (1947) ({\em in Russian}); 
Usp. Mat. Nauk {\bf 7}, 171 (1952) ({\em in Russian}).

\bibitem{Luu}   T. Luu, M. J. Savage, A. Schwenk and J. P. Vary, 
Phys. Rev. C {\bf 82}, 034003 (2010).

\bibitem{Nollett} R. M. Nollett, Phys. Rev. C {\bf 86}, 044330 (2012).

{\colour{red}\bibitem{Timofeyuk} N. K. Timofeyuk, Phys. Rev. C {\bf 92}, 034330 (2015).}

\bibitem{PLB644} A. M. Shirokov, J. P. Vary, A. I. Mazur and T. A. Weber,
Phys. Lett. B  {\bf 644},  
33 (2007).

\bibitem{JISP16-web} A Fortran code generating the  JISP16 matrix elements  is available
         at \url{http://nuclear.physics.iastate.edu}.

\bibitem{RP2013} A. I. Mazur, A. M. Shirokov, J. P. Vary, P. Maris and I.~A.~Mazur,
 in {\em Proc. Int. Workshop Nucl. Theor. Supercomputing Era (NTSE-2012), Khabarovsk, Russia, June 18--22, 2012}, eds.
  A. M. Shirokov and A.~I.~Mazur. Pacific National University, Khabarovsk, 2013,
 p.~146,\linebreak 
\url{http://www.ntse-2012.khb.ru/Proc/A_Mazur.pdf}.

{\colour{red}
\bibitem{NNandNNN} A.~M.~Shirokov, V.~A.~Kulikov, P.~Maris and J.~P.~Vary, in {\em $NN$ and $3N$ Interactions,  Chap.~8}, eds. 
L.~D.`Blokhintsev and I.~I.~Strakovsky. Nova Science, Hauppauge, NY (2014), p. 231, see 
\url{https://www.novapublishers.com/catalog/product_} \url{info.php?products_id= 50945}.

\bibitem{QMC} K. M. Nollett, S.~C.~Pieper,  R.~B.~Wiringa, J.~Carlson and G.~M.~Hale, Phys. Rev. Lett. {\bf 99}, 022502 (2007).

\bibitem{NavQua09} S. Quaglioni and P. Navr\'atil, Phys. Rev. C {\bf 79}, 044606 (2009).

\bibitem{NavQua10} P.~Navr\'atil,  R.~Roth and S.~Quaglioni, Phys. Rev. C {\bf 82}, 034609 (2010).

\bibitem{NavQua13} G.~Hupin. J.~Langhammer, P.~Navr\'atil,   S.~Quaglioni, A.~Calci and R.~Roth, Phys. Rev. C {\bf 88}, 054622 (2013).
}

\bibitem{Arab} A. D. Alhaidari, E. J. Heller, H. A. Yamani and M. S. Abdelmonem (eds.), 
{\em The $J$-matrix method. Developments and applications.} Springer, 2008.

\bibitem{true} A. M. Shirokov, Yu. F. Smirnov
and S. A. Zaytsev, in {\em Modern problems in quantum theory},  
eds. V.~I.~Savrin and O.~A.~Khrustalev.
Moscow, 1998, p.~184. 
\bibitem{trueZ} S.~A.~Zaytsev, Yu.~F.~Smirnov and A.~M.~Shirokov, 
Teoret. Mat. Fiz. {\bf 117}, 227 (1998) [Theor. Math. Phys. {\bf 117}, 1291 (1998)].


\bibitem{PRC04} A. M. Shirokov, A. I. Mazur, S. A. Zaytsev, J. P. Vary and  T. A. Weber,
        Phys. Rev. C {\bf 70}, 
        044005 (2004);   
in {\em  The $J$-matrix method. Developments and applications,} eds. A.~D.~Alhaidari, E.~J.~Heller, H.~A.~Yamani and M.~S.~Abdelmonem.
Springer, 2008, p.~219.


\bibitem{Baz} A. I. Baz', Ya. B. Zel'dovich and A. M. Perelomov, {\em Scattering, reactions and decay  in non-relativistic
quantum mechanics.} Israel Program for Scientific Translation, Jerusalem, 1969.

\bibitem{Taylor}  R. G. Newton, {\em Scattering theory of waves and
particles, 2nd. ed.} Springer-Verlag, New York, 1982.


\bibitem{WSB} J. Bang and C. Gignoux, Nucl. Phys. A {\bf 313}, 119 (1979).


\bibitem{mamazfdn1} P.~Maris, M.~Sosonkina, J.~P.~Vary, E.~G.~Ng and  C.~Yang, Proc.~Comput.~Sci. {\bf 1}, 97 (2010).

\bibitem{mamazfdn2} H.~M. Aktulga, C.~Yang, E.~G.~Ng, P.~Maris and J.~P.~Vary,
Concurrency Computat. Pract. Exper. {\bf 26}, 2631 (2014).


\bibitem{naExpB} J. E. Bond and F. W. K. Firk, Nucl. Phys. A {\bf 287}, 317 (1977).

\bibitem{Hale} { A.~Cs\'ot\'o and G.~M.~Hale}, 
Phys. Rev. C {\bf 55}, 536 (1997).


\bibitem{4n} A. M. Shirokov, G. Papadimitriou, A. I. Mazur, I.~A.~Mazur, R. Roth and J.~P.~Vary,
{\colour{red}Phys. Rev. Lett. {\bf 117}, 182502 (2016)}.


\end{thebibliography}
\end{document}